\documentclass[10pt,aps,prd,nofootinbib,superscriptaddress,twocolumn]{revtex4-1}
\usepackage[utf8]{inputenc}
\usepackage{geometry}
\usepackage[dvipsnames]{xcolor}

\usepackage{mathtools}
\usepackage{amsfonts}
\usepackage{dsfont}
\usepackage{CJKutf8}
\usepackage{mathrsfs}
\usepackage{bbm}
\usepackage[normalem]{ulem}
\usepackage{slashed}
\usepackage{tensor}

\usepackage{pifont}

\usepackage{adjustbox}

\usepackage{graphicx}
\usepackage{array}[=2016-10-06]
\usepackage{orcidlink}

\usepackage{placeins}
\usepackage{makecell}
\usepackage{xspace}
\usepackage{xfrac}
\usepackage{hyperref}
\usepackage[nameinlink]{cleveref}
\usepackage{appendix}
\usepackage{siunitx}

\Crefname{appsec}{Appendix}{Appendices}

\usepackage{booktabs}
\usepackage{multirow}

\usepackage[hyphenation]{impnattypo}
\usepackage[all]{nowidow}
\usepackage{microtype}

\usepackage{xifthen}
\usepackage{multirow}
\usepackage{acronym}
\hypersetup{
	colorlinks,
	linkcolor={red!75!black},
	citecolor={blue!75!black},
	urlcolor={blue!75!black},
	pdftitle={Diffeomorphism-invariant Approach to Asymptotically Safe Quantum Gravity},
	pdfauthor={Ihssen, Knorr, Mezger, Pawlowski, Sprenger},
}
\usepackage{enumerate}

\usepackage{tikz}
\usetikzlibrary{decorations.pathmorphing,calc}\usepackage{tikz}
\usetikzlibrary{decorations.pathmorphing,calc}

\graphicspath{{./figures/}}

\newcommand{\PIG}{\textrm{\begin{CJK}{UTF8}{gbsn}猪\end{CJK}}}

\newcommand{\gettitle}{Diffeomorphism-invariant Approach to Asymptotically Safe Quantum Gravity}

\begin{document}
	
	\title{\gettitle}

	\author{Friederike~Ihssen\,\orcidlink{0000-0002-1550-3423}}
	\affiliation{Theoretische Physik III, Ruhr-Universit{\"a}t Bochum, Universit{\"a}tsstraße 150, 44801 Bochum, Germany}
	\affiliation{Institut für Theoretische Physik, Universität Heidelberg, Philosophenweg 16, 69120 Heidelberg, Germany}
	\author{Benjamin~Knorr\,\orcidlink{0000-0001-6700-6501}}
	\affiliation{Institut für Theoretische Physik, Universität Heidelberg, Philosophenweg 16, 69120 Heidelberg, Germany}
	\author{Silas~Mezger\,\orcidlink{0009-0007-6845-071X}}
	\affiliation{Institute of Microstructure Technology, Karlsruhe Institute of Technology, Hermann-von-Helmholtzplatz 1, 76344 Eggenstein-Leopoldshafen, Germany}
	\affiliation{Institut für Theoretische Physik, Universität Heidelberg, Philosophenweg 16, 69120 Heidelberg, Germany}
	\author{Jan~M.~Pawlowski\,\orcidlink{0000-0003-0003-7180}}
	\affiliation{Institut für Theoretische Physik, Universität Heidelberg, Philosophenweg 16, 69120 Heidelberg, Germany}
	\affiliation{ExtreMe Matter Institute EMMI, GSI, Planckstr. 1, 64291 Darmstadt, Germany}
	\author{Paul~P.~Sprenger\,\orcidlink{0009-0008-7107-0504}}
	\affiliation{Institut für Theoretische Physik, Universität Heidelberg, Philosophenweg 16, 69120 Heidelberg, Germany}

%%%%%%%%%%%%%%%%	

\begin{abstract}

We provide a novel diffeomorphism-invariant approach to asymptotically safe metric quantum gravity. It is based on physics-informed renormalisation group flows with a renormalisation group kernel that guarantees diffeomorphism invariance at each renormalisation group step. Importantly, it also allows us to control and maintain the relevance counting of operators in metric quantum gravity. The resulting effective action is quantum diffeomorphism-invariant and background-independent. As a first non-trivial application, we compute the Reuter fixed point in a manifestly diffeomorphism-invariant way. This computation is augmented with a detailed discussion of regularisation dependence and a systematic estimate of the errors arising from approximations. 

\end{abstract}
%%%%%%%%%%%%%%%%%%%%%%%%
	
	\maketitle
	
%%%%%%%%%%%%%%%%%%%%
\section{Introduction}
\label{sec:Introduction}

Asymptotically safe quantum gravity \cite{Weinberg:1980gg, Reuter:1996cp} is a promising candidate for a consistent quantum theory of gravity. Within the functional renormalisation group 
framework, substantial evidence has been accumulated for the existence of a non-trivial ultraviolet (UV)
fixed point -- the Reuter fixed point -- across increasingly sophisticated truncations. For a comprehensive review on the functional renormalisation group and its applications see \cite{Dupuis:2020fhh}, for recent reviews on asymptotically safe quantum gravity see \cite{Bonanno:2020bil, Pawlowski:2020qer, Reichert:2020mja, Basile:2024oms, Eichhorn:2026uqj} and the handbook of quantum gravity \cite{Bambi:2023jiz}, chapters \cite{Knorr:2022dsx, Eichhorn:2022gku, Morris:2022btf, Wetterich:2022ncl, Martini:2022sll, Saueressig:2023irs, Pawlowski:2023gym, Platania:2023srt, Bonanno:2024xne}.

The method of choice for the resolution of asymptotically safe gravity is the functional renormalisation group (fRG). As a renormalisation group method, it is tailor-made for resolving the scaling properties of asymptotic safety close to its ultraviolet fixed point. Its versatility also allows us to bridge the many orders of magnitude in momenta between the ultraviolet scaling regime beyond the Planck scale and the experimentally accessible infrared regime ruled by general relativity and small quantum fluctuations of the metric. To date, these enormous advantages are paid for with the necessity to gauge-fix, the requirement of a background metric, and the breaking of diffeomorphism invariance due to the infrared regularisation of the theory. 

There are diffeomorphism-invariant and background-independent formulations of the functional renormalisation group approach such as the commonly used background-field approximation \cite{Reuter:1996cp}, proper time flows, e.g.~\cite{Bonanno:2004sy, Bonanno:2019ukb}, or the simplified flow equation \cite{Wetterich:2024ivi}. At their core, these formulations obtain diffeomorphism invariance and background independence by elevating the background metric in the regulator and the gauge-fixing action into a dynamical field. 
This elevates an auxiliary symmetry (background diffeomorphism invariance) to a dynamical one. Such a procedure changes the dynamics of the theory, and the respective functional obtained from the flow ceases to be the 1PI effective action. Moreover, the relation to the latter is unclear, as is the relation of the auxiliary, now dynamical, background diffeomorphism invariance to physical diffeomorphism invariance. For a comprehensive discussion see \cite{Pawlowski:2020qer}. Another unwelcome by-product of these seemingly diffeomorphism-invariant approximations is the potential change of the relevance counting of operators at the ultraviolet fixed point. This can not only change the number of relevant operators and hence relevant couplings at the UV fixed point, but it also triggers qualitative changes of the matter-gravity dynamics, see e.g.~\cite{Meibohm:2015twa, Pawlowski:2020qer}. 

In the present work, we resolve these intricacies within the \textit{physics-informed renormalisation group} (PIRG) \cite{Ihssen:2023nqd, Ihssen:2024ihp, Ihssen:2025cff, Ihssen:2025hyl}. The PIRG is based on the generalised flow equation for the effective action \cite{Pawlowski:2005xe} which also underlies the \textit{essential RG} \cite{Baldazzi:2021ydj, Knorr:2022ilz, Knorr:2023usb, Baldazzi:2023pep, Knorr:2024yiu, Falls:2025sxu, Ohta:2025xxo}. The generalised flow equation can be obtained from a Legendre transform of the Wegner equation \cite{Wegner_1974}, see e.g.~\cite{Ihssen:2022xjv}, and accommodates general reparametrisations of the microscopic fields, see e.g.~\cite{Wetterich:2024uub}. 

In the PIRG framework, the standard flow with a (covariant) momentum scale is augmented with a general flowing reparametrisation of the theory via the \textit{flowing field} or \textit{physics-informed kernel} (PIK). This reparametrisation can be chosen such that the flow steps are diffeomorphism-invariant at each step \cite{Ihssen:2025cff}. We use the PIRG framework for gauge theories \cite{Ihssen:2025cff} to construct a fully diffeomorphism-invariant and background-independent setup to investigate asymptotically safe gravity. 

Furthermore, the present approach allows us to control the above-mentioned relevance counting of operators 
within a diffeomorphism-invariant setting. In short, it allows us to use the background-field approach and powerful heat-kernel methods while keeping the correct dynamics and number of relevant couplings. 
This important property allows us, for the first time, to construct a diffeomorphism-invariant flow with the correct relevance counting of metric quantum gravity. 

We demonstrate the potential of this setup within the Einstein-Hilbert approximation. 
The explicit results provide a diffeomorphism-invariant Reuter fixed point with the correct relevance counting of the volume operator $\sqrt{g}$ and the curvature scalar $\sqrt{g} R$. 
These computations also illustrate the practical feasibility of the approach.

We close the introduction with a bird's-eye view of the work. In \Cref{sec:DiffInvariantFlows} we set up the background-independent diffeomorphism-invariant functional renormalisation group approach within the framework of the physics-informed renormalisation group. A chiefly important aspect of this approach is the control of the relevance counting of diffeomorphism-invariant operators, which allows us to maintain the relevance counting of metric quantum gravity. In \Cref{sec:DiffinFlowsRelevance} and \Cref{sec:Enforce-RP-min} we 
construct diffeomorphism-invariant PIRG flows for effective actions with the same relevance counting of quantum gravity. In \Cref{sec:DiffPIRGatWork} we discuss the practical implementation of the approach. We show in particular that the approach allows for the use of powerful heat-kernel techniques. Finally, in \Cref{sec:Results} we use the novel approach for a first computation of the diffeomorphism-invariant Reuter fixed point and the phase structure of asymptotically safe gravity. In \Cref{sec:Conclusions} we briefly summarise our results and outline further steps in this promising approach. We have deferred many technical details as well as conceptual discussions and results to the Appendices to provide concise access to the main properties and results.

%%%%%%%%%%%%%%%%%%%%
\section{Diffeomorphism-invariant flows}
\label{sec:DiffInvariantFlows}

The standard functional renormalisation group approach to asymptotically safe gravity is constructed from a few basic components. To begin with, it builds upon the metric $\hat g_{\mu\nu}$, where the hat indicates variables integrated over in the path integral. The second component is the quantum field theoretical setting via the (general) fRG flow for the effective action $\Gamma$ of metric gravity. The third, pivotal ingredient of general flows is the propagator of the dynamical degree of freedom, whose definition requires a gauge fixing or explicit coordinate frame in field space. This enforces the introduction of a metric background $\bar g_{\mu\nu}$ and a fluctuation field $\hat h_{\mu\nu}$. For now, we restrict ourselves to a linear split, 
\begin{align}
	\hat g_{\mu\nu} = \bar g_{\mu\nu} +\sqrt{32 \pi Z_h \,G_{N}}\, \hat h_{\mu\nu}\, , 
	\label{eq:linSplit}
\end{align}
where $\sqrt{Z_h G_N}$ gives the fluctuation fields mass dimension one and expands the metric with a renormalisation group invariant fluctuation field $Z_h^{1/2} h$. This ensures the correct RG-scaling of the field operator $\hat g_{\mu\nu}$. Moreover, the scaling of the respective two-point function $\langle \hat h(x) \hat h(y)\rangle$ is given by the anomalous dimension $\eta_h$ of the fluctuation field. The rescaling 	\labelcref{eq:linSplit} also relates to the standard expansion in powers of the Newton coupling, matching that in the fluctuation field, $h^2$ with $G_N^0$ and $h^n$ with $G_N^{(n-2)/2}$, see e.g.~\cite{Pawlowski:2020qer, Pawlowski:2023gym}. General fluctuation fields and splits will be considered in \Cref{sec:PIGs}.

Common choices for the gauge fixing are linear in the fluctuation field, 
\begin{align}
	F_\mu[\bar g, \hat h] =
	\bar{\nabla}^\nu \hat h_{\mu \nu} -\frac{1+ \beta}{4} \bar{\nabla}_\mu \hat h^{\nu}_{~\nu} \, , 
	\label{eq:gfCovariant}
\end{align}
where $\beta$ is a gauge-fixing parameter. Details of the gauge fixing procedure are deferred to \Cref{app:Approximation+GaugeFixing}. The complete gauge fixing sector is a combination of the gauge fixing term \labelcref{eq:gfGeneral} and a Faddeev-Popov term \labelcref{eq:Sghost}, 
\begin{align} 
	S_\textrm{gauge}[\bar g, \bar g+\hat h,\hat c,\hat{\bar c}]=S_{\text{gf}}[\bar g, \hat h]+S_{\text{gh}}[\bar g, \hat h,\hat c,\hat{\bar c}]\,, 
	\label{eq:Sgauge}
\end{align}
see also \labelcref{eq:SgaugeApp}. The quantum effective action $\Gamma$ depends on the expectation values of fields, which we denote without a hat.
Importantly, it depends on both metrics individually, i.e., the background metric $\bar g_{\mu\nu}$ and the full metric $g_{\mu\nu}(\bar g,h)$. Nevertheless, observables derived from it are gauge- and background-independent.

%%%%%%%%%%%%%%%%%%%%%%%%%%%%%
\subsection{Functional flow in quantum gravity}
\label{sec:FunFlowsGravity}
	
We now discuss the quantisation of metric gravity within the fRG approach. It is based on infrared cutoff terms for the dynamical fields $\hat\Phi$ with 
\begin{align} 
 \hat\Phi =\left(\hat h_{\mu\nu}\,,\,\hat c_\mu\,,\,\hat{\bar c}_\mu\right)\,,\qquad  \Phi=\langle \hat\Phi\rangle\,. 
	\label{eq:Phi}
\end{align} 
$\Phi$ is the mean superfield. In what follows, we will consider the one-particle irreducible effective action to depend on the background metric and the full field, $\Gamma[\bar g,\bar g+\Phi]$, where the second argument is to be read as the dependence on the full fields. In other words, we explicitly indicate the artificial dependence on the background metric in the first argument. This allows for a more concise discussion of the isolated $\bar g$-dependences.

The fRG approach typically uses infrared cutoff terms that are quadratic in the fluctuation fields $\Phi$ irrespective of the nature of the split, be it linear or non-linear,
\begin{align} 
	\Delta S_k[\bar g,\Phi]= \frac12 \int {\rm{d}}^4 x \,\Phi \cdot R_k[\bar g] \cdot \Phi \, . 
	\label{eq:RegTerm}
\end{align} 
In this, $R_k$ is the infrared regulator. Note that in contrast to the standard literature, we have stored the measure $\sqrt{\bar g}$ of the space-time integration in the regulator matrix $R_{k}$, which is thus a density. 
The quadratic constraint is in one-to-one correspondence with the one-loop exact form of the functional flow equation for the scale-dependent effective action $\Gamma_k[\bar g,\bar g+\Phi]$, the Wetterich equation~\cite{Wetterich:1992yh, Ellwanger:1993mw, Morris:1993qb} for quantum gravity \cite{Reuter:1996cp}, 
\begin{subequations} 
\label{eq:FunFlow} 
\begin{align}
	\partial_t \Gamma_k[\bar g,\bar g+ \Phi] = \frac{1}{2} \text{Tr} \,G_k[\bar g,\bar g+\Phi]\,\partial_t R_k[\bar g]  \, .
	\label{eq:wetterichEq}
\end{align}
Here, $t=\ln k/k_0$ with $k_0$ being an arbitrary reference scale, $G_k$ is the propagator of the fluctuation fields, 
\begin{align} 
	G_k[\bar g,\bar g+\Phi] = \frac{1}{\Gamma_k^{(0,2)}[\bar g, \bar g+\Phi]  + R_k[\bar g]} \,,
\label{eq:PropFluc} 
\end{align}
\end{subequations}
and the trace sums over space-time or momenta, Lorentz indices, internal indices and species of fields. The second-order field derivatives in \labelcref{eq:wetterichEq} are those with respect to the fluctuation field, and we use the general notation 
\begin{align}
	\Gamma_k^{(m,n)}[x,y] = \frac{\delta^{m+n} \Gamma_k[x,y]}{ \delta x^m \delta y^n} \, ,
\label{eq:DerivativesGamma}
\end{align}
for mixed derivatives. In particular, a $\bar g$-derivative is to be understood as acting on the artificial background dependence only, not on the combined dependence on $\bar g+\Phi$. Note again that in our convention, these derivatives $\Gamma^{(m,n)}$ contain the factor $\sqrt{\bar g}$ and are thus densities. The cutoff terms source yet another separate dependence on $\bar g$. Consequently, the effective action has a genuine separate dependence on $\bar g$ and $\bar g+\Phi$. For all cutoff scales, it satisfies modified Slavnov-Taylor identities (mSTIs) and Nielsen identities (mNIs) that flow into the standard ones~\cite{Nielsen:1975fs,Fukuda:1975di} for $k\to 0$. The mSTIs are obtained from BRST-transformations of the fluctuation fields that encode \textit{physical} diffeomorphism invariance and its breaking through the cutoff terms. The mNIs encode the difference of $\bar g$- and $\Phi$-derivatives. 
Their general structure is given by \cite{Pawlowski:2020qer}, 
\begin{align}  \nonumber 
\Gamma_k^{(1,0)}[\bar g,\bar g+\Phi]=  & \, \left \langle  S^{(1,0)}_\textrm{gauge}[\bar g,\bar g+\Phi]\right\rangle\\[1ex]
&\hspace{1cm}+ \left \langle  \Delta S^{(1,0)}_k[\bar g,\bar g+\Phi]\right\rangle\, .
	\label{eq:mNI}
\end{align}
The mNIs encode the fact that the dependence on the isolated background metric only comes from the necessity of introducing a background metric to the gauge fixing sector and the regulator term. In turn, the classical or fixed point action of metric quantum gravity is diffeomorphism-invariant and background-independent and hence only depends on $\bar g+\Phi$. Moreover, the contribution from the regulator vanishes in the physical limit $k\to 0$ and the mNIs turn into NIs. The contribution from the gauge fixing vanishes on-shell, i.e., on the solution of the equations of motion. \Cref{eq:mNI} allows us to map fluctuation field derivatives to background metric ones plus the expectation values in \labelcref{eq:mNI}. Moreover, in combination, the STIs and NIs guarantee diffeomorphism invariance and background independence of the setup. 

The flow equation \labelcref{eq:FunFlow} for $\Gamma_k[\bar g, \bar g+\Phi]$ with its dynamical fluctuation field $\Phi$ (in the second argument) and the isolated background metric dependence (first argument) defines metric quantum gravity. In particular, the relevance counting of operators in metric quantum gravity is encoded in that of the correlation functions of the dynamical fluctuation field $\Phi$. Moreover, correlation functions of the background metric are not independent but follow from those of the fluctuation field. Accordingly, any approach to metric quantum gravity has to explicitly or implicitly encode this relevance counting. In short, they will serve as the baseline of our background-independent and diffeomorphism-invariant approach. The Euclidean and more recently also Minkowski flows of the correlation functions of $\Phi$ have been studied intensely over the past one and a half decades in the bimetric approach \cite{Manrique:2009uh, Manrique:2010mq, Manrique:2010am, Becker:2014qya} and mainly in the fluctuation approach~\cite{Christiansen:2012rx, Codello:2013fpa, Christiansen:2014raa, Christiansen:2015rva, Meibohm:2015twa, Meibohm:2016mkp, Denz:2016qks, Christiansen:2017cxa, Christiansen:2017bsy, Knorr:2017fus, Knorr:2017mhu, Eichhorn:2018akn, Eichhorn:2018ydy, Eichhorn:2018nda, Burger:2019upn, Bonanno:2021squ, Knorr:2021niv, Fehre:2021eob, Pastor-Gutierrez:2022nki, Saueressig:2023tfy, Korver:2024sam, Pastor-Gutierrez:2024sbt, Saueressig:2025ypi, Kher:2025rve, Pawlowski:2025etp, Chiesa:2026tlz, Knorr:2026jcg, Assant:2026dca}. For reviews see \cite{Pawlowski:2020qer, Pawlowski:2023gym}. 

The fluctuation approach also includes a background-independent and diffeomorphism-invariant effective action, 
the \textit{background} effective action, 
\begin{align} 
	\Gamma_k[g] = \Gamma_k[\bar g=g,\bar g+\Phi=g] \,. 
	\label{eq:BackEffAct}
\end{align} 
For all cutoff scales $k$, it is diffeomorphism-invariant, and at $k=0$ it is background-independent. More details on these core relations can be found in \cite{Pawlowski:2020qer, Pawlowski:2023gym}. 

The flow of $\Gamma_k[g]$ is given by \labelcref{eq:wetterichEq} evaluated for $\Phi=0$,
\begin{align}
	\partial_t\Gamma_k[g] = \frac{1}{2} \text{Tr} \, G_k[g,g]\, \partial_t R_k[g]  \, . 
	\label{eq:Backflow}
\end{align}
 This equation is manifestly diffeomorphism-invariant, but it is \emph{not} closed: it depends on the fluctuation propagator $G_k[g,g]$, which cannot be obtained directly from $\Gamma_k[g]$, but can be computed from the Nielsen identities. In practice, however, solving the triple of flow equation, modified Nielsen identity and modified Slavnov-Taylor identity in parallel is a challenging task in any given approximation. In non-perturbative approximations, this is generically not possible.  

The common choice for an approximate solution is the \textit{background-field approximation}: one \emph{assumes} that the dynamical physics or quantum part $\Gamma_{\textrm{qu},k}$ of the effective action, or rather its second derivative, is a function of the full metric $g$. This implies that the separate dependence on $\bar g$ is restricted to the ``classical'' gauge-fixing and regulator part. In its minimal form, this assumption applies to the two-point function of the fluctuation field $\Phi$ at $\Phi=0$ which enters the background flow \labelcref{eq:Backflow} via the fluctuation propagator. We thus require 
\begin{align}
	\Gamma_k^{(0,2)}[{g},g] \approx \Gamma_{\textrm{qu},k}^{(2)}[g] +S_{\mathrm{gauge}}^{(0,2)}[g,g] \, ,
\label{eq:CoreBackgroundApprox}
\end{align}
which is the core relation in the background-field approximation. For the sake of a concise presentation, we suppressed the $c_\mu,\bar c_\mu$-dependence in \labelcref{eq:CoreBackgroundApprox}.  This relation implies a trivial Nielsen identity since $\Gamma_{\textrm{qu},k}$ is a one-field functional, and it ignores its violation via the gauge fixing and regulator term on the right-hand side of \labelcref{eq:mNI}. It also elevates the modified Slavnov-Taylor identities to Ward identities of diffeomorphism invariance. 

The core relation \labelcref{eq:CoreBackgroundApprox} of the background-field approximation leads us to a diffeomorphism-invariant, background-independent and \textit{closed} flow equation, 
\begin{align}
	\partial_t &\Gamma_{\textrm{qu},k}[g] \approx \nonumber \\ & \quad \frac{1}{2} \mathrm{Tr}\,\frac{1}{\Gamma_{\textrm{qu},k}^{(2)}[g] + S_\mathrm{gauge}^{(0,2)}[g,g] + R_k[g]}\partial_t{R}_k[g] .
\label{eq:Flow-BFA}
\end{align}
In contradistinction to \labelcref{eq:Backflow} it is closed, since the background-field approximation \labelcref{eq:CoreBackgroundApprox} relates fluctuation and background-field derivatives. Moreover, the one-field nature of \labelcref{eq:Flow-BFA} allows us to use powerful heat kernel techniques~\cite{Vassilevich:2003xt} for its resolution in non-trivial approximations, including those with a general dependence on curvature invariants such as the Ricci scalar $R$, or an expansion in metric derivatives. Due to its diffeomorphism invariance and technical accessibility, the background-field approximation is widely used in asymptotically safe gravity, for a selection see~\cite{Reuter:1996cp, Reuter:2001ag, Lauscher:2002sq, Machado:2007ea, Benedetti:2010nr, Falls:2014tra, Demmel:2015oqa, Gies:2016con, Falls:2017lst, Knorr:2021slg, Kluth:2022vnq, Knorr:2023usb, Baldazzi:2023pep}. 

However, its intriguingly simple structure comes at a high price: \Cref{eq:Flow-BFA} makes it explicit that the trivial NIs and STIs are achieved by elevating the dependence on the background metric $\bar{g}$ in the gauge-fixing part $S^{(0,2)}_\textrm{gauge}[\bar g,\bar g]$ and in the regulator $R_k[\bar g]$ to one on the full dynamical metric, $\bar g\to g$.  Hence, the dynamical fluctuation field is introduced into both, leading to further (relevant) vertices that are not present originally. It has been shown that this approximation (or more specifically, the additional vertices) triggers \textit{qualitative} failures. In Yang-Mills theory, even the one-loop $\beta$-function in the background-field approximation is regulator-dependent and one-loop universality is lost~\cite{Litim:2002ce}. More generally, one can show that the approximation impacts the relevance counting of operators \textit{qualitatively}~\cite{Litim:2002ce, Litim:2002xm, Litim:2002hj}. For a discussion of these issues in asymptotically safe matter-gravity systems, see \cite{Folkerts:2011jz, Meibohm:2015twa}. An even simpler example is given by the Wilson-Fisher fixed point of the Ising model in three dimensions. In the background-field approximation, it can be made to disappear \cite{Bridle:2013sra}. For more details on these aspects of the background-field approximation and the importance of resolving the dynamics of the fluctuation field, we refer to \cite{Pawlowski:2020qer,Pawlowski:2023gym}.

%%%%%%%%%%%%%%%%%%%%%%
\subsection{Physics-informed quantum gravity}	
\label{sec:PIGs}

The analysis of the last section leaves us with a predicament: trying to reinstate diffeomorphism invariance and background independence via an approximation such as \labelcref{eq:CoreBackgroundApprox} is seemingly successful but actually violates both. Moreover, it comes with the additional price of changing the relevance counting of operators.

A way out is the geometric, or Vilkovisky-DeWitt, approach to quantum field theories, more specifically gauge theories~\cite{Vilkovisky:1984st,DeWitt:1985sg}. There, a reparametrisation-invariant path integral is defined by coupling the current to a geometric field, instead of the fundamental field. For a discussion within the fRG approach see \cite{Branchina:2003ek, Pawlowski:2003sk, Pawlowski:2005xe, Donkin:2012ud}. In gravity, this approach comes with non-localities which are difficult to accommodate, for the related relational and dressed approaches see e.g.~\cite{Falls:2025tid, Aguilar-Gutierrez:2026svf}. 

General reparametrisations of quantum gravity are realised with the generalised flow equation \cite{Pawlowski:2005xe}. It includes the geometrical approach as a specific case, but it also allows for more general scale-dependent reparametrisations. The field composites are collected into a composite superfield 
\begin{align}
	\hat\Phi = \left\{ \hat\Phi_{h}, \hat \Phi_{c}, \hat\Phi_{{\bar c}}	\right\} \, , \qquad \hat\Phi_i=\hat\Phi_i[\hat h,\hat c,\hat{\bar c}]\,, 
\label{eq:CompositehatPhi}
\end{align}
where it is understood that the composite depends on the cutoff scale $k$. The mean of the composite field is then defined as  
\begin{align}
	\Phi = \langle \hat\Phi \rangle \equiv \left\{ \Phi_{h}, \Phi_{c}, \Phi_{\bar c}	\right\} \, , 
\label{eq:CompositeMeanField}
\end{align}
and it is the independent variable (i.e., it carries no $k$-dependence) of the effective action $\Gamma_\Phi[\bar g,\bar g+\Phi]$. We emphasise that the effective action $\Gamma_\Phi$ is 1PI with respect to the composite $\Phi$, and in general does not agree with the 1PI effective action $\Gamma_k$ of the fundamental fields \cite{Ihssen:2024ihp}.

\begin{figure}[t]
	\centering
	\includegraphics[width=0.45\textwidth]{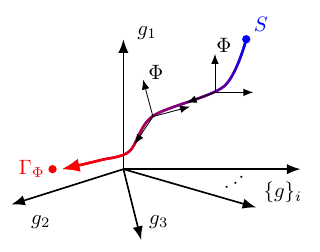}
	\caption{The flow of the action functional $\Gamma_\Phi$ is shown in theory space from the ultraviolet to the infrared. In the PIRG framework, the microscopic field $\hat\Phi$ transforms in a $k$-dependent manner. We use this field transformation, here visualised as a rotation in theory space, to reinstate gauge invariance and background independence.}		
	\label{fig:FlowingCoordinateSystem}
\end{figure}

All components of the composite carry a general dependence on the fundamental fields, and the subscript indicates the underlying fundamental field. The generalised flow equation for the effective action $\Gamma_{\Phi}$ is given by \cite{Pawlowski:2005xe}, 
\begin{subequations}
\label{eq:GenRG}
\begin{align}\nonumber 
	\bigg(\partial_t + \dot\Phi \frac{\delta}{\delta \Phi} \bigg) \Gamma_{\Phi}[\bar{g}, \bar g+\Phi] &\,   \\[1ex]
		&\hspace{-3.5cm}=\frac{1}{2} \textrm{Tr}\,\left[G_{\Phi}[\bar g,\bar g+\Phi]\bigg(\partial_t + 2 \frac{\delta \dot\Phi}{\delta \Phi} \bigg) R_k[\bar{g}] \right] \,,
		\label{eq:GenFlow}
\end{align}
with the propagator of the fluctuation composite 
\begin{align} 
	G_{\Phi}[\bar g,\bar g+\Phi] = \frac{1}{\Gamma_{\Phi}^{(0,2)}[\bar g, \bar g+\Phi]  + R_k[\bar g]} \, .
	\label{eq:PropFlucComposite} 
\end{align}
The functional $\dot\Phi$ encodes the change of the coordinate system in mean field space depicted in \Cref{fig:FlowingCoordinateSystem}. It is the expectation value of the flow of the superfield in operator space, 
\begin{align}
		\dot \Phi [\bar g,\bar g+\Phi] = \langle \partial_t \hat\Phi \rangle \,.
		\label{eq:dotPhi}
\end{align}
\end{subequations}
\Cref{eq:dotPhi} will be the key player in our setup below: it allows us to absorb the \textit{local} violations of diffeomorphism invariance and background independence that originate in the gauge fixing sector and the infrared cutoff term.

The constraints for $\dot\Phi$ are discussed in \cite{Ihssen:2024ihp} in the context of the recently put forward \textit{physics-informed renormalisation group} (PIRG). The PIRG is based on a simple but powerful change of perspective in the fRG approach. Instead of viewing \labelcref{eq:GenRG} as a flow equation for the effective action, we view it as a differential equation for the PIRG pair of the \textit{target action} $\Gamma_\Phi$ and the flowing field or \textit{physics-informed kernel} (PIK) $\dot\Phi$, 
\begin{align}
	\left( \Gamma_{\Phi}[\bar g,\bar g+\Phi]\,,\, \dot\Phi [\bar g, \bar g+\Phi] \right) \, .
\label{eq:PIRG-Pair}
\end{align}
Consequently, the physics of the system at hand is not only contained in the target action, but also in the PIK $\dot\Phi$. It is this shift in perspective that we will leverage in the next subsection to formulate a diffeomorphism-invariant flow.

It is chiefly important to be aware of the difference between the 1PI effective action of the fundamental field and a target action. The former is uniquely defined, and carries all the physics of the theory in its correlation functions. In turn, the latter is part of the PIRG pair \eqref{eq:PIRG-Pair}, and its physics content can vary between full physics and no physics -- the ``remainder'' being stored in the PIK $\dot \Phi$. For more details on the PIRG see \cite{Ihssen:2022xjv, Ihssen:2023nqd, Ihssen:2024ihp, Ihssen:2025cff, Ihssen:2025hyl, Bonanno:2025mon, Ihssen:2025ybn, Ihssen:2026njd}. The PIRG has a direct relation to the \textit{essential RG} developed and used in \cite{Baldazzi:2021ydj, Baldazzi:2021orb, Knorr:2022ilz, Knorr:2023usb, Baldazzi:2023pep, Knorr:2024yiu, Falls:2024noj, Ohta:2025xxo, Falls:2025sxu, Falls:2026nuh}. The latter can be understood as a special case of the former with reparametrisations $\dot\Phi$ which remove operators that vanish on-shell.

%%%%%%%%%%%%%%%%%%%%%%%%%
\subsection{Diffeomorphism-invariant flows}
\label{sec:DiffinFlows} 

In the present work, we leverage the versatility of the PIRG approach to derive a background-independent and diffeomorphism-invariant flow equation for an effective action $\Gamma_{\Phi}[\bar g+\Phi]$ which \emph{only} depends on the combination $\bar g+\Phi$. As a consequence, the modified Nielsen identity is trivial. This also allows us to make use of the powerful heat-kernel methods as in the background-field approximation. 

More specifically, we require the following three properties for the effective action $\Gamma_{\Phi}[\bar g+\Phi]$:  
\begin{itemize} 
\item[(1)] diffeomorphism invariance,
\item[(2)] background independence, 
\item[(3)] relevance counting of metric quantum gravity.
\end{itemize} 
For a first discussion of these properties, see \cite{Ihssen:2025cff}. Before we set up this approach, we add a few remarks. In its weakest form, diffeomorphism invariance is achieved for an effective action that satisfies the standard STIs rather than modified STIs. Here, we aim for full diffeomorphism invariance of the effective action for which the STIs reduce to Ward identities. This already implies that the effective action only depends on the combination $\bar g+\Phi$, and hence satisfies (2). 

Property (3) is key to an easy access to the physics of metric quantum gravity: there is only a small subset of all flows with (1) and (2) for which the effective action $\Gamma_\Phi$ carries the whole physics of metric quantum gravity, or at least the dominant or relevant part of it. Then, it is possible to view the respective effective action as an approximation to the effective action of the fundamental fields, $\Gamma_\Phi\approx \Gamma$. For all other flows, it is vital to also unearth the part of the physics buried in $\dot\Phi$, and one has to consider the PIRG pair. Property (3) defines a subset of the flows for which the relevance counting of operators is the same as that in metric quantum gravity, and the latter is defined by the relevance counting in the effective action $\Gamma[\bar g, \bar g+h,c,\bar c]$ and its flow. These flows will be called \textit{relevance-preserving}. How they can be constructed is discussed in \Cref{sec:DiffinFlowsRelevance}.

\begin{table*}[t]
	\centering
	
	\begingroup
	\renewcommand{\arraystretch}{1.5}
	\setlength{\tabcolsep}{14pt}
	
	\begin{tabular}{@{}l
			r@{\hspace{0.35em}}c@{\hspace{0.35em}}l
			r@{\hspace{0.35em}}c@{\hspace{0.35em}}l
			r@{\hspace{0.35em}}c@{\hspace{0.35em}}l@{}}
		\hline
		\textbf{Setup}
		& \multicolumn{3}{c}{\textbf{Field operators} $\hat{\Phi}$}
		& \multicolumn{3}{c}{\textbf{Mean fields} $\langle\hat{\Phi}\rangle$}
		& \multicolumn{3}{c}{\textbf{PIK} $\dot{\Phi}$}
		\\
		\hline
		
		Fundamental fields
		& $\hat{\Phi}$ & $=$
		& $\bigl(\hat h,\,\hat c,\,\hat{\bar c} \bigr)$
		& $\Phi$ & $=$
		& $\bigl(h,\,c,\,\bar c\bigr)$
		& $\dot{\Phi}$ & $\equiv$
		& $0$
		\\
		
		Composite fields
		& $\hat{\Phi}[\hat h,\hat c,\hat{\bar c}]$ & $=$
		& $\bigl(\hat\phi_h,\,\hat\phi_c,\,\hat\phi_{\bar c}\bigr)$
		& $\Phi$ & $=$
		& $\bigl(\phi_h,\,\phi_c,\,\phi_{\bar c}\bigr)$
		& $\dot{\Phi}$ & $\neq$
		& $0$
		\\
		\hline
	\end{tabular}
	\endgroup
	\caption{Overview of the different PIRG ingredients when using
		fundamental or composite fields.}
	\label{tab:FieldVariables}
\end{table*}

%%%%%%%%%%%%%%%%%%%%%%%%%
\subsubsection{Diffeomorphism-invariant flows with the PIRG}
\label{sec:DiffinFlowsDerivation} 

The properties (1) and (2) are readily achieved within a bootstrap approach: we demand $\partial_t \Gamma_\Phi$ to only depend on $\bar g+\Phi$ and choose its form. This leaves us with the non-trivial constraint (3), and the following analysis concentrates on its formal and in particular practical realisation. We start with the PIRG pair \labelcref{eq:PIRG-Pair} and define the set of $\dot\Phi$ that lead to a flow of the effective action that depends only on the full metric $g$, 
\begin{align} 
\dot\Phi\in \left\{\dot \Phi[\bar g,\bar g+\Phi]\,|\, \partial_t \Gamma_\Phi[\bar g, \bar g+\Phi] \stackrel{!}{=} \partial_t \Gamma_\Phi[ \bar g+\Phi]\right\}\,.
\label{eq:SetofdotPhi}
\end{align}
The set defined in \labelcref{eq:SetofdotPhi} is rather large. Different PIKs lead to different diffeomorphism-invariant flows, and hence different physical effective actions at $k=0$. This already hints at the fact that $\Gamma_\Phi$ should not be viewed as the only object that carries the physics at hand; it is carried by the PIRG pair \labelcref{eq:PIRG-Pair}. Moreover, the remaining freedom of choosing a PIK $\dot\Phi$ in the set \labelcref{eq:SetofdotPhi} allows us to maximise the physics content of $\Gamma_\Phi$. 

We proceed with the \textit{physics-informed gravity} setup with diffeomorphism-invariant flows.  
\Cref{eq:SetofdotPhi} suggests the following general parametrisation of the full inverse two-point function including the regulator term, 
\begin{align}\nonumber 
	\Gamma_{\Phi}^{(0,2)}[\bar g,\bar g+\Phi]+R_k[\bar g]  & \\[1ex] 
	&\hspace{-1.5cm}:= \Gamma_{\Phi,\textrm{qu}}^{(2)}[\bar g+\Phi] + \PIG_k[\bar g,\bar g+\Phi] \, .
	\label{eq:CoregravityPIG}
\end{align}
\Cref{eq:CoregravityPIG} has the form \labelcref{eq:CoreBackgroundApprox}, but it is \emph{exact}. It comes with a respective PIK $\dot\Phi[\bar g,\bar g+\Phi]$ which is yet to be determined. $\Gamma_{\Phi,\textrm{qu}}^{(2)}$ singles out the diffeomorphism- and background-independent part of the effective action that satisfies the generalised flow equation, 
\begin{align} 
	 \partial_t \Gamma_{\Phi,\textrm{qu}}[\bar g+ \Phi] =\partial_t  \Gamma_{\Phi}[\bar g+ \Phi]\,.
\end{align} 
The other part $\PIG_k$ contains the regulator and the gauge fixing term.  The latter is required for the invertibility of the two-point function. It may or may not have a $k$-dependence, which would be triggered by the PIK $\dot\Phi$. In the remainder of this work, we only consider two different choices, 
\begin{align}\nonumber 
\PIG_k^{(a)}[\bar g,\bar g+\Phi]=&\, S^{(0,2)}_\textrm{gauge}[\bar g,\bar g+\Phi] +R_k[\bar g]\,,\\[1ex]
\PIG_k^{(b)}[\bar g,\bar g+\Phi]=&\, S^{(0,2)}_\textrm{gauge}[\bar g+\Phi, \bar g+\Phi]+R_k[\bar g]\,.
\label{eq:Choiceab}
\end{align} 
The $\bar g$-dependence enters both through the regulator term with $R[\bar g]$ and parts of the gauge fixing sector.  

In the first case, we have combined the full gauge fixing sector with the cutoff term into $\PIG_k^{(a)}$. 
In the second case, we have simply lifted all isolated $\bar g$-dependences in the gauge fixing sector to ones on the full metric. Note in particular that $S^{(0,2)}_\textrm{gf}[\bar g+\Phi,\bar g+\Phi]$ cannot be derived from a diffeomorphism-invariant action and hence cannot be part of $\Gamma_{\Phi,\textrm{qu}}[\bar g+\Phi]$. These two cases will be discussed in detail in \Cref{sec:DiffinFlowsRelevance} in the context of the relevance counting in metric quantum gravity. 
 
With the parametrisation \labelcref{eq:CoregravityPIG}, the composite field propagator \labelcref{eq:PropFlucComposite} reads
 \begin{align} 
 	G_\Phi[\bar g,\bar g+\Phi] =\frac{1}{\Gamma_{\Phi,\textrm{qu}}^{(2)}[\bar g+\Phi] + \PIG_k[\bar g,\bar g+\Phi]} \, .
 	\label{eq:PropFlucCompositeOneField} 
 \end{align}
With these preparations, we can formulate the general background-independent and diffeomorphism-invariant PIRG approach: inserting \labelcref{eq:CoregravityPIG} into the generalised flow equation \labelcref{eq:GenFlow} leads us to the flow equation for \\[-1ex]

(1) diffeomorphism-invariant, and \\[-2ex]

(2) background-independent \\[-1ex] 
 
\noindent target actions $\Gamma_{\Phi,\textrm{qu}}$ in metric quantum gravity \cite{Ihssen:2025cff}, 
\begin{subequations} 
\label{eq:GenFlowPIG} 
\begin{align}\nonumber 
\partial_t \Gamma_{\Phi,\textrm{qu}}[\bar g+\Phi] =&\, \frac{1}{2} \textrm{Tr}\,G_{\Phi}[\bar g+\Phi]\,\partial_t R_k[g]\\[1ex] 
&+F_\Phi[\bar g+\Phi]
 \,,
	\label{eq:GenFlowDiffIn}
\end{align}
with the diffeomorphism-covariant and background-independent propagator 
\begin{align}
\hspace{-.1cm}	G_\Phi[\bar g+\Phi] = \frac{1}{\Gamma_{\Phi,\textrm{qu}}^{(2)}[\bar g+\Phi]  + \PIG_k[\bar g+\Phi,\bar g+\Phi]} \,. 
	\label{eq:GenFlowDiffinProp}
\end{align}
The term $F_\Phi[\bar g+\Phi]$ in the second line of \labelcref{eq:GenFlowDiffIn} is a general background-independent and diffeomorphism-invariant \textit{driving force}. Its choice determines the physics content of the diffeomorphism-invariant target action $\Gamma_{\Phi}$. It enters the differential equation for the PIK $\dot\Phi$ of the flow \labelcref{eq:GenFlowDiffIn}, 
\begin{align}\nonumber 
F_\Phi[\bar g+\Phi] =&\, \Delta\textrm{Flow}[\bar g,\bar g+\Phi] \\[1ex] 
&\hspace{-1.9cm}+   \textrm{Tr}\,G_{\Phi}[\bar g,\bar g+\Phi]\,\frac{\delta \dot\Phi}{\delta \Phi} \,R_k[\bar{g}] -  \dot\Phi \frac{\delta\Gamma_{\Phi,\textrm{qu}}[\bar g+\Phi]}{\delta \Phi}  \,,
	\label{eq:FPhi}
\end{align}
with 
\begin{align}\nonumber 
	\Delta\textrm{Flow}[\bar g,\bar g+\Phi] = &\,\frac{1}{2} \textrm{Tr}\,\Bigl[ G_{\Phi}[\bar g,\bar g+\Phi]\,\partial_t R_k[\bar g]\\[1ex]
	& - G_{\Phi}[\bar g+\Phi]\,\partial_t R_k[g] \Bigr]\,.
	\label{eq:DeltaFlow}
\end{align} 
\end{subequations}
\Cref{eq:DeltaFlow} is the difference of the background-independent and diffeomorphism-invariant flow in \labelcref{eq:GenFlowDiffIn} and the standard (Wetterich) flow term. Together with the first term in the second line in \labelcref{eq:FPhi} and the loop term in \labelcref{eq:GenFlowDiffIn}, it is the right-hand side of the generalised flow equation \labelcref{eq:GenFlow}. The last term in the second line of \labelcref{eq:DeltaFlow} is the coordinate transformation term on the right-hand side of  \labelcref{eq:GenFlow}. In short, apart from requiring the properties (1,2), we have slightly reorganised the terms in the generalised flow equation. 

\begin{figure*}[t]
	\centering
	\hspace{.5cm}%\subfloat{
		\includegraphics[width=0.43\textwidth]{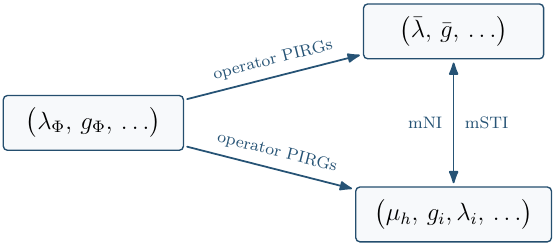}
		\label{fig:couplings}
	%}
	%
	\hfill
	%\subfloat{
		\includegraphics[width=0.43\textwidth]{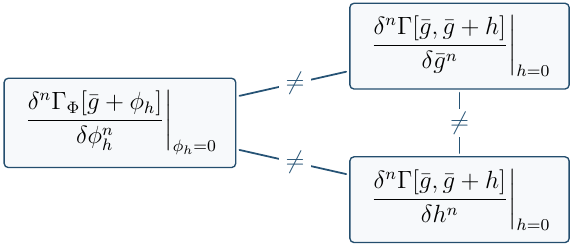}
		\label{fig:correlationFunctions}
	%}
	\hspace{.5cm}
	\caption{
		\label{fig:couplingsAndCorrelationFunctions}%
		Overview of the coupling sectors and the corresponding parametrisations of the effective action.
		\textbf{Left panel:} Three distinct sets of couplings are considered: the background couplings, denoted by an over-bar $\bar g_i$; the PIRG couplings, denoted by a subscript, such as $g_\phi$ and the fluctuation couplings $(\mu_h,g_3,\ldots)$. Both the PIRG and fluctuation sectors encode information about gauge invariance and can be used to reconstruct the background couplings. The PIRG couplings are distinct because the fundamental field is transformed during the flow. 
		\textbf{Right panel:} The effective action can be parametrised in terms of either $(\bar g,h)$ or $(\bar g,\phi)$. Consequently, the background-field and PIRG two-point functions are distinct. The relation between the macroscopic field associated with a flowing microscopic field and the fundamental field is non-trivial; see \cite{Ihssen:2025hyl} for details and the corresponding reconstruction using the \textit{operator PIRGs}. The relation between background and fluctuation correlation functions is governed by the modified Nielsen identity (mNI) and is discussed in \cite{Pawlowski:2020qer}. 
	}
	\label{fig:PIRGWrapup}
\end{figure*}

\Cref{eq:FPhi} defines the set of admissible $F_\Phi$ as those with a solution $\dot\Phi$ of \labelcref{eq:PIKequation}. Within this large set, $F_\Phi$ can be chosen freely, and hence warrants the name driving force. Note also that the propagator $G_{\Phi}[\bar{g}, \bar g+\Phi]$ only depends on $\bar g$ individually through the combined regulator and gauge-fixing sector $\PIG_k[\bar{g}, \bar g+\Phi]$. By construction, this does not feed back into the dynamical part $\Gamma_{\Phi,\textrm{qu}}^{(2)} [\bar{g}+\Phi]$ of the two-point function. Importantly, both terms in $\Delta\textrm{Flow}[\bar g, \bar g+\Phi]$ can be computed from the two-point function $\Gamma_{\Phi,\textrm{qu}}^{(2)}[\bar g+\Phi]$, and \labelcref{eq:GenFlowDiffIn} is a closed equation for a given $F_\Phi$. \Cref{eq:GenFlowPIG} has to be compared with the Wetterich equation in the fluctuation approach, the flow of the background field effective action,  \labelcref{eq:Backflow} and that in the background-field approximation \labelcref{eq:Flow-BFA}. The different field variables are compiled in  \Cref{tab:FieldVariables}.

%%%%%%%%%%%%%%%%%%%%%%%%%
\subsubsection{Effective action, physics-informed kernel and observables with the PIRG}
\label{sec:DiffinFlowsObservables} 

With \labelcref{eq:GenFlowDiffIn}, we have achieved two of the three goals set in the beginning of this section: it is (1) diffeomorphism-invariant, and (2) background-independent. (1) is guaranteed for regulators $R_k[g]$ that transform as tensors under diffeomorphism transformations, whereas (2) is achieved by its dependence on $\bar g+\Phi=g,c, \bar c$.

For a given $F_\Phi$ and the respective solution $\Gamma_{\Phi}$ of the flow equation, \labelcref{eq:FPhi} is a constraint equation for the physics-informed kernel $\dot\Phi[\bar g,\bar g+\Phi]$ with given coefficients, for a respective discussion see \cite{Ihssen:2025ybn}. Importantly, it is a linear functional differential equation and not a flow equation. Hence, it has to and can be solved independently for all $t$. Its general form is given by 
\begin{align} 
	{\cal C}_{1,i} \dot \Phi_i+ {\cal C}_{2,ij} \frac{\delta \dot \Phi_i}{\delta \Phi^j} = {\cal C}_3\,, 
\label{eq:PIKequation}
\end{align} 
with given coefficients ${\cal C}_{1,2,3}$. In the present application, they can be read off from \labelcref{eq:FPhi}, 
\begin{align} \nonumber 
	{\cal C}_1[\bar g+\Phi]=&\, \frac{\delta\Gamma_{\Phi}[\bar g+\Phi] }{\delta \Phi}\,, \\[1ex]\nonumber 
	{\cal C}_2[\bar g,\bar g+\Phi]=&\,R_k[\bar{g}] \,G_{\Phi}[\bar g,\bar g+\Phi]\,,\\[1ex]
	{\cal C}_3[\bar g,\bar g+\Phi]= &\,\Delta\textrm{Flow}[\bar g,\bar g+\Phi] - F_\Phi[\bar g+\Phi]\,.
\label{eq:PIKCoeffs}
\end{align}
The two terms on the left-hand side in \labelcref{eq:PIKequation} encode the reparametrisation of the effective action triggered by the PIK $\dot\Phi$ (first term), and that of the flow (second term). This term can be mapped to the Jacobian of the change of the measure in the path integral, see \cite{Ihssen:2025ybn, Ihssen:2026njd}. The term on the right-hand side of \labelcref{eq:PIKequation} constitutes the ``desired'' flowing change of the theory. In the present case, it is the change from a flow without diffeomorphism invariance and background independence to one with both. In view of diffeomorphism-invariant formulations of quantum gravity such as the geometric approach, it generates the respective field variables in a \textit{local} way within the flow. In short, it avoids or rather controls the inevitable non-localities of such formulations.

\Cref{eq:GenFlowPIG} provides us with a general approach to diffeomorphism-invariant flows in metric quantum gravity and, more generally, to gauge theories. From the practical point of view, it seems to be viable to simply \textit{assume} the existence of a $\dot\Phi$ for a given $F_\Phi$, and only solve the flow equation \labelcref{eq:GenFlowDiffIn} for the effective action $\Gamma_{\Phi}$. Then, one reads off the physics from $\Gamma_{\Phi}$ at $k=0$. However, the above analysis makes it abundantly clear that there is no restriction on $F_\Phi$ except the locality of the change that is encoded in ${\cal C}_3$ in \labelcref{eq:PIKCoeffs}. 

An extreme choice is given by $\Gamma_{\Phi,\textrm{qu}}=S_\textrm{EH}$, i.e., the target action is simply the classical action, complemented with a gauge-fixing term. The respective $F_\Phi$ reads 
\begin{align} 
	F_\Phi[\bar g+\Phi]=  - \frac{1}{2} \textrm{Tr}\,G_{\Phi}[\bar g+\Phi]\,\partial_t R_k[g]\,,
\label{eq:ClassicalPIK}
\end{align}
and the derived coefficients \labelcref{eq:PIKCoeffs} are given by 
\begin{align} \nonumber 
	{\cal C}_1[\bar g+\Phi]=&\, \frac{\delta S_\textrm{EH}[\bar g+\Phi]}{\delta \Phi} \,, \\[1ex]\nonumber 
	{\cal C}_2[\bar g,\bar g+\Phi]=&\,R_k[\bar{g}] \,G_\textrm{cl}[\bar g,\bar g+\Phi]\,,\\[1ex]
	{\cal C}_3[\bar g,\bar g+\Phi]= &\, \frac{1}{2} \textrm{Tr}\,G_\textrm{cl}[\bar g,\bar g+\Phi]\,\partial_t R_k[\bar g] \,, 
	\label{eq:PIKCoeffsClassical}
\end{align}
where the classical propagator $G_\textrm{cl}$ is given by \labelcref{eq:PropFlucCompositeOneField} with $\Gamma_{\Phi,\textrm{qu}}^{(2)}\to S_\textrm{EH}^{(2)}$. Hence, the locality of ${\cal C}_3$ is satisfied for all $t$ as the flow itself is local. This example has been studied for a scalar field theory in \cite{Ihssen:2024ihp}. It is very peculiar as for such a choice, all physics is stored in the PIK $\dot\Phi$ itself. While an extreme choice, it is a viable one. Evidently, the relevance counting of the full theory is now  hidden in the PIK, or rather divided between the canonical counting of the classical target action and the quantum scaling of the functional $\dot\Phi$. 

This example highlights the important property of the PIRG that the effective action $\Gamma_\Phi$ only covers part of the physics. The latter has to be extracted from the PIRG pair \labelcref{eq:PIRG-Pair}. It has been discussed in \cite{Ihssen:2024ihp} and in particular in \cite{Ihssen:2025hyl} how this is practically done: the correlation functions of the fundamental fields satisfy an operator flow equation. Structurally it is given by \cite{Ihssen:2025hyl}, 
\begin{align} 
	\partial_t {\cal O} = - \frac12 \textrm{Tr}\, G_\phi \partial_t R_k G_\phi {\cal O}^{(2)}+\dot\Phi\textrm{-terms}\,,
\label{eq:OpFlow}
\end{align}
and the set of observables ${\cal O}$ that satisfy \labelcref{eq:OpFlow} includes all correlation functions $\langle \hat h^n \hat c^m \hat{\bar c}^m\rangle$.  This concludes the setup of physics-informed gravity, or more generally gauge theories. For the convenience of the reader, we have wrapped up the relations of the couplings in the different effective actions in \Cref{fig:PIRGWrapup}: the PIRG effective action  $\Gamma_\Phi[\bar g+\Phi]$, the fluctuation effective action $\Gamma[\bar g,\bar g+\Phi]$, and the background effective action $\Gamma[\bar g+\Phi]=\Gamma[\bar g+\Phi,\bar g+\Phi]$ via the operator PIRG.

%%%%%%%%%%%%%%%%%%%%%%%%%
\section{Relevance-preserving PIRG}
\label{sec:DiffinFlowsRelevance} 

In \Cref{sec:DiffinFlows} we have set up (1) diffeomorphism-invariant and (2) background-independent flows. These flows can be characterised by the driving force $F_\Phi$ whose choice, roughly speaking, determines the amount of physics that is stored in the PIK $\dot\Phi$.  Generically, this changes the relevance counting of operators, as is clear from our extreme classical target action example discussed around \labelcref{eq:ClassicalPIK} and \labelcref{eq:PIKCoeffsClassical}. 

In the following, we also minimise the change of relevance counting of operators, hence achieving the goal (3). This serves a twofold purpose: if the relevance counting and ordering of metric quantum gravity is maintained exactly, the PIK provides the direct map from the gauge-fixed correlation functions to the respective diffeomorphism-invariant correlation functions or form factors. In turn, if the relevance counting of operators is deformed, the non-linearity of the map not only accounts for the diffeomorphism-invariant dressing of gauge-fixed correlation functions, but also reshuffles the dynamics. In this case, $\dot\Phi$ comprises part of the physics.

%%%%%%%%%%%%%%%%%%%%%%%
\subsection{Background-field approximation as a PIRG} 
\label{sec:PIRGBack}

Before we set up the relevance-preserving PIRG, we illustrate the distribution of the physics dynamics in the PIRG pair \labelcref{eq:PIRG-Pair} within the standard background-field approximation. This approximation has been and remains widely used in various forms in asymptotic safety and gauge theories as an approximation of the Wetterich equation. Within the PIRG, it is a legitimate target action, and we can monitor what part of the dynamics is stored in the PIK. Its flow is given by 
\begin{align}
	\partial_t \Gamma^{(\textrm{back})}_{\Phi}[\bar g+\Phi] = \frac{1}{2} \textrm{Tr}\,G_{\Phi}[\bar g+\Phi]\,\partial_t R_k[g]\,,
	\label{eq:PIRG-BackFlow}
\end{align}
and is obtained from \labelcref{eq:GenFlowDiffIn} with the driving force 
\begin{align} 
	F_\Phi\equiv 0\,. 
	\label{eq:FPhi-BackApprox}
\end{align} 
\Cref{eq:FPhi-BackApprox} entails that the PIK $\dot\Phi^{(\textrm{back})}$ absorbs the difference between the diffeomorphism-invariant flow on the right-hand side of \labelcref{eq:PIRG-BackFlow}, and the dynamical flow with a background-metric dependent gauge-fixing and regulator term $\PIG_k[\bar g,\bar g+\Phi]$ and its flow $\partial_t R_k[\bar g]$. The respective PIK is the solution of the PIK equation \labelcref{eq:PIKequation} with the coefficients 
\begin{align} \nonumber 
	{\cal C}^{(\textrm{back})}_1[\bar g+\Phi]=&\, \frac{\delta \Gamma_{\Phi}^{(\textrm{back})}[\bar g+\Phi]}{\delta \Phi} \,, \\[1ex]\nonumber 
	{\cal C}^{(\textrm{back})}_2[\bar g,\bar g+\Phi]=&\,R_k[\bar{g}] \,G_{\Phi}[\bar g,\bar g+\Phi]\,,\\[1ex]
	{\cal C}^{(\textrm{back})}_3[\bar g,\bar g+\Phi]= &\, \Delta\textrm{Flow}^{(\textrm{back})}[\bar g,\bar g+\Phi] \,. 
	\label{eq:PIKCoeffsBackApprox}
\end{align}
The coefficient ${\cal C}^{(\textrm{back})}_{1}$ only depends on the diffeomorphism-invariant solution $\Gamma^{(\textrm{back})}_\Phi$ of \labelcref{eq:PIRG-BackFlow}, while the coefficients ${\cal C}^{(\textrm{back})}_{2,3}$ also depend on the choice of $\PIG_k^{(a,b)}[\bar g,\bar g+\Phi]$ in \labelcref{eq:Choiceab}. We shall come back to this important fact later. In either case, the coefficient ${\cal C}^{(\textrm{back})}_3$ includes relevant operators in metric quantum gravity. This set is given by 
\begin{align} 
	\Bigl\{{\cal O}_\textrm{rel}\Bigr\} =	\Bigl\{ \sqrt{g}, \sqrt{g}\, { R},\sqrt{g}\, { R}^2,...\Bigr\}\,, 
	\label{eq:SetofRelO}
\end{align} 
where the dots refer to potential additional relevant or marginal operators, for respective discussions see e.g.~ \cite{Denz:2016qks}, for reviews see e.g.~\cite{Pawlowski:2020qer, Pawlowski:2023dda}. Consequently, the $\dot\Phi$-terms include relevant operators, and hence part of their dynamics is carried by the PIK. 

Once again, we emphasise that it is only the PIRG pair $(\Gamma^{(\textrm{back})}_{\Phi}, \dot\Phi^{(\textrm{back})})$ that retains the complete information about the physics. We conclude that the relevance counting of operators in $\Gamma^{(\textrm{back})}_\Phi$ is not that of metric quantum gravity. This can seemingly trigger \textit{qualitative} changes if one only concentrates on the target action and not the PIRG pair.  A relevant example is given by the stability analysis of matter-gravity systems with the target action $\Gamma_\Phi$ with \labelcref{eq:FPhi-BackApprox} of the background-field approximation. If compared with that obtained in metric quantum gravity, computed in the fluctuation approach for the effective action of the fundamental degrees of freedom, they seemingly behave qualitatively differently. More details are provided in \cite{Meibohm:2015twa, Meibohm:2016mkp, Christiansen:2017cxa} and the reviews \cite{Pawlowski:2020qer, Pawlowski:2023gym}.  This strongly hints at the fact that if the background-field approximation target action is considered, it is essential to assess the complete PIRG pair and not only the target action.  

This structural analysis of the PIRG for the background-field approximation corroborates and complements that in \cite{Litim:2001ky, Litim:2002hj, Litim:2002xm, Folkerts:2011jz}, see also the reviews \cite{Dupuis:2020fhh, Pawlowski:2020qer, Pawlowski:2023gym}. It leads us to the following conclusions:\\[-1ex]

(i) The PIRG approach elevates the background-field approximation flow, including proper time flows, to exact flows with a systematic error control within a three-step workflow:  \\[-1ex] 

($\alpha$) computation of the target action $\Gamma_\Phi$, \\[-1ex] 

($\beta$) analysis of the dynamics shifted to the PIK, \\[-1ex]  

($\gamma$) computation of the operator flows. \\[-1ex] 

In step ($\alpha$), one solves the background-field approximation flow for the target action $\Gamma_\Phi^{(\textrm{back})}$. In step ($\beta$), one solves the PIK equation \labelcref{eq:PIKequation} for $\dot\Phi^{(\textrm{back})}$ with the input $\Gamma_\Phi^{(\textrm{back})}$. This allows us to monitor the part of the physics which is stored in the PIK $\dot\Phi^\textrm{(back)}$. In step ($\gamma)$, one solves the operator PIRG with the PIRG pair $(\Gamma_\Phi^{(\textrm{back})}, \dot\Phi^\textrm{(back)})$ as the sole input. In short, the PIRG approach completes the background-field approximation. \\[-2ex]
 
(ii) The flow in the background-field approximation is one-particle irreducible, and hence also the target action  $\Gamma_\Phi^{(\textrm{back})}$ has this property (in terms of $\hat\Phi$-correlation functions). It follows from (i) that the relevance counting of the background-field approximation is \textit{not} that of the one-particle irreducible action of metric quantum gravity as the respective PIK $\dot\Phi^\textrm{(back)}$ includes relevant operators. Its target action only includes part of the dynamics of metric quantum gravity, and the relevance counting of metric quantum gravity is revealed in the three-step workflow described under (i).\\[-2ex]

(iii) Intriguingly, with (i) the PIRG setup gives access to a systematic stability analysis including a systematic error estimate
within the background-field approximation. This allows us to fully and directly exploit the \textit{powerful heat-kernel techniques} used and developed within this approximation over the past three decades~\cite{Benedetti:2010nr, Groh:2011dw, Codello:2012kq, Kluth:2019vkg, Knorr:2023usb}.  \\[-1ex]

This concludes our analysis of the background-field approximation PIRG. We emphasise that the above analysis is readily extended to most of the suggestions for gauge-invariant flows: The flow equations in \cite{Demmel:2014hla, Dietz:2015owa, Safari:2016gtj, Morris:2016nda, Wetterich:2016ewc, Falls:2020tmj, Wetterich:2024ivi, Falls:2025tid, Falls:2026nuh} can be formulated as PIRGs with \textit{target actions} $\Gamma_\Phi$ that are obtained from the respective closed flow equations. The target actions can then be used to compute their PIKs $\dot\Phi$. This provides a systematic error estimate and also, potentially, allows for a direct interpretation of the correlation functions obtained from the target actions. Finally, the operator PIRG allows for a direct computation of the correlation functions and observables of metric quantum gravity. 

We hasten to add that the geometrical flows discussed in \cite{Branchina:2003ek, Pawlowski:2003sk, Donkin:2012ud} are exact and provide diffeomorphism-invariant effective actions. In view of the PIRG approach this is achieved by introducing the composites from the outset: the geometrical fields are composites of the fundamental ones that are constructed from the demand of reparametrisation invariance. The details of the geometrical construction also entail that the relevance counting is not changed. As one can understand the PIRGs as a local way to introduce geometrical fields, it is interesting to compare the respective results. In conclusion, the explicit results in \cite{Donkin:2012ud} serve as a natural benchmark.

%%%%%%%%%%%%%%%%%%%%%%%
\subsection{Constructing the relevance-preserving PIRG} 
\label{sec:PIRG-RP}

The structural analysis of the background-field approximation begs the question whether we can shorten the workflow $(\alpha, \beta,\gamma)$ discussed above. 
While all of these steps are well-defined and computationally accessible, it is advantageous to maximise the physics content in step ($\alpha$). Put differently, for an easy access to physics one should minimise the physics stored in $\dot\Phi$ while keeping the properties (1,2). Apart from convenience, it also limits the potential approximation artefacts in ($\gamma$). For example, if part of the scaling of the Reuter fixed point is stored in the PIK, it requires a self-consistent scaling solution of the PIK equation \labelcref{eq:PIKequation} in step ($\beta$), and the operator flow \labelcref{eq:OpFlow} in step ($\gamma$) of the workflow. 

Formally, the maximisation of the dynamics content of the target action is a minimisation problem for the PIK. The respective discussion resembles that of functional optimisation \cite{Pawlowski:2005xe} and that of the PIRG \cite{Ihssen:2024ihp, Ihssen:2025hyl}. While the optimisation theory of fRG flows is a chiefly important subject, the optimisation task at hand requires a practical solution, in particular in simple approximations. 

For the analysis of the minimal PIK, we have to first discuss the (re)construction problem of diffeomorphism-invariant effective actions in the gauge-fixed approach to metric quantum gravity: via the LSZ-formula, scattering amplitudes and S-matrix elements are given as a sum of tree-level diagrams of 1PI fluctuation correlation functions in a background that solves the quantum equations of motion. For this purpose, it is convenient to recast the fluctuation effective action in terms of an expansion in diffeomorphism-covariant $n$-point functions or form factors. We use the notation $\Gamma_\Phi^{(\textrm{diff}),(a)}[g]$ for this effective action as we shall construct it from the regulator and gauge-fixing term $\PIG_k^{(a)}$ in \labelcref{eq:Choiceab}. It is diffeomorphism-invariant but gauge-dependent. In turn, the solution of the quantum equations of motion is obtained from the background field effective action \labelcref{eq:BackEffAct}. We shall construct it from the regulator and gauge-fixing term $\PIG_k^{(b)}$ in \labelcref{eq:Choiceab} and use the notation $\Gamma_\Phi^{(\textrm{diff}),(b)}[g]$. This effective action is both background-independent and diffeomorphism-invariant, and its equations of motion are the quantum equations of motion. For further discussions on the respective actions and reconstruction problems we refer to e.g.~\cite{Pawlowski:2023dda} and the reviews \cite{Pawlowski:2020qer, Pawlowski:2023gym}. 

In both cases, the effective action admits an expansion in diffeomorphism-invariant operators, 
\begin{align} 
	\Gamma_\Phi^{(\textrm{diff}),(a,b)}[g]=\sum_{n=0}^\infty C^{(a,b)}_n {\cal O}_n[g]\,,  
\label{eq:GammaDiffin}
\end{align}
with the ``couplings'' $\boldsymbol{C}=(C_0,...)$ and their dimensionless counterparts $c_n=C_n k^{d_{{\cal O}_n}}$. Here, $d_{{\cal O}_n}$ is the canonical mass dimension of the operator ${\cal O}_n$. The scaling of the $c_n$ in the vicinity of a fixed point $\boldsymbol{c}^*$ is captured by the critical exponents $\boldsymbol{\theta}$, i.e.~the negative of the eigenvalues of the stability matrix.
Note that the stability matrix is not diagonal and its eigendirections are typically not aligned with specific operators. 

In case $(a)$, the relevance counting of the operators is that of the underlying fluctuation correlation functions or operators. By contrast, in case $(b)$, the counting is that of the background effective action. A prominent example of the difference is the two-point function: its anomalous dimension is simply that of the fluctuation effective action $\Gamma^{(\textrm{diff}),(a)}$, while it is the $\beta$-function of the Newton coupling in the background field effective action $\Gamma^{(\textrm{diff}),(b)}$.  This is discussed further in \Cref{sec:Enforce-RP}, and is illustrated in \Cref{fig:Gamma-ab}. 

In view of the above, we formulate the map from the gauge-fixed formulation of metric quantum gravity with the effective action $\Gamma_k[\bar g,\bar g+h,c,\bar c]$ and the flow \labelcref{eq:FunFlow} as a two-step process. The first step consists of the substitution of the dynamical part $\Gamma_{k,\textrm{qu}}^{(0,2)}$ of the two-point function in the propagator \labelcref{eq:PropFluc} by its diffeomorphism-covariant counterpart \labelcref{eq:CoregravityPIG}, and we assume that no dynamics has been reshuffled, 
\begin{subequations} 
	\label{eq:2StepDiffIN}
\begin{align} 
	\Gamma_{k,\textrm{qu}}^{(0,2)}[\bar g,\bar g+\Phi]\to  \Gamma_{\Phi,\textrm{qu}}^{(2)}[\bar g+\Phi]\,.
	\label{eq:Step1}
\end{align} 
As discussed above, this step is not unique, and the two different  parametrisations in \labelcref{eq:Choiceab} entail two different definitions of the dynamical part $\Gamma_{k,\textrm{qu}}^{(a,b)}$ of the effective action. 

The second step consists of the substitution of the background metric in the combined gauge-fixing and regulator term $\PIG_k$ and the scale derivative of the regulator $\partial_t R_k$ with the full dynamical metric, 
\begin{align} \nonumber 
	\PIG_k[\bar g,\bar g+\Phi] \to &\, \PIG_k[\bar g+\Phi, \bar g+\Phi]\,,\\[1ex]
	\partial_t R_k[\bar g] \to &\,\partial_t R_k[g]\,.
	\label{eq:Step2}
\end{align} 
\end{subequations}

The first step \labelcref{eq:Step1} is buried in the $\dot\Phi$-terms in the PIK equation \labelcref{eq:FPhi}, and supposedly does not change the dynamics. The second step \labelcref{eq:Step2} is captured explicitly in the $\Delta \textrm{Flow}$-term within \labelcref{eq:DeltaFlow}. This simply substitutes the background metric in the regulator, and in case (a) also the gauge-fixing sector, with the full dynamical metric. Clearly, this changes the dynamics of the theory, and hence its relevance counting, as it introduces additional vertices. In the current PIRG setup, this step is easily tracked down, and the change of the relevance counting can be undone systematically.

%%%%%%%%%%%%%%%%%%%%%%%
\subsection{Enforcing relevance preservation in the PIRG} 
\label{sec:Enforce-RP}

With these preparations, we are now in the position to discuss the property (3) and how to obtain it in the PIRG approach. In \Cref{sec:ParticularRP} we discuss some general important properties of the diffeomorphism-invariant target actions and their relevance counting and in \Cref{sec:RP-DerivingForce} we derive the relevance-preserving driving force at leading order.

%%%%%%%%%%%%%%%%%%%%%%%%%
\subsubsection{Particularities of relevance preservation}
\label{sec:ParticularRP} 

Relevance preservation implies in particular that the critical exponents at the UV fixed point are unchanged by the PIK. The respective constraint is given by 
\begin{align} 
\frac{\delta{\boldsymbol{\theta}}}{\delta \dot\Phi} =0\,, 
\label{eq:NoChange}
\end{align} 
in the second step \labelcref{eq:Step2}, which is a fixed point condition. We emphasise that the first step already comes with a potential change of the critical exponents as the underlying effective actions of metric quantum gravity are different: for $(a)$ it is the fluctuation field effective action, for $(b)$ it is the background field effective action. 

This entails that \labelcref{eq:NoChange} comes with several challenges. In particular it requires that the relevance-preserving target action is that of metric quantum gravity in disguise. The critical exponents of the latter are those of the fluctuation operators. Let us briefly discuss this requirement for the two diffeomorphism-invariant target actions $\Gamma_\Phi^{(a,b)}$ based on \labelcref{eq:Choiceab}, see also \cite{Pawlowski:2023dda, Pawlowski:2023gym}:\\[-2ex] 

As discussed in \Cref{sec:PIRG-RP}, in case $(a)$ the target action $\Gamma_\Phi^{(a)}$ is the diffeomorphism-invariant completion of $\Gamma[\bar g,\bar g+\Phi]$ and the form factors are the diffeomorphism-invariant completions of the fluctuation correlation functions. In particular this entails that the two-point $\Gamma_\Phi^{(2)}$ scales with  $\eta_h$ and has to be seen as a (field-dependent) anomalous dimension. The latter reflects that neither of the $\Gamma_\Phi^{(a),(n)}$ is RG-invariant. However, matrix elements of scattering processes are computed from tree-level diagrams with these form factors.\\[-2ex]

In case $(b)$, the target action $\Gamma_\Phi^{(b)}$ is the background field effective action, and the form factors are diffeomorphism- and RG-invariant. It can directly be used to compute the quantum equations of motion. \\[-2ex]

\begin{figure}[t]
	\includegraphics[width=0.44\textwidth]{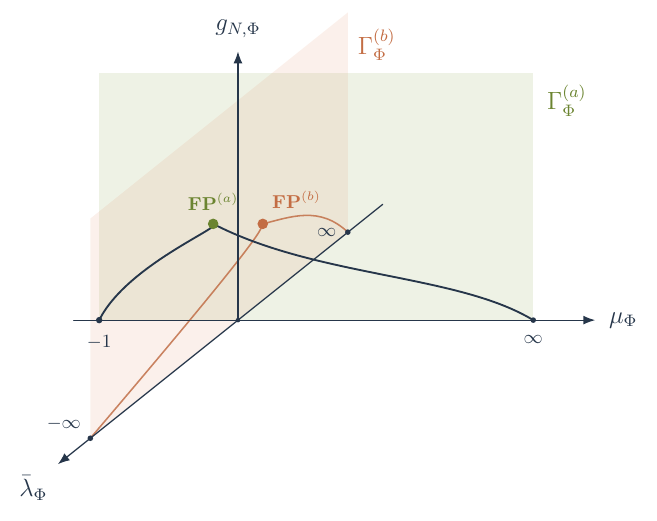}
	\caption{A sketch of theory space and the different choices of $\PIG_k^{(a,b)}$. Each choice leads to a diffeomorphism-invariant effective action $\Gamma_\Phi[\bar g+ \Phi]^{(a,b)}$, but they live in different spaces of theory space with broken diffeomorphism symmetry. The choice $\PIG_k^{(a)}$ leads to the fluctuation effective action, whereas  $\PIG_k^{(b)}$ yields the background effective action.}
	\label{fig:Gamma-ab}
\end{figure}

In particular this entails that the two target actions are not identical, and there are non-linear transformations between the form factors of $\Gamma_\Phi^{(a,b)}$. This is illustrated in  \Cref{fig:Gamma-ab} using the example of $\lambda_\Phi^{(a,b)}$ and $g_N^{(a,b)}$: While for the relevance-preserving target action $\Gamma_\Phi^{(a)}$, the parameter $\lambda_\Phi^{(a)}$ is related to the mass parameter $\mu_h$ of the fluctuating graviton, $\mu_h=-2 \lambda_\Phi^{(a)}$, the parameter  $\lambda_\Phi^{(b)}$ is the cosmological constant. Consequently, $\lambda_\Phi^{(a)}< 1/2$ and this constraint is similar to that of e.g.~$m_\phi^2> -1$ in a scalar $\phi^4$-theory where  $m_\phi^2<0$ simply signals the broken phase. Importantly, this is not a constraint for the cosmological constant which still can take any value, see \cite{Denz:2016qks}. 

In turn, $\lambda_\Phi^{(b)}\in \mathbbm{R}$ is the cosmological constant and is not constrained. However, the background-field approximation flow obtained with $F^\textrm{(back)}_\Phi=0$ experiences a singularity at $\lambda_\Phi^{(b)}=1/2$ while the flow of the background-field effective action in metric gravity does not. Hence, the background flow PIK $\dot \Phi^\textrm{(back)}$ is singular at this point and the map breaks down, as does the background-field approximation. Moreover, the relevance-preserving PIK $\dot\Phi^{(b)}$ of the background field effective action has to induce a driving force $F_\Phi^{(b)}$ that cancels this singularity and allows for the smooth limit $\lambda_\Phi^{(b)}\to +\infty$. Consequently, in general the flow for $\Gamma_\Phi^{(b)}$ is split into two singular parts whose singularities cancel each other. 

In summary, we expect structural differences in the construction of the diffeomorphism-invariant target actions $\Gamma_\Phi^{(a,b)}$. In particular, within the construction of $\Gamma_\Phi^{(b)}$ we have to overcome the singularity at $\lambda_\Phi^{(b)}=1/2$, which occurs in the background approximation flow. There it is a bug, while the same singularity for $\lambda_\Phi^{(a)}=1/2$ is a feature. Moreover, the critical exponents, directly computed from either effective action, do not necessarily agree. In short, the constraint \labelcref{eq:NoChange} comes with the potential necessity of also deriving non-trivial maps between form factors.

%%%%%%%%%%%%%%%%%%%%%%%%%
\subsubsection{Relevance-preserving driving force}
\label{sec:RP-DerivingForce}

A directly related practical constraint is given by eliminating all additional artificial vertices that are potentially generated by the PIK in the second step \labelcref{eq:Step2}. These vertices are generated by elevating the background metric to a dynamical one, specifically in the regulator. The ensuing constraint for the admissible relevance-preserving $F_{\Phi}$ is given by 
\begin{align} 
	\dot\Phi[F_\phi]: \quad 	\left[	{\cal C}_{1,i} \dot \Phi_i+ {\cal C}_{2,ij} \frac{\partial \dot \Phi_i}{\partial \Phi^j} \right]_\textrm{art}={\cal C}_{3,\textrm{art}}=0\,, 
	\label{eq:RelevancePreservation}
\end{align} 
with the coefficients ${\cal C}_{1,2,3}$ in \labelcref{eq:PIKCoeffs}. \Cref{eq:RelevancePreservation} can be implemented for all $k$, not only at the fixed point. The coefficient ${\cal C}_3$ is the difference of the driving force $F_\Phi$ and $\Delta\textrm{Flow}$, and the constraint \labelcref{eq:RelevancePreservation} translates into 
\begin{align} 
F_{\Phi,\textrm{art}}^{(n)}[\bar g+\Phi]=-\Delta\textrm{Flow}^{(n,0)}[\bar g, \bar g+\Phi ]+\cdots \,,   
\label{eq:ArtificialVertices}
\end{align}  
where the dots indicate different contributions from the $\dot\Phi$ terms that render $F^{(n)}_{\Phi,\textrm{art}}$ diffeomorphism-covariant and ensure that the right-hand side is indeed an $n$-th derivative. 

The complement of $F_{\Phi,\textrm{art}}$ in the driving force $F_\Phi$ is its physics part 
\begin{align} 
	F_{\Phi,\textrm{phys}}^{(n)}[\bar g+\Phi]=\Delta\textrm{Flow}^{(0,n)}[\bar g,\bar g+\Phi ]+\cdots \,, 
	\label{eq:PhysicalVertices}
\end{align}  
and the sum of the two terms is the complete driving force, 
\begin{align}
	F_\Phi= F_{\Phi,\textrm{art}}+F_{\Phi,\textrm{phys}}\,.
	\label{eq:SumFPhi}
\end{align} 
The requirement of \labelcref{eq:ArtificialVertices,eq:PhysicalVertices} being $n$-th derivatives together with \labelcref{eq:SumFPhi} determines the distributions of the $\dot\Phi$-terms in \labelcref{eq:ArtificialVertices,eq:PhysicalVertices}. Finally, relevance preservation is the constraint that the PIK does not contain relevance-changing contributions, and hence the physics part of the driving force vanishes, 
\begin{align} 
F_{\Phi,\textrm{phys}}\stackrel{!}{=}0\,. 
\end{align} 
Putting everything together, we arrive at the relevance-preserving PIRG flow 
\begin{subequations} 
	\label{eq:RG-PIRG}
\begin{align} 
\partial_t \Gamma_{\Phi,\textrm{rp}}[\bar g+\Phi] =\frac12 \textrm{Tr}\, G_{\Phi}[\bar g+\Phi] \partial_t R_k[g] +F_{\Phi}[\bar g+\Phi]\,, 
\label{eq:RP-Flow}
\end{align}
with the relevance-preserving force term $F_{\Phi}$, 
\begin{align} 
	F_{\Phi}[\bar g+\Phi]=F_{\Phi,\textrm{art}}[\bar g+\Phi]\,,
	\label{eq:RP-FPhi}
\end{align}  
with $F_{\Phi,\textrm{art}}$ derived from \labelcref{eq:ArtificialVertices}.
\end{subequations} 
\Cref{eq:NoChange} or \labelcref{eq:RP-FPhi} are very severe constraints on the admissible $\dot\Phi$ or $F_\Phi$. First, by implementing them we would lose much of the flexibility of the PIRG approach. This flexibility is instrumental in facilitating practical computations which may give access to hitherto unresolved phenomena, see e.g.~\cite{Bonanno:2025mon}. After all, \labelcref{eq:OpFlow} gives us a practical access to all correlation functions of metric quantum gravity. Second, the definition of $F_{\Phi}[\bar g+\Phi]$ via \labelcref{eq:ArtificialVertices,eq:PhysicalVertices,eq:SumFPhi} comes with a heavy reconstruction problem, namely, how to distribute diffeomorphism-covariant parts of mixed derivatives of $\Delta\textrm{Flow}$ to the physical and artificial parts.

In view of these two obstructions, we remind the reader that one of the possibilities is to use e.g.~$F_\Phi\approx 0$, settling in the class of the target action of the background-field approximation with $(i-iii)$ and the workflow $(\alpha - \beta)$.

%%%%%%%%%%%%%%%%%%%%%%%
\section{Minimal relevance preservation} 
\label{sec:Enforce-RP-min}

Instead of either strictly enforcing relevance preservation or not at all, we may choose 
a middle path: we only apply relevance preservation successively for the most relevant operators and use the workflow $(\alpha - \gamma)$. The more operators we include, the better the additional approximation of dropping $\gamma$ works. 

Practically, we reduce the difference of $\Gamma_\Phi$ and (one of the) diffeomorphism-invariant effective actions of metric quantum gravity $\Gamma_\Phi^{(\textrm{diff}),(a,b)}$ within a minimal implementation or approximation of \labelcref{eq:RP-FPhi}. This is done by using \labelcref{eq:RP-FPhi}, or rather \labelcref{eq:ArtificialVertices}, only for the set (or even a subset) of relevant and  marginal operators \labelcref{eq:SetofRelO}, i.e., $F_{\Phi,\textrm{art}} \to F_{\Phi,\textrm{rel}}$ with  
\begin{align} 
\hspace{-.1cm}	F_{\Phi,\textrm{rel}}^{(n)}[\bar g+\Phi]=\Bigl[ -\Delta\textrm{Flow}^{(n,0)}[\bar g, \bar g+\Phi ]+\cdots\Bigr]_\textrm{rel} .   
	\label{eq:RelVertices}
\end{align}  
Note also that this entails that the fixed-point constraint \labelcref{eq:NoChange} is only approximately satisfied even for the relevant and marginal critical exponents $\theta_i\geq 0$. \Cref{eq:RelVertices} leaves us with a finite number of constraints which can be practically implemented in terms of closed operator traces that can be computed with the straightforward application of the powerful heat-kernel techniques used and further developed in the context of asymptotically safe quantum gravity~\cite{Vassilevich:2003xt, Benedetti:2010nr, Groh:2011dw, Codello:2012kq, Kluth:2019vkg, Knorr:2023usb}. We will demonstrate this within a first simple application in the next section. 

We illustrate this with two structural applications of the relevance-preserving condition. The first one concerns the optimal choice of regulators that minimise  $F_{\Phi,\textrm{art}}$. The second application concerns the simplification of the computation of $F_{\Phi,\textrm{art}}$ and consistency checks.

%%%%%%%%%%%%%%%%%%%%%%%
\subsection{Optimal regulators for relevance preservation} 
\label{sec:ChoiceRegulator}

Evidently, contributions to the relevant terms \labelcref{eq:SetofRelO} that come from the field dependence of the regulator are part of $F_{\Phi,\textrm{rel}}$. We have parametrised a general regulator  $R_i$ in \labelcref{eq:RegMatrix} with $i=\textrm{TT}, \textrm{Tr}, c$ in terms of the respective spin $s$ background metric Laplacian $ \bar \Delta_s= (\bar \Delta_0 + \bar{\cal E}_s) \Pi_i $ and a separate dependence on the background metric endomorphism $\bar{\cal E}_s$, see 	\labelcref{eq:R_sApp} in \Cref{app:Approximation+GaugeFixing}, 
\begin{align} 
	R_i=\sqrt{\bar g}\,R_{s_i}(\bar \Delta_s,\bar{\cal E}_s)\,.
	\label{eq:R_s}
\end{align} 
The endomorphisms in \labelcref{eq:R_s} can only contribute to 
\begin{align} 
\sqrt{\bar g} \bar R\,,\qquad  \sqrt{\bar g} R\,,
\label{eq:Endocont} 
\end{align} 
as well as higher order terms. Both parts have to be included into $F_\Phi$: the first term in \labelcref{eq:Endocont} is a pure regulator term and the second term is a contribution to the background metric dependence of the cosmological constant which is not considered here. These terms are easily identified in an expansion of the regulators in 	\labelcref{eq:R_s} about 
$\bar{\cal E}_s=0$. For the sake of simplicity we restrict ourselves to covariantly constant metric and arrive at 
\begin{align} \nonumber 
\sqrt{\bar g}\,	\Pi_{i}  R_{s}(\bar \Delta_s,\bar{\cal E}_s) 	\Pi_{i}  =&	\sqrt{\bar g}\, \Pi_{i}  \Biggl[  R_{s}(\bar \Delta_0,0)	  \\[1ex] \nonumber 
	&\hspace{-2cm} +  \left(R_s^{(1,0)}(\bar \Delta_0,0)+R_s^{(0,1)}(\bar \Delta_0,0)\right) \bar{\cal E}_s\\[1ex]
	& \hspace{-2cm}  +O(\bar{\cal E}_s^2)\Biggr]\,	\Pi_{i} \,.  
	\label{eq:R_sEndoExpand}
\end{align} 
All the higher order terms in the second and third line will be subtracted by a relevance-preserving driving force and hence can be left out to begin with. This defines a class of regulators that maximises relevance-preservation in the background-field approximation, 
 \begin{align} 
 	R^\textrm{(rp)}_i=\sqrt{\bar g}\,R_{s_i}(\bar \Delta_0,0)\,.
 	\label{eq:R_sRP}
 \end{align} 
In the classification \textit{Type I,II} and \textit{III} of regulators in the background-field approximation, see e.g.~\cite{Codello:2008vh}, this singles out \textit{Type I} regulators as those that minimise the difference between the target action $\Gamma^{(\textrm{back})}$ and the effective action of metric quantum gravity. It has already been observed that this regulator minimises the differences between results in the fluctuation approach and the background-field approximation and the PIRG-approach provides the conceptual argument.

%%%%%%%%%%%%%%%%%%%%%%%
\subsection{Computing the minimal driving force} 
\label{sec:RPComputation}

Here we use the specific form of the propagator after the first step \labelcref{eq:Step1}. It entails that all dependences on $g$ are encoded in $\Gamma_{\Phi,\textrm{qu}}^{(2)}[\bar g+\Phi]$ and the gauge-fixing part $S^{(0,2)}$ and the dependence of the latter carries the differences between $\Gamma_\Phi^{(a)}$ and $\Gamma_\Phi^{(b)}$. This entails that we can either trace the $\bar g$-dependences that are elevated to $g$-dependences and subtract them with an appropriately chosen $F_{\Phi,\textrm{art}}$. Alternatively, we may simply compute $F_{\Phi,\textrm{art}}$ from the constraint 
\begin{align} 
 \frac{\delta \partial_t \Gamma_\Phi}{\delta g} \!\stackrel{!}{=} \! - \frac12 \textrm{Tr}\, G_\Phi\,\dot R_k\,G_\Phi\frac{\delta }{\delta g}\left(\Gamma^{(2)}_{\Phi,\textrm{qu}}[g] + S^{(0,2)}_\textrm{gauge}[\bar g, g]\right),
\label{eq:FPhiConstraint}
\end{align}
and higher order ones. \Cref{eq:FPhiConstraint} allows for a simple access to the most relevant operators $\sqrt{g}$ and $\sqrt{g} R$, and we shall use it below.

%%%%%%%%%%%%%%%%%%%
\section{Gravity PIRG at work}
\label{sec:DiffPIRGatWork}
	
In this section, we put the relevance-preserving PIRG setup in \Cref{sec:DiffinFlowsRelevance} to work. We concentrate on the correlations of the full metric $g$, and evaluate the flow for the target action $\Gamma_\Phi[\bar g+\Phi]$ at $\Phi=0$,  
\begin{align}
	\partial_t \Gamma_{\Phi}[g] =&\, \frac{1}{2} \textrm{Tr}\,G_{\Phi}[g]\,\partial_t R_k[g]+F_{\Phi,\textrm{rel}}[g]
	\,.
	\label{eq:PIGFlowg}
\end{align}
The first term on the right-hand side of \labelcref{eq:PIGFlowg} is the diffeomorphism-invariant flow term also present in the background-field approximation flow \labelcref{eq:Flow-BFA}, with the target action $\Gamma_\Phi^{(\textrm{back})}$. However, we emphasise again that $\Gamma_\Phi[g]\neq \Gamma_k[g,g]$ for a general $F_\Phi$. It is a target action, and part of the physics is stored in the PIK $\dot\Phi$ except for the choice \labelcref{eq:RP-FPhi} for $F_\Phi$. In our practical implementation, we use the relevance-preserving driving force $F_{\Phi,\textrm{rel}}$, leading to $\Gamma_\Phi[g]\approx \Gamma_k[g,g]$ as discussed in the last section.

%%%%%%%%%%%%%%%%%%%
\subsection{Relevance-preserving driving force at leading order}
\label{sec:RelPreservingDriving}

In the following, we develop a practical solution strategy which leaves us with an $F_{\Phi,\textrm{rel}}$ that can be computed with heat-kernel techniques. Then, the computational effort for solving the relevance-preserving PIRG is in the same ballpark as that for solving the background-field approximation, but we resolve the qualitative difference between the action in the background-field approximation and the effective action.

We start with the PIK equation \labelcref{eq:FPhi} for $F_\Phi$ and use the locality of all its ingredients. Then, $F_\Phi$ has an expansion in local invariants, 
\begin{align} 
	F_\Phi[g]=\int {\textrm{d}}^4 x\,\Bigl[f_{1}  \sqrt{g} + f_{R} \sqrt{g} R +\cdots \Bigr]\,.
	\label{eq:ExpandFPhi}
\end{align} 
This is mirrored in a similar expansion of $\Delta \textrm{Flow}$ and the $\dot \Phi$-terms in \labelcref{eq:FPhi}. However, in these expansions we have to take care of diffeomorphism-invariant terms that cancel out in the sum, but also cancel at $\bar g=g$. The difference flow is built from the set ${\cal D}$ of invariants and covariant tensors 
\begin{align} 
	{\cal D} =\bigl\{ \textrm{d}^4 x \sqrt{g}, R, R_{\mu\nu},...\bigr\}\,,
\label{eq:cal-D}
\end{align} 
in either $g$ or $\bar g$. Further tensor elements are $g_{\mu\nu}, \bar g_{\mu\nu}$ which will occur in anti-symmetric combinations that reflect $\Delta\textrm{Flow}[ g,g]=0$, 
\begin{align} 
	{\cal B}=\bigl\{ \textrm{tr} \,\bar g g^{-1} - \textrm{tr}\,   g \bar g^{-1}, g_{\mu\nu} - \bar g_{\mu\nu},...\bigr\}\,. 
\label{eq:cal-B}
\end{align}  
The elements in ${\cal B}$ originate in the fact that the traces in \labelcref{eq:DeltaFlow} only contain covariant tensors. Accordingly, we split $\Delta\textrm{Flow}$ as follows, 
\begin{align} 
	\Delta\textrm{Flow}[\bar g,g]=\Delta\textrm{Flow}_d[\bar g,g]+ \Delta\textrm{Flow}_b[\bar g,g]\,,
	\label{eq:DeltaFlowSplit}
	\end{align} 
where $\Delta\textrm{Flow}_{d}[\bar g,g]$ comprises all terms that are products of elements of the set ${\cal D}$ only. The other term, $\Delta\textrm{Flow}_{b}[\bar g,g]$, contains the rest and hence is built from products of elements of ${\cal D},{\cal B}$ with at least one element from ${\cal B}$. We concentrate on the leading terms and find 
\begin{align}\nonumber  
	&\Delta\textrm{Flow}_d[\bar g,g]= \int \textrm{d}^4 x\,\Biggl\{d_{1} \,\left( \sqrt{\bar g}-  \sqrt{g} \right)\\[1ex] 
	 & \hspace{.0cm}+d_{21} \left(\sqrt{\bar g} \bar R -\sqrt{ g} R\right) +d_{22} \left( \sqrt{\bar g} R -\sqrt{ g} \bar R\right) + \cdots \Biggr\}\,, 
	\label{eq:ExpandDeltaFlowd}
\end{align} 
where the $\cdots$ stands for terms that only reduce to higher order invariants at $\bar g=g$. The second term reads  
\begin{align}\nonumber  
	\Delta\textrm{Flow}_b[\bar g,g]= &\int \textrm{d}^4 x\,\Biggl\{b_1 \sqrt{\bar g}\, {\cal B}_1+b_2 \sqrt{g}\, {\cal B}_2 \\[1ex]\nonumber 
&\hspace{-1cm} + 	b_3 \sqrt{\bar g}\, \bar R_{\mu\nu\rho\sigma}\, {\cal B}_3^{\mu\nu\rho\sigma}+ b_4 \sqrt{\bar g}\, R_{\mu\nu\rho\sigma} \,{\cal B}_4^{\mu\nu\rho\sigma}\\[1ex] 
&\hspace{-1cm} + b_5 \sqrt{g}\, R_{\mu\nu\rho\sigma} \,{\cal B}_5^{\mu\nu\rho\sigma}+\cdots \Biggr\} \,, 
	\label{eq:ExpandDeltaFlowB}
\end{align} 
with the coefficients ${\cal B}_i(\bar g, g)$ where 
\begin{align} 
{\cal B}_i(g,g)=& 0\,,\qquad \frac{\delta {\cal B}_{1,2}}{\delta \bar g_{\mu\nu} }(g,g)\propto g_{\mu\nu}\,,\cdots \,.
\end{align} 
From the form of the regulator we conclude that the terms in  \labelcref{eq:ExpandDeltaFlowB} can only be generated by  
\begin{align} 
	\frac{\delta \bar \Delta_0}{\delta \bar g_{\mu\nu}} = \bar \Delta_0\,\bar g^{\mu\nu}+ \bar \Xi^{\mu\nu}\,, 
\label{eq:DiffDelta0}
\end{align} 
in both parts $S^{(0,2)}_\textrm{gauge}$ and $R_k$ of $\PIG^{(a,b)}_k$ defined in \labelcref{eq:Choiceab}. For $\Gamma_\Phi^{(a)}$ there are further terms induced by $S^{(0,2)}_\textrm{gauge}[\bar g,\bar g +\Phi]$ related to the projection properties of the gauge fixing. These terms require specific attention and care as they carry the difference between the target actions $\Gamma_\Phi^{(a,b)}$ and in particular that between $\lambda_\Phi^{(a,b)}$. 

For the relevance-preserving PIK, we have to identify the relevance-changing terms in $\Delta\textrm{Flow}$ and annihilate them with the driving force. This leaves us with $\Delta\textrm{Flow}_d$ in \labelcref{eq:ExpandDeltaFlowd} and we conclude 
\begin{align} 
	f_1 = - d_1 \,.
	\label{eq:FphiDeltaFlow1stOrder}
\end{align}
\Cref{eq:FphiDeltaFlow1stOrder} allows us to set up a relevance-preserving PIRG for  the $\sqrt{g}$-term which is a crucial first step: to begin with, the flow of the cosmological constant is directly proportional to the regulator and hence is dominated by regulator artefacts. For higher order operators, these artefacts lose their dominance progressively, and this argument has been corroborated within explicit computations in \cite{Meibohm:2015twa}. There, it was found in an investigation of minimally coupled matter-gravity systems that the main ``culprit'' for the qualitative failure of the background-field approximation is the difference in the relevance counting in the flow of the background cosmological constant to that of the graviton mass parameter in metric quantum gravity. Recast in the language of the PIRG, it is the flow of the ``cosmological constant'' in the target action  $\Gamma_\Phi^{(\textrm{back})}$ which differs qualitatively from that of the graviton mass parameter in metric quantum gravity. 

For $\Gamma_\Phi^{(a)}$, further constraints fix the anomalous dimension of $\Gamma_\Phi^{(a),(2)}$ and the flow of $g_N$ to that of the fluctuation field effective action and we shall use the fluctuation flows for $\eta_h$ and $g_3$, the latter being the coupling of the graviton three-point function. The derivation of these higher order terms from the PIRG will be considered elsewhere.

For $\Gamma_\Phi^{(b)}$, the situation is more complicated as discussed in \Cref{sec:PIRG-RP} and \Cref{sec:Enforce-RP}, see \Cref{fig:Gamma-ab}. Specifically, the driving force has to remove the singularity at $\lambda_\Phi^{(b)}=1/2$ or the PIK itself runs into a singularity. 

In \Cref{sec:Results} we show results for the phase structure of metric quantum gravity in the Einstein-Hilbert truncation for both cases $\PIG_k^{(a,b)}$ in \labelcref{eq:Choiceab}. Choice (a) with \labelcref{eq:RP-FPhi} provides us with a target action that approximately has the same relevance counting as the fluctuation effective action $\Gamma_\Phi^{(\textrm{diff}),(a)}$. In turn, choice (b) with \labelcref{eq:RP-FPhi} provides us with a target action that approximately has the same relevance counting as the background effective action $\Gamma_\Phi^{(\textrm{diff}),(b)}$, see the discussion at the beginning of \Cref{sec:PIRG-RP}.

%%%%%%%%%%%%%%%%%%%%%%%%%%%%
\subsection{Systematic error estimate}
\label{sec:integrability+Nielsen_Strategy}

We close this section with a discussion of a systematic error estimate for the linear order implementation of the relevance-preserving PIRG setup in the last section. Apart from a check of the relevance preservation, it also checks the approximation as such. This is done by using the  \emph{integrability condition}
\begin{align}
	\left[\frac{\delta}{\delta \bar g}, \, \partial_t\right] \Gamma_{\Phi,\textrm{qu}}[\bar{g}+\Phi] = 0 = 	\left[\frac{\delta}{\delta \Phi_i}, \, \partial_t\right] \Gamma_{\Phi,\textrm{qu}}[\bar{g}+\Phi] \, .
	\label{eq:Integrability}
\end{align}
Performing the $t$-derivative first leads us to the generalised flow equation \labelcref{eq:GenFlow} and the subsequent derivatives w.r.t.~the background metric and the fluctuation composite lead to the linear order implementation of relevance preservation. 

If performing the field derivatives first, we are led to functional Dyson-Schwinger equations (DSEs) for the background metric and the fluctuation fields. Its dynamics is carried by the expectation value of the classical equation of motion, and for illustration we quote the fluctuation DSE for background metric and the fundamental fluctuation fields (no composites), 
\begin{align} \nonumber 
\frac{ \delta \Gamma_\Phi }{\delta h} =  \left\langle  \frac{\delta S_\textrm{EH}}{\delta \hat g }\right\rangle+\left\langle  \frac{\delta \left(S_\textrm{gauge} +\Delta S_k\right)}{\delta \hat h}\right\rangle\,, \\[2ex] 
\frac{ \delta \Gamma_\Phi }{\delta \bar g} =  \left\langle  \frac{\delta S_\textrm{EH}}{\delta \hat g }\right\rangle+\left\langle  \frac{\delta \left(S_\textrm{gauge} +\Delta S_k\right)}{\delta \bar g}\right\rangle\,,  
	\label{eq:FunDSEflucFund}	
\end{align} 
where the $\bar g$-derivatives hit the isolated $\bar g$ as well as that in $\bar g+h$. 
We have also used that the Einstein-Hilbert action is a function of $\hat g=\bar g + \hat h$. 

In short, the quantum \textit{equations of motion} (EoM) are the expectation values of the classical EoM. They consist of two parts, and roughly speaking the dynamics is carried by the expectation value of $\delta S_\textrm{EH}/\delta \hat g$. The expectation value of the derivative of gauge and cutoff action takes care of the parametrisation and regularisation. Due to the non-polynomial nature of the Einstein-Hilbert action, the first term on the right-hand side contains all loop orders in full propagators and vertices and one classical vertex.  

The Nielsen identity is the difference of the two DSEs in \labelcref{eq:FunDSEflucFund}:  it carries the regularisation and gauge-fixing specifics that trigger the difference between $g$ and $\bar g$. In turn, the expectation value of $\delta S_\textrm{EH}/\delta \hat g$ that carries the dynamics cancels out. The $t$-derivative leads to the differential form of the Nielsen identity and can be used to compute the driving force $F_\Phi$, for more details and the inclusion of the composite see \Cref{app:DSE+NI}. 

The two procedures agree due to the integrability condition \labelcref{eq:Integrability}, but build on different functional hierarchies, the fRG- and DSE-hierarchies. Comparing results obtained with both hierarchies provides a systematic error estimate for the linear order implementation of the relevance preservation but also on the approximation itself. This systematic error estimate based on the different hierarchies has been used very successfully in QCD, see \cite{Fischer:2026uni}. We shall compare the two procedures in \Cref{sec:Results}.

%%%%%%%%%%%%%%%%%%%%
\section{Results}
\label{sec:Results}

In this section, we present the phase diagram of the gauge-invariant flow of metric quantum gravity, for both choices $(a,b)$ for $\PIG_k$.

We first discuss the phase diagram within the choice of a background gauge sector for the fluctuation field, option (a), followed by the gauge-covariant gauge sector, option (b). For both options we compute the beta functions within both calculation methods: directly via the flow, and via the flow of the Nielsen identity, given in \Cref{sec:RPComputation} and \Cref{sec:integrability+Nielsen_Strategy}, respectively. Here, we show the results from the flow due to their better systematics, whereas the results via flow of the Nielsen identity are deferred to the appendix as part of our systematic error estimate. 

We define the critical exponents to be \textit{minus} the eigenvalues of the stability matrix 
\begin{align}
	{\cal B}_{ij} = \frac{\partial \beta_{c_i}}{\partial c_j} \Bigg|_{\mathbf{c} = \mathbf{c}^\ast} \, , 
	\label{eq:stabilityMatrix}
\end{align}
for the dimensionless couplings $\mathbf{c} = \{c_i\}_i$. In this convention, relevant operators have a \textit{positive} critical exponent. The technical setup, especially the gauge-fixing and regulator choice, is given in \Cref{app:Approximation+GaugeFixing}.

%%%%%%%%%%%%%%%%%%%%%%%%%%%
\subsection{Phase diagrams for the diffeomorphism-invariant target actions}
\label{sec:PhaseStructure}

\begin{figure*}[!t]
		\centering
		\includegraphics[width=0.48\textwidth]{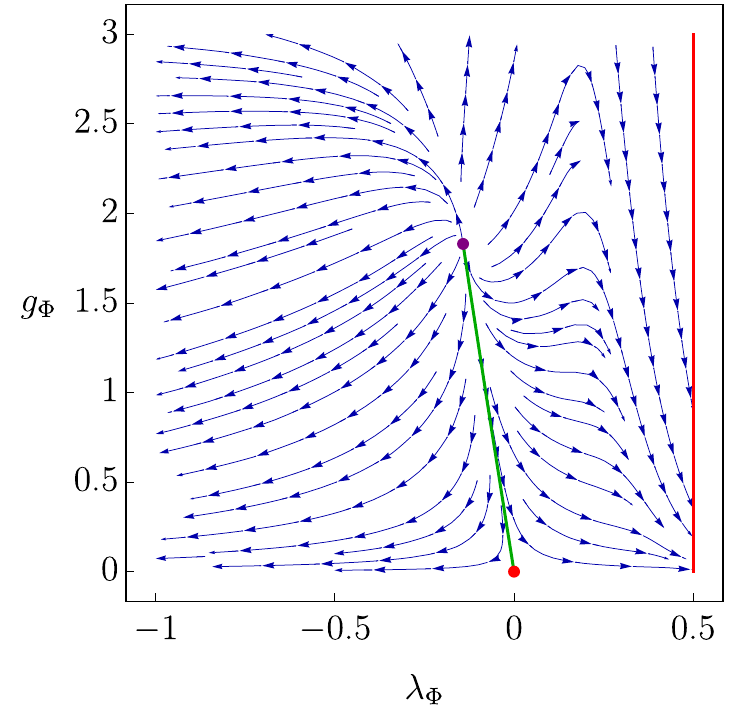}
		\includegraphics[width=0.48\textwidth]{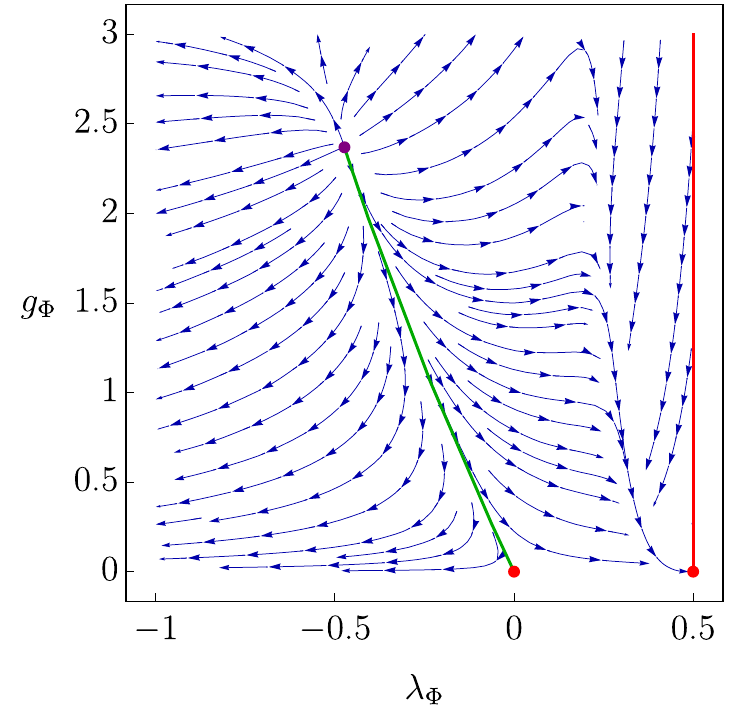}
		\caption{Phase diagram for the PIRG effective action $\Gamma_{\Phi, \textrm{qu}}^{\textrm{(diff)}}$ for the dimensionless couplings $g_\Phi$ and $\lambda_\Phi$, for options (a) (left panel) and (b) (right panel) of $\PIG_k$. The red dots indicate the IR fixed points, the violet dot is the Reuter fixed point. The green line connects the Reuter fixed point with the Gaussian fixed point, whereas the red line indicates the contour of a divergence of the flow.}	
		\label{fig:phaseDiagramDirect}
\end{figure*}

We discuss the target actions $\Gamma_\Phi^{(a,b)}$ where the target action $\Gamma_\Phi^{(a)}$ is the PIRG-proxy for the diffeomorphism-invariant envelope of the fluctuation effective action and the target action $\Gamma_\Phi^{(b)}$ is a PIRG-proxy for the background field effective action. 

%%%%%%%%%%%%%%%%%%%%%%%%%%%
\subsubsection{PIRG Phase diagram for the diffeomorphism-invariant envelope of the fluctuation field effective action}
\label{sec:PhaseStructurea}

We start our analysis with $\Gamma_\Phi^{(a)}$. After computing the driving force $F_\Phi$, we find the following beta function for $\lambda_\Phi$
\begin{align}
			\beta_{\lambda_\Phi}^{(a)}=&
			-2 \lambda_\Phi + \left( \frac{\beta_{g_{\Phi}}}{g_\Phi} -2 \right) \lambda_\Phi  \nonumber \\ 
				&+  \frac{-3(8-\eta_h) + 20 \lambda_\Phi(6-\eta_h)}{24\pi(1-2 \lambda_\Phi)^2} g_\Phi \, ,
			\label{eq:betaDirectA}
\end{align}
where $\eta_h$ is the anomalous dimension of the graviton. The latter and the anomalous dimension of the ghost, $\eta_c$, are provided in \Cref{app:anomDim}. The beta function \labelcref{eq:betaDirectA} is augmented by the beta function for the Newton constant in the background-field approximation, see \Cref{app:betaFuncs}. This yields a minimal setup to study the phase portrait of metric quantum gravity in our approach. 
	
The corresponding phase diagram is shown in the left panel of \Cref{fig:phaseDiagramDirect}. We find the Reuter fixed point at 
\begin{align*}
		(\lambda_\Phi^\ast\,,\, g_\Phi^\ast) = (-0.142,1.830) \, , 
\end{align*}
with the critical exponents
\begin{align*}
		\theta_{1,2} = \{3.86, \, 1.95 \} \, . 
\end{align*}
The critical exponents show near-canonical scaling of the operators $\sqrt{g}$ and $\sqrt{g} \, R$, with canonical scaling dimensions $4$ and $2$, respectively, for the diffeomorphism-invariant envelope of the fluctuation diffeomorphism-invariant action.  

%%%%%%%%%%%%%%%%%%%%%%%%%%%
\subsubsection{PIRG Phase diagram for the background field effective action}
\label{sec:PhaseStructureb}

Analogously, we obtain the beta functions for the diffeomorphism-invariant flow equation for option (b). This leads us to the background diffeomorphism-invariant effective action. In this case, the beta function for $\lambda_\Phi$ reads
\begin{align}
		\beta_{\lambda_\Phi}^{(b)}=&
		-2 \lambda_\Phi + \left( \frac{\beta_{g_{\Phi}}}{g_\Phi} -2 \right) \lambda_\Phi  - \frac{8-\eta_c}{6\pi}g_\Phi\nonumber \\ 
		&+  \frac{5(-8+4 \lambda_\Phi(6-\eta_h)+\eta_h)}{24\pi(1-2 \lambda_\Phi)^2} g_\Phi\, .
		\label{eq:betaDirectB}
\end{align}
The phase diagram is shown in the right panel of \Cref{fig:phaseDiagramDirect}. We find the Reuter fixed point at 
\begin{align*}
		(\lambda_\Phi^\ast, g_\Phi^\ast) = (-0.473,2.369) \, , 
\end{align*}
with the critical exponents
\begin{align*}
		\theta_{1,2} = \{4.66, 1.89\} \, . 
\end{align*}
The critical exponents also show near-canonical scaling of the Reuter fixed point for the background diffeomorphism-invariant effective action.

%%%%%%%%%%%%%%%%%%%%%%%%%%%
\subsubsection{Global properties of the flows}
\label{sec:singularityStructure}

While we do not aim for the UV-IR phase structure within the present approximation, we can cautiously study the global properties of the phase diagram, see \Cref{fig:phaseDiagramDirect}. We note that there is a singular line at $\lambda_\Phi =1/2$, which is inherited from the pole of the fluctuation propagator for the graviton. This line reflects the convexity restoring pole at $\mu_h = -1$ and hence is expected to be a feature of the flow. The UV-IR flows converge into this singularity, and this property is also present in the direct computation of the fluctuation field phase diagram. 

The PIRG-proxy of the background field effective action is given by choice (b) for $\PIG_k$. In this case the infrared value of the cosmological constant can take any value. The background cosmological constant has been computed in the fluctuation approach in \cite{Denz:2016qks} and the explicit computation confirms this expectation. However, the propagator necessarily shows a singularity at $\lambda_\Phi = 1/2$ which has to be cancelled exactly by a respective one in the driving force. In the present approximation, we see an IR attractive fixed point at $\lambda_\Phi=1/2$ and $g_\Phi=0$ similarly to the phase structure with the choice $(a)$. This similarity is largely due to the form of the anomalous dimension $\eta_h$ in the beta function. The required cancellation of the singularity in the flow of the target action $\Gamma_\Phi^{(b)}$ with the driving force $F_\Phi$ is not seen in the present approximation. However, it is intricate as mentioned above. The existence and properties of the singularity and a possible IR fixed point are related to the graviton propagator and the anomalous dimensions, which will be discussed in a future work.  
\begin{table}[t]
\renewcommand{\arraystretch}{1.1}
		
\begin{ruledtabular}
			\begin{tabular}{l | c c}
				$\PIG_k^{(a,b)}$
				& $(\lambda_\Phi^\ast,g_\Phi^\ast)$
				& $\{\theta_i\}$ \\ 
				
				\midrule
				\midrule
					$\Gamma_{\Phi, \textrm{qu}}^{\textrm{(a)}}[g]$
				& $(-0.142,\,1.830)$
				& $\{3.86, \, 1.95\}$\\[1ex]
				
					$\Gamma_{\Phi, \textrm{qu}}^{\textrm{(b)}}[g]$
				& $(-0.473,\,2.369)$
				& $\{4.66\, , 1.89\}$
				\\

\end{tabular}
\end{ruledtabular}
		
\caption{Summary of the fixed-point coordinates,
			$(\lambda_\Phi^\ast,g_\Phi^\ast)$, and critical exponents,
			$\theta_i$, for the fluctuation and background diffeomorphism-invariant actions, corresponding to the choices (a) and (b) for $\PIG_k$, respectively. The PIRG-corrected beta function for $\lambda_{\Phi}$ is augmented by the background-field approximation beta function for $g_N$.}
\label{tab:mainResults}
\end{table}
%		

%%%%%%%%%%%%%%%%%%%%%%%%%%%%%
\subsection{Comparison and error estimate}
\label{eq:Comparison+Error}

The integrability condition discussed in \Cref{sec:integrability+Nielsen_Strategy} allows us to put an error estimate on the results for the fixed point values. As a reminder, the alternative computation strategy takes the (modified) Nielsen identity as a starting point and hence we call it the \textit{Nielsen strategy}. We collect and compare the results for both the background and fluctuation PIRG actions $\Gamma_{\Phi, \textrm{qu}}^{\textrm{(diff)}}$ in \Cref{tab:comparResults}. 

The fixed point values for the different choices of $\PIG_k^{(a,b)}$ show a trend but vary throughout the different strategies and choices of completion with $\beta_{g_\Phi}$, which we take as either a background or a fluctuation beta function for the Newton constant. In particular, the fixed point value for $g_\Phi$ seems to cluster according to the option -- option (a) seems to favour smaller values of $g_\Phi^\ast$, whereas option (b) seems to favour larger values. Furthermore, the fixed point value for the cosmological constant appears to be more negative for the background effective action when compared to the fluctuation effective action. When comparing the completions with a background or fluctuation beta function for $g_\Phi$ this can be traced back to the remaining differences between the fluctuation and background formalism beyond the cosmological constant. Furthermore, we took the beta function from \cite{Denz:2016qks}, which computed the running of $g_3$ in a different gauge. 

A further error estimate is given by the comparison of the direct computation with the Nielsen strategy. The fixed point values are smaller in magnitude within this strategy when compared to the direct computation for the respective choice of $\PIG_k^{(a,b)}$. The rough agreement of the fixed points and phase portraits is a promising first outcome of the error estimate for the rather simple truncation of the driving force $F_\Phi$. 

Next, we compare the critical exponents. We find that although the fixed point values vary throughout the different strategies, the critical exponents are in good qualitative agreement. The critical exponents signal near-canonical scaling of the operators $\sqrt{g}, \sqrt{g}R$, apart from the Nielsen strategy for the choice (a) of $\PIG_k$. This may be traced back to the choice of anomalous dimension, see the discussion in \Cref{app:phaseDiagrams}. The stability of the critical exponents may be traced back to the relevance preservation property of the flowing field, \labelcref{eq:NoChange}. By construction, the critical exponents are those of metric quantum gravity. In particular, if the same operators are considered, the critical exponents of choice (a) and (b) for the diffeomorphism-invariant effective action must match although the fixed point values may differ. Note that this statements comes with several layers of complexity as the map of operators from $\Gamma_\Phi^{(a)}$ (diffeomorphism-invariant but gauge-dependent) to $\Gamma_\Phi^{(b)}$ (diffeomorphism-invariant and potentially gauge-independent) has not been worked out, for a more comprehensive discussion see \cite{Pawlowski:2020qer}. Hence, we consider our findings in this  rather simple truncation of $F_\Phi$ as a promising first hint on a rather simple one-to-one mapping of operators between the two effective actions. 

\begin{table}[t]
	\renewcommand{\arraystretch}{1.1}
	
	\begin{ruledtabular}
		\begin{tabular}{l | l c c}
			$\PIG_k^{(a,b)}$
			& Procedure 
			& $(\lambda_\Phi^\ast,g_\Phi^\ast)$
			& $\{\theta_i\}$ \\
			
			\midrule
			\midrule
			\multirow{3}{*}{%
				$\Gamma_{\Phi, \textrm{qu}}^{\textrm{(a)}}[g]$}
			& Flow, $\beta_{g_\Phi}^{\textrm{(back)}}$
			& $(-0.142,\,1.830)$
			& $\{3.86, \, 1.95\}$\\[1.0ex] 
			
			& Flow, $\beta_{g_\Phi}^{\textrm{(fluc)}}$
			& $(-0.224,\,2.956)$
			& $\{\,4.00,\, 2.05\}$\\
			
			\cmidrule{2-4}
			& Nielsen, $\beta_{g_\Phi}^{\textrm{(back)}}$
			& $(0.135,\,1.130)$
			& $\{\,6.68, \, 1.87\}$ \\[1.0ex] 
			
			\midrule
			\midrule
			\multirow{3}{*}{%
				$\Gamma_{\Phi, \textrm{qu}}^{\textrm{(b)}}[g]$}
			& Flow, $\beta_{g_\Phi}^{\textrm{(back)}}$
			& $(-0.473,\,2.369)$
			& $\{4.66\, , 1.89\}$
			\\[1.0ex]
			
			& Flow, $\beta_{g_\Phi}^{\textrm{(fluc)}}$
			& $(-0.638,\,3.339)$
			& $\{4.83, \, 2.07\}$
			\\ 
			
			\cmidrule{2-4}
			& Nielsen, $\beta_{g_\Phi}^{\textrm{(back)}}$
			& $(-0.031,\,1.551)$
			& $\{3.70, \, 1.81\,\}$
			\\[1.0ex]

		\end{tabular}
	\end{ruledtabular}
	
	\caption{Summary of the fixed-point coordinates,
		$(\lambda_\Phi^\ast,g_\Phi^\ast)$, and critical exponents,
		$\theta_i$, obtained with the direct procedures, Flow and Nielsen, for both the (a) fluctuation and (b) background and diffeomorphism-invariant effective action. The beta functions for $\lambda_\Phi$ are augmented with the background or fluctuation beta function for $g_\Phi$.}
	\label{tab:comparResults}
\end{table}
%	

%%%%%%%%%%%%%%%%%%%%%%%%%%%%%%%%%%%%%
\subsection{Proper time flows as PIRGs and relevance preservation}
\label{sec:PTRG-PIRG}

We close this section with a brief discussion of PIRGs for proper time flows, e.g.~\cite{Bonanno:2004sy, Bonanno:2019ukb}, and the simplified flow equation \cite{Wetterich:2024ivi}. The PIRG approach elevates these flows to exact ones for the respective target actions $\Gamma_\Phi^{(\textrm{PT})}$. The respective PIKs, or rather their driving forces $F_\Phi^{\textrm{(PT)}}$, have been derived in \Cref{app:NielsenPractitioner}. A chiefly relevant question is the relation of the proper time driving force to the relevance-preserving driving forces $F_\Phi^{(a,b)}$. In \Cref{app:NielsenPractitioner} we compute the difference $F_\Phi^{(a,b)}-F_\Phi^{\textrm{(PT)}}$. For this computation we have used the driving force as derived from the Nielsen identity and the result is given in \labelcref{eq:DeltaFphiApp}. We recall the result here, 
\begin{align}  \nonumber 
	\frac{\delta F_\Phi^\textrm{(a,b)}[g]}{\delta g}-\frac{\delta F_\Phi^\textrm{(PT)}[g]}{\delta g}& \\[1ex]
&\hspace{-2.5cm}= - \frac{1}{2} \, \textrm{Tr}\, \partial_t \left[ G_{\Phi}[g, g]\,  S_\textrm{gauge}^{(a,b),(1,2)}[g,g]\right] \,.
\label{eq:DeltaFphi} 
\end{align} 
For the choice $(b)$, the gauge-fixing part is vanishing, $S_\textrm{gauge}^{(b),(1,2)}[g,g]=0$ and the proper time driving force overlaps with the relevance-preserving driving force \labelcref{eq:RP-FPhi}. Note, however, that the Nielsen identity only provides an approximation for the relevance-preserving driving force, and the dropped $\partial_t\Gamma_\Phi^{(2)}$-term triggers sizeable contributions. This requires further investigations which will be published elsewhere.

%%%%%%%%%%%%%%%%%%%%
\section{Conclusions}
\label{sec:Conclusions}

In the present work, we have applied the physics-informed renormalisation group (PIRG) approach for gauge theories \cite{Ihssen:2025cff} to metric quantum gravity, leading to a background-independent and diffeomorphism-invariant flow equation \labelcref{eq:GenFlowPIG} for an effective action $\Gamma_\Phi$, the target action. This approach comes at the price that the physics of metric quantum gravity is now stored in the PIRG pair $(\Gamma_\Phi,\dot\Phi)$ of the target action and the physics-informed kernel $\dot\Phi$ (PIK). The flow differs from the commonly used flow in the background-field approximation by an additional diffeomorphism-invariant 'driving force' $F_\Phi$. Importantly, practical computations can directly use the powerful heat-kernel techniques used and further developed in the background-field approximation. Moreover, while the effective action is not the standard 1PI effective action of the fundamental degrees of freedom, observables can be computed directly from operator flows \labelcref{eq:OpFlow}, see \cite{Ihssen:2025hyl}. 

The driving force $F_\Phi$ can be chosen such that the diffeomorphism-invariant flow has the same relevance counting as metric quantum gravity, see the discussion in \Cref{sec:DiffinFlowsRelevance} and \Cref{sec:DiffPIRGatWork}. Relevance-preserving PIKs and the respective driving forces $F_\Phi$ maximise the physics content of the target action, minimising the difference between the target action and the one-particle irreducible effective action. This allows us to study important questions such as the stability of gravity-matter systems directly from the flow of the effective action. If, on the other hand, the relevance counting of the diffeomorphism-invariant flow differs from that of metric quantum gravity, one has to study the PIRG pair and the operator flow for conclusive statements. This is the case for the background-field approximation as $F_\Phi$ is relevance-changing. 

Moreover, the present general approach allows us to classify \textit{all} suggestions for diffeomorphism-invariant flows in terms of the PIK $\dot\Phi$ which renders these flows exact. This augments these flows, and more generally, any approximation with a systematic error estimate as one can compute the respective PIK and assess the physics stored in $\dot\Phi$. For the important question of the relevance of operators we can classify all suggestions for diffeomorphism-invariant flows in terms of the change of relevance counting that goes with them. We have discussed this classification using the example of the proper time flow, e.g.~\cite{Bonanno:2004sy, Bonanno:2019ukb} and the simplified flow equation \cite{Wetterich:2024ivi}. We hope to report on a more detailed extension of this study in the near future. 

In the present work we have implemented the relevance-preserving constraint at leading order and computed the Reuter fixed point as well as the phase structure of metric quantum gravity in the Einstein-Hilbert approximation. We have augmented this computation with two different computation schemes for the relevance-preserving constraint that relate to two different diffeomorphism-invariant effective actions: one is the diffeomorphism-invariant closure of the fluctuation effective action, the second is the background field effective action. A comprehensive analysis requires at least an upgrade of the current computation with the curvature scalar and the curvature-squared invariants. We hope to report on this matter soon, including an analysis of essential and inessential operators \cite{Baldazzi:2021ydj, Baldazzi:2021orb} which can be readily done within the PIRG approach. 

Finally, this approach is tailor-made for the discussion of general gravity-matter systems in a diffeomorphism-invariant setup: to begin with, relevance-preserving $F_\Phi$'s resolve the deficiencies of the background-field approximation and specifically lead to a \textit{qualitative} change of the flow of the cosmological constant. The fixed point analysis of gravity-matter systems with these flows allows for a comparison with the results \cite{Meibohm:2015twa} in the fluctuation approach, which has the relevance-counting of metric quantum gravity by definition. Preliminary results show a qualitative agreement with the results in the fluctuation approach and we hope to report on this exciting development soon.

%%%%%%%%%%%%%%%%%%%%
\begin{acknowledgments}
		We thank Astrid Eichhorn, Manuel Reichert and Christof Wetterich for discussions. This work is funded by the Deutsche Forschungsgemeinschaft (DFG, German Research Foundation) under Germany’s Excellence Strategy EXC 2181/1 - 390900948 (the Heidelberg STRUCTURES Excellence Cluster), the Collaborative Research Centre SFB 1225 - 273811115 (ISOQUANT) and the Walter Benjamin Programme with project number 574031240.
\end{acknowledgments}

%%%%%%%%%%%%%%%%%%%%
\appendix
\crefalias{section}{appsec}

%%%%%%%%%%%%%%%%%
\section{Gauge fixing and truncation}
\label{app:Approximation+GaugeFixing}

The classical diffeomorphism-invariant Einstein-Hilbert action $S_\textrm{EH}$ is given by 
\begin{align}
	S[g] = \frac{1}{16\pi G_N} \int \text{d}^4x \sqrt{g} \, \left(-R + 2 \Lambda\right) \,, 
	\label{eq:EH-Action}
\end{align}
with the two couplings $\Lambda$ (cosmological constant) and $G_N$ (Newton constant). A minimal approximation for the flowing effective action $\Gamma_k$ is given by \labelcref{eq:EH-Action} with 
flowing couplings $(\Lambda_k,G_{
N,k})$, 
\begin{align}
		\Gamma_k[g] =S[g;\Lambda_k,G_{N,k}] \, .
		\label{eq:EHtruncation}
\end{align}
As discussed in \Cref{sec:FunFlowsGravity}, we rely on linear gauge fixings in the fluctuation field, see	\labelcref{eq:gfGeneral}. In the general case this is the fluctuation $\hat \Phi[\bar g,\bar g+\hat h]$ but for the sake of simplicity we only discuss $\hat\phi_h=\hat h$ and drop the hat in the expressions below.
The complete gauge-fixing sector is a combination of the gauge-fixing term \labelcref{eq:gfGeneral} and 
Faddeev-Popov term, $S_\textrm{gauge} = S_\textrm{gf} + S_\textrm{gh}$ with 
the gauge-fixing term $S_\textrm{gf}$ and the Faddeev-Popov term $S_\textrm{gh}$, 
\begin{subequations} 
	\label{eq:GaugeFixing}
\begin{align}
		S_{\text{gf}}[\bar g, h]=\frac{1}{2 \alpha} \int \!\mathrm{d}^4x
		\sqrt{\bar{g}}\; \bar{g}^{\mu \nu} F_\mu F_\nu \,. 
		\label{eq:gfGeneral} 
\end{align}
The gauge-fixing term is necessarily a functional of the background metric and the fluctuation. A typical gauge-fixing choice in \labelcref{eq:gfGeneral} is the linear covariant 
gauge-fixing condition for the fluctuation field $h_{\mu\nu}$, 
\begin{align}
		F_\mu[\bar g, h] =
		\left(\bar{\nabla}^\nu h_{\mu \nu} -\frac{1+ \beta}{4} \bar{\nabla}_\mu h^{\nu}_{~\nu}\right) \,, 
		\label{eq:gfCovariantApp}
\end{align}
where $\bar\nabla$ is the covariant derivative with the Levi-Civita connection of the background metric $\bar g_{\mu\nu}$. The gauge-fixing sector $S_\textrm{gauge}$ in \labelcref{eq:Sgauge} is completed by the respective Faddeev-Popov term,  
\begin{align}
		S_{\text{gh}}[\bar g, h,c,\bar c]= \int \!\mathrm{d}^4x
		\sqrt{\bar{g}}\; \bar c^\mu M_{\mu}^{\phantom{\mu}\nu} c_\nu\,, 
		\label{eq:Sghost}
\end{align}
with the Faddeev-Popov operator 
\begin{align}
		M_{\mu\nu}= \bar\nabla^\rho\! \left(g_{\mu\nu} \nabla_\rho +g_{\rho\nu} \nabla_\mu\right) -\frac{1+\beta}{2} \bar g^{\sigma\rho} \bar\nabla_\mu g_{\nu\sigma} \nabla_\rho\,,
	\label{eq:OpFP}
\end{align}
\end{subequations} 
which is linear in the fluctuation field, for more discussions see e.g.~\cite{Pawlowski:2023gym}. The sum of the gauge-fixing term \labelcref{eq:gfGeneral} and the ghost term \labelcref{eq:Sghost} is the gauge-fixing action, 
\begin{align} 
	S_\textrm{gauge}[\bar g, \bar g+h,c,{\bar c}]=S_{\text{gf}}[\bar g,  h]+S_{\text{gh}}[\bar g,  h, c,{\bar c}]\,.
	\label{eq:SgaugeApp}
\end{align}
The gauge-fixing parameters $\alpha,\beta$ flow with the RG scale $k$. The Landau-DeWitt limit $\alpha \to 0$ constitutes a line of fixed points for $\beta\neq 3$~\cite{Litim:2002ce, Knorr:2017fus}. In the PIRG framework, the gauge fixing is kept classical and the gauge-fixing parameters do not run. More accurately, their running is absorbed in the reparametrisation of the theory via $\dot\Phi$. Many computations are done in the harmonic gauge with 
\begin{align}
		\alpha = 1 \, , \qquad  \beta= 1 \, . 
		\label{eq:harmonicGF}
\end{align}
This choice of gauge fixing facilitates the computations as it disentangles the graviton propagator into its traceless and trace mode, i.e.~there is no kinetic mixing between them. 
The propagator matrix has the entries
\begin{align}
	(G_{ij}) = \left( \begin{array}{ccc} 
		G_{hh} & 0  &0\\[1ex]
		 0 & 0 & G_{\bar c c}  \\[1ex] 
		0 & -G_{\bar c c}  &  0 
	\end{array} \right) \, .
\end{align}
The regulator matrix $R_{ij}$ with $i=h,c,\bar c$ in \labelcref{eq:RegTerm} (in harmonic gauge) is given by 
\begin{align} 
	(R_{ij})  &= \sqrt{\bar g}\,\Bigl[\Pi_\textrm{TT} {R}_\textrm{TT} \Pi_\textrm{TT} 
	+\Pi_\textrm{Tr} {R}_\textrm{Tr} \Pi_\textrm{Tr} +2 \Pi_c {R}_c  \Pi_c\Bigr] \nonumber \\[2ex]
	&= \sqrt{\bar g} \,\left( \begin{array}{cccc} 
		R_\textrm{TT} & 0 & 0 &0\\[1ex]
           0         & R_\textrm{Tr} & 0 &0\\[1ex] 
           0 &  0 & 0 & {R}_{\textrm{c}}  \\[1ex] 
           0 &0 & -{R}_{\textrm{c}}  &  0 
           \end{array} \right)\, .
\label{eq:RegMatrix}
\end{align}
It is block-diagonal and has entries in the spin two \textit{T}ransverse-\textit{T}raceless subspace, the scalar spin zero \textit{Tr}ace subspace  and the ghost subspace. 
In general, the regulator for each mode is a function of its respective spin $s$ background metric Laplacian $\bar \Delta_s$ and the spin $s$ background metric endomorphism $\bar{{\cal E}}_s$, 
	\begin{align}
		R_i = Z_{s_i} \, R_{i}\left(\bar{\Delta}_{s_i}, \bar{\cal E}_{s_i}\right) \qquad \text{for}   \quad  i = \text{TT}, \text{Tr}, \text{c} \, ,
		\label{eq:R_sApp}
	\end{align}
with $s_\textrm{TT}=2, s_\textrm{Tr}=0, s_\textrm{c}=1$ and the wave-function renormalisation $Z_{s_i}$.   
The spin $s$ Laplacian is given by 
\begin{align}
	\bar{\Delta}_s = - \bar{\nabla}_\mu \bar{\nabla}^\mu + \bar{\mathcal{E}}_s \, ,
\end{align}
where $\bar{\mathcal{E}}_s$ is the canonical curvature endomorphism of the mode $s$. In the harmonic gauge, it is given by
\begin{align}
	\bar{\mathcal{E}}_{\text{c}}
	&= \bar{\mathcal{R}}^{\mu}{}_{\nu},
	\qquad
	\bar{\mathcal{E}}_{\text{Tr}} = 0,
	\nonumber\\
	\bar{\mathcal{E}}_{\text{TT}}
	&= \frac{2}{3}\bar{\mathcal{R}}\,
	{\Pi^{\mathrm{TT}}}_{\mu\nu}{}^{\rho\sigma}
	- \bar{C}^{\mu\rho\nu\sigma}
	- \bar{C}^{\mu\sigma\nu\rho},
	\label{eq:endomorphisms}
\end{align}
with $C$ being the Weyl tensor. In the present work we restrict ourselves to the choice $R_i(\bar{\Delta}_{s_i},0)$ unless explicitly stated otherwise.  
The resulting flow equations within the background-field approximation within this truncation, gauge-fixing choice and regulator are given in \cite{Basile:2024oms}. 
	
In the present work we choose the Litim regulator for all spins $s$
\begin{align}
		R(z) = \left(k^2-z\right) \theta\left(1-\frac{z}{k^2}\right)  \,,
		\label{eq:litimReg}
\end{align}
to find the analytic beta functions.

%%%%%%%%%%%%%%%%%%%%%%%%%%%%%%
\section{Dyson-Schwinger equations for metric quantum gravity and the Nielsen identity}
\label{app:DSE+NI} 

The functional Dyson-Schwinger equation (DSE) for metric quantum gravity requires the specification of the gauge-fixed classical action and we resort to the sum of the Einstein-Hilbert action \labelcref{eq:EH-Action} and the gauge fixing action in \labelcref{eq:SgaugeApp}. This leads us to 
\begin{widetext} 
	\begin{align} 
		\left[\frac{ \delta}{\delta \bar g} +  \int \left\langle  \frac{\delta \hat\Phi_i}{\delta \bar g} \right\rangle \frac{ \delta}{\delta  \Phi_i}\right] \left( \Gamma_\Phi+\Delta S_k\right) =  \left\langle  \frac{\delta S_\textrm{EH}[\bar g+\hat h,\hat c,\hat{\bar c}]}{\delta \bar g}\right\rangle+\left\langle  \frac{\delta \left(S_\textrm{gauge}[\bar g,\bar g+\hat \Phi] +\Delta S_k[\bar g,\bar g+\hat \Phi]\right)}{\delta \bar g}\right\rangle\,. 
		\label{eq:FunDSEback}	
	\end{align} 
	The first term on the right-hand side has infinite loop orders in full propagators and full and classical vertices. This term also occurs in the fluctuation field DSE which reads 
	\begin{align} 
		\int \left\langle  \frac{\delta \hat\phi_h}{\delta \hat h} \right\rangle \frac{ \delta}{\delta \phi_h} \left( \Gamma_\Phi+\Delta S_k\right) =  \left\langle  \frac{\delta S_\textrm{EH}[\bar g+\hat h,\hat c,\hat{\bar c}]}{\delta \hat h }\right\rangle+\left\langle  \frac{\delta \left(S_\textrm{gauge}[\bar g,\bar g+\hat \Phi] +\Delta S_k[\bar g,\bar g+\hat \Phi]\right)}{\delta \hat h}\right\rangle\,. 
		\label{eq:FunDSEfluc}	
	\end{align} 
	We subtract the two functional DSEs which leads us to the Nielsen identity, 
	\begin{align} \nonumber 
		\left[ \frac{ \delta}{\delta \bar g} +  \int \left\langle\left( \frac{\delta }{\delta \bar g} -  \frac{\delta }{\delta \hat h} \right) \hat\phi_h \right\rangle \frac{ \delta}{\delta \phi_h} \right] \Gamma_\textrm{qu}[\bar g, \bar g+\Phi] & \\[2ex]
		&\hspace{-2cm}= \left\langle   S^{(1,0)}_\textrm{gauge} +\Delta S^{(1,0)}_k+\left[ \left( \frac{\delta }{\delta \bar g} - \frac{\delta }{\delta \hat h} \right) \hat\phi_h \right] \left(  S^{(0,1)}_\textrm{gauge} +\Delta S^{(0,1)}_k \right)\right\rangle_\textrm{1PI}\,, 
		\label{eq:NI}	
	\end{align} 
	where we have restricted ourselves to the case $\hat\phi_c= \hat c, \hat\phi_{\bar c}= \hat{\bar c}$. We also have shifted the mean field terms to the right-hand side, leading to $\Gamma_\Phi\to \Gamma_\textrm{qu}$. 
\end{widetext} 
The right-hand side of \labelcref{eq:NI} is two-loop exact, and the only two-loop term originates in the ghost action which contains up to three fluctuation fields. Now we drop higher loop order including the higher order terms in $\hat\phi_h$, 
\begin{align} 
	\left( \frac{\delta }{\delta \bar g} - \frac{\delta }{\delta \hat h} \right) \hat\phi_h[\bar g,\bar g+\Phi] \approx  \hat\phi_h^{(1,0)}\,. 
	\label{eq:Approxphih}
\end{align} 
This leads us to 
\begin{align} \nonumber 
	\left[ \frac{ \delta}{\delta \bar g} +  \int \left\langle  \hat\phi^{(1,0)}_h[\bar g,\bar g+\hat \Phi] \right\rangle \frac{ \delta}{\delta \phi_h} \right]  \Gamma_\textrm{qu} & \\[1ex] 
	&\hspace{-6cm} =\frac12 \textrm{Tr} \,G_k\,\PIG^{(1,0)}_k + \textrm{Tr} \,G_{\Phi}\, \left\langle  \hat\phi^{(1,1)}_h \right\rangle  \,\PIG_k\,.
	\label{eq:NIApprox}	
\end{align} 
Now we take the $t$-derivative in order to obtain $\partial_t \Gamma^{(1,0)}$ in the integrability condition \labelcref{eq:Integrability}. We drop higher order terms in the expectation values in \labelcref{eq:NIApprox}, 
\begin{align} 
	\left\langle  \hat\phi_h^{(1,0)} \right\rangle \approx -\mathbbm{1} \,,\qquad 
	\left\langle  \hat\phi_h^{(1,1)} \right\rangle  \approx 0\,,
	\label{eq:ApproxdotPhiExp}	
\end{align} 
and 
\begin{align} 
	\partial_t  \left\langle  \hat\phi_h^{(1,0)} \right\rangle \approx \dot \phi_h^{(1,0)}\,,\qquad 
	\partial_t  \left\langle  \hat\phi_h^{(1,1)} \right\rangle  \approx \dot \phi_h^{(1,1)}\,.
	\label{eq:ApproxdotPhiExpflow}	
\end{align} 
This leads us to our final approximation of the flow of the Nielsen identity 
\begin{align} \nonumber 
	\partial_t  \Gamma_\textrm{qu}^{(1,0)}+ \dot\phi_h^{(1,0)}\Gamma_\textrm{qu}^{(0,1)}& \\[1ex] 
	&\hspace{-3.5cm} =\frac12 \textrm{Tr} \,\partial_t \Bigl[G_{\Phi}\,\PIG^{(1,0)}_k\Bigr] + \textrm{Tr} \,G_{\Phi}\, \dot \phi_h^{(1,1)}  \,\PIG_k\,.
	\label{eq:NIFinalApprox}	
\end{align} 
This closes our derivation of the Nielsen identity. Apart from the two-loop ghost term we have applied the approximations 	\labelcref{eq:ApproxdotPhiExpflow}. These terms augment the $\Delta \textrm{Flow}$ terms in order to guarantee diffeomorphism invariance of the flow and we shall drop them in the following comparison of the driving force derived from the generalised flow equation and the Nielsen identity.  

For that purpose we consider the $\bar g$-derivative of the generalised flow \labelcref{eq:GenFlow}, 
\begin{align}\nonumber 
	\partial_t  \Gamma_\textrm{qu}^{(1,0)} + \dot\phi_h^{(1,0)} \Gamma_\textrm{qu}^{(0,1)} + \dot\phi_h  \Gamma_\textrm{qu}^{(1,1)} &\,   \\[1ex]\nonumber 
	&\hspace{-5cm}=\frac{1}{2} \textrm{Tr}\,\left[G_{\Phi}\,\partial_t R_k \right]^{(1,0)} + \textrm{Tr}\,G_{\Phi}\,\left[ \dot\phi_h^{(0,1)} R_k \right]^{(1,0)} \\[1ex]  
	&\hspace{-5cm}= \Delta\textrm{Flow}^{(1,0)} + \dot\phi_h^{(1,0)}-\textrm{terms}\,. 
	\label{eq:GenFlow-dbarg}
\end{align}
The difference of the left-hand sides of \labelcref{eq:NIFinalApprox} and \labelcref{eq:GenFlow-dbarg} vanishes  identically as $\Gamma_\textrm{qu}^{(1,m)}\equiv 0$ for all $m$. Accordingly, up to the $\dot\phi$-terms the right-hand sides are simply given by  $ \Delta\textrm{Flow}^{(1,0)}$. Rewriting the right-hand side of \labelcref{eq:NIFinalApprox} as a sum of $ \Delta\textrm{Flow}^{(1,0)}$, $\dot\phi_h$-terms and a remainder, this remainder has to be a higher order term. We find 
\begin{align} 
	\Delta\textrm{Flow}^{(1,0)} = \Delta\textrm{Flow}_\textrm{NI}^{(1,0)} +\frac12 \textrm{Tr}\,G_{\Phi}\PIG_k^{(1,0)}G_{\Phi}\,\partial_t\Gamma^{(2)}_\textrm{qu} \,. 
	\label{eq:DeltaFlowNI-Flow}
\end{align}
As expected, the difference term is of higher order and we shall use the results obtained with the $F_\Phi$ from the Nielsen identity as a systematic error estimate. We hasten to add that this is not a comprehensive analysis of the systematic error which requires far more work. 

We close this Appendix with the final expression for the minimal relevance-preserving flow or rather its first derivative. For the driving force we find 
\begin{align}
	\frac{\delta F_\Phi[g]}{\delta g} = - \frac{1}{2} \, \textrm{Tr}\, \partial_t \left[ G_{\Phi}[g, g]\, \PIG_k^{(1,0)}[g,g]\right] \,.
	\label{eq:DrivingNielsen} 
\end{align}
If we add this term to the $g$-derivative of the flow term, we obtain  
\begin{align}\nonumber 
	\partial_t \Gamma^{(1)}_\textrm{qu}[g] = &\,- \frac{1}{2} \, \textrm{Tr}\, G_\Phi\,\left(  \Gamma^{(3)}_\textrm{qu}+S_\textrm{gauge}^{(0,3)}\right) \, G_\Phi \, \partial_t R_\Phi \\[1ex]
	& \,+\frac{1}{2} \, \textrm{Tr}\, G_\Phi\,\PIG^{(1,0)}_k \, G_\Phi \,\partial_t \Gamma^{(2)}_\textrm{qu}\,. 
	\label{eq:FinalDrivingNIelsen} 
\end{align}
The first term on the right-hand side of \labelcref{eq:FinalDrivingNIelsen} is precisely \labelcref{eq:FPhiConstraint}. The term in the second line is proportional to $\partial_t \Gamma^{(2)}_\textrm{qu}$ and hence is a two-loop correction term as are the $\dot\phi$-terms. We use  	the comparison of the results derived from \labelcref{eq:FPhiConstraint} and \labelcref{eq:FinalDrivingNIelsen} for our systematic error estimate.

%%%%%%%%%%%%%%%%%%%%%%%
\section{Extended results}
\label{app:extendedResults}

\subsection{Anomalous dimensions}
\label{app:anomDim}

The anomalous dimensions for the graviton and ghost are defined 
\begin{align}
	\eta_c = -\partial_t \ln Z_c\, , \qquad \eta_h = -\partial_t \ln Z_h \, ,
\end{align}
with the wave-function renormalisation $Z$ defined in \labelcref{eq:R_sApp}. For the projection we choose a TT projector for the graviton and a transverse projector for the ghost. We approximate the anomalous dimensions at one-loop, i.e. we ignore the anomalous dimensions on the right-hand side of the Wetterich equation. We find
\begin{align}
	\eta_c &= - \frac{5 g_\Phi}{6\pi(1-2\lambda_\Phi)} - \frac{13 g_\Phi}{6\pi(1-2\lambda_\Phi)^2} \, , \nonumber \\ 
	\eta_h &= - \frac{2 g_\Phi}{3\pi(1-2\lambda_\Phi)^3} + \frac{11 g_\Phi}{4\pi(1-2\lambda_\Phi)^4} \, .
\end{align}

\subsection{Beta functions}
\label{app:betaFuncs}

%%%%%%%%%%%%%%%%%%%%%%%%%%%
\subsubsection{Beta functions for the Newton constant}
\label{sec:Beta}

The beta function for $g$ in the background-field approximation gauge fixing and regulator as given in \Cref{app:Approximation+GaugeFixing}  is given by
\begin{align}
	\beta_{g_\Phi}^{\text{BFA}} =& 2g_\Phi - \frac{14-3\eta_c}{6\pi} g_\Phi^2  \nonumber \\ 
	&+ \frac{5(4-\eta_h)}{12\pi(1-2\lambda_\Phi)}g_\Phi^2 - \frac{6-\eta_h}{2\pi(1-2\lambda_\Phi)^2} g_\Phi^2 \, .
\end{align}
We also complete the flow of $\lambda_\Phi$ with the beta function obtained within the fluctuation approach. For this we choose the running of the avatar $g_3$ from \cite{Denz:2016qks}. We identify $g_i = g_\Phi$, $\lambda_i = \lambda_\Phi$ and in particular $\mu_h = - 2 \lambda_\Phi$. Within these identifications, we obtain

\begin{align}
		&\beta_{g_\Phi}^{\text{fluc}} = 2g_\Phi + \frac{g_\Phi^2}{19\pi} \bigg( \frac{16 (1 - 3 \lambda_\Phi) \lambda_\Phi}{(1 - 2 \lambda_\Phi)^4}\nonumber \\  
		&-  \frac{2 (229 - 1780 \lambda_\Phi + 3640 \lambda_\Phi^2 - 2336 \lambda_\Phi^3)}{15 (1 - 2 \lambda_\Phi)^5} \nonumber \\ 
		& + \frac{1}{10} \bigl(480 + 53 (-10 + \eta_{c}{})\bigr) + \frac{47 (-6 + \eta_{h}{})}{6 (1 - 2 \lambda_\Phi)^2} \nonumber \\ 
		& -  \frac{45 (-8 + \eta_{h}{}) + 232 \lambda_\Phi (-6 + \eta_{h}{}) - 360 \lambda_\Phi^2 (-4 + \eta_{h}{})}{18 (1 - 2 \lambda_\Phi)^3} \nonumber \\ 
		&-  \frac{147 (-10 + \eta_{h}{}) - 1860 \lambda_\Phi (-8 + \eta_{h}{})}{80 (1 - 2 \lambda_\Phi)^4} \nonumber \\ 
		& -  \frac{3380 \lambda_\Phi^2 (-6 + \eta_{h}{}) + 25920 \lambda_\Phi^3 (-4 + \eta_{h}{})}{80 (1 - 2 \lambda_\Phi)^4}\bigg) \, .
\end{align}

\begin{figure*}[!t]
	\centering
	\includegraphics[width=\textwidth]{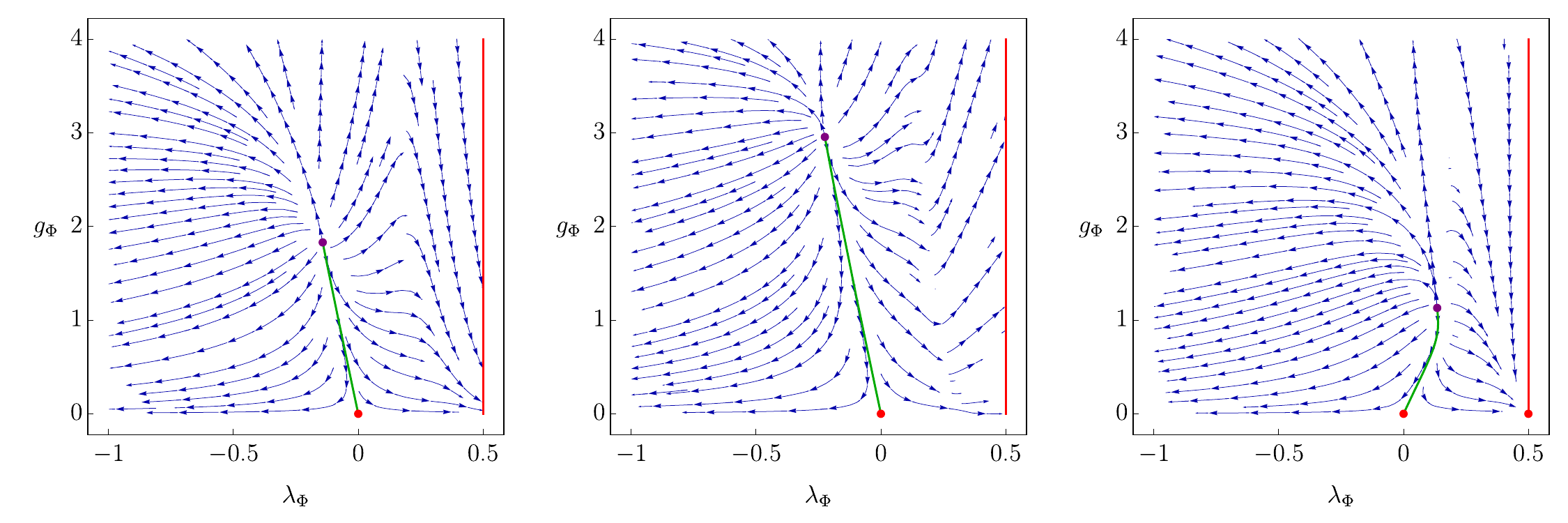}
	\caption{Phase diagram for the PIRG effective action $\Gamma_{\Phi, \textrm{qu}}^{\textrm{(a)}}$ for the dimensionless couplings $g_\Phi$ and $\lambda_\Phi$, for the direct computation augmented with $\beta_{g_\Phi}^{\textrm{(back)}}$ (left panel), with $\beta_{g_\Phi}^{\textrm{(fluc)}}$ (middle panel) and with the Nielsen strategy with $\beta_{g_\Phi}^{\textrm{(back)}}$ (right panel).  The red dots indicate IR fixed points, the violet dot is the Reuter fixed point. The green line connects the Reuter fixed point with the Gaussian fixed point, whereas the red line indicates the contour of a divergence of the flow.}	
	\label{fig:phaseDiagramOptionA}
\end{figure*}
%

%%%%%%%%%%%%%%%%%%%%%%%%%%%%%%%%%%%%%
\subsubsection{Beta function for the Nielsen strategy}
\label{eq:BetaNielsen}

The beta function for $\lambda_\Phi$ in the Nielsen strategy is given by
\begin{align}
	&\beta_{\lambda_\Phi}^{\text{(a), NI}} = -2\lambda_\Phi + \left(\frac{\beta_{g_\Phi}}{g_\Phi} - 2 \right) \lambda_\Phi + \frac{2g_\Phi}{3\pi} \nonumber \\ 
	& \, + \frac{-40g_\Phi^2 - 60\beta_{g_\Phi} \lambda_\Phi \pi + 3g_\Phi\pi(36+40\lambda_\Phi - 7 \eta_h)}{12\pi(5g_\Phi+ 6\pi(1-2\lambda_\Phi)^2)} \, ,
\end{align}
for the choice (a) in $\PIG_k$. For choice (b), we obtain
\begin{align}
	\beta_{\lambda_\Phi}^{\text{(b), NI}} =& -2\lambda_\Phi + \left(\frac{\beta_{g_\Phi}}{g_\Phi} - 2 \right) \lambda_\Phi - \frac{4-\eta_c}{6\pi}g_\Phi\nonumber \\ 
	& + \frac{5 (-2 (-4 + \eta _{c}{}) g_\Phi^2 -12 \beta _{g_\Phi}{} \lambda_\Phi  \pi )}{12 \pi (5 g_\Phi+ 6 \pi(1  -2 \lambda_\Phi )^2)} \nonumber \\ 
	&  + \frac{15 g_\Phi (4 + 8 \lambda_\Phi  - \eta _{h}{})}{12 (5 g_\Phi+ 6 \pi (1  -2 \lambda_\Phi )^2)} \, .
\end{align}

The beta function from the Nielsen strategy has a different structure from the one obtained from the direct computation, cf.~\labelcref{eq:betaDirectA} and \labelcref{eq:betaDirectB}.  This is due to the different structure generated by the flow of the Nielsen identity. The RG-time derivative acts upon the full trace, and not only on the regulator as is the direct strategy. This generates a further $\beta_{\lambda_\Phi}$ in $F_\Phi$, leading to terms proportional to $\beta_{\lambda_\Phi}$ on both sides of the gauge-invariant flow equation.

%%%%%%%%%%%%%%%%%%%%%%%%%%%%%%%%
\subsection{Phase diagrams}
\label{app:phaseDiagrams}

\begin{figure*}[!t]
	\centering
	\includegraphics[width=\textwidth]{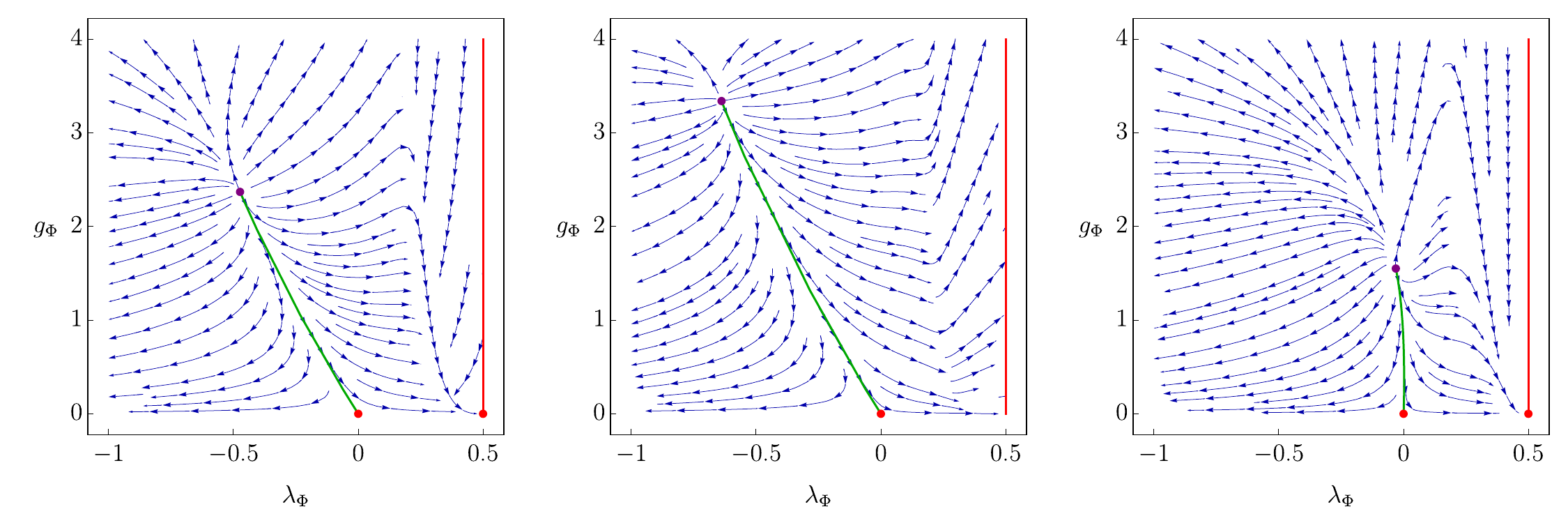}
	\caption{Phase Diagram for the PIRG effective action $\Gamma_{\Phi, \textrm{qu}}^{\textrm{(b)}}$ for the dimensionless couplings $g_\Phi$ and $\lambda_\Phi$, for the direct computation augmented with $\beta_{g_\Phi}^{\textrm{(back)}}$ (left panel), with $\beta_{g_\Phi}^{\textrm{(fluc)}}$ (middle panel) and with the Nielsen strategy with $\beta_{g_\Phi}^{\textrm{(back)}}$ (right panel). The red dots indicate IR fixed points, the violet dot is the Reuter fixed point. The green line connects the Reuter fixed point with the Gaussian fixed point, whereas the red line indicates the contour of a divergence of the flow.}	
	\label{fig:phaseDiagramOptionB}
\end{figure*}

We collect the phase diagrams for different procedures for option (a) and (b) in \Cref{fig:phaseDiagramOptionA} and \Cref{fig:phaseDiagramOptionB}, respectively. 
The fixed point for the Nielsen strategy for $\Gamma^{\text{(a)}}_\Phi$ has larger critical exponents than the direct strategy, \Cref{tab:comparResults}. This may be due to the large anomalous dimension $\eta_h$, which we obtained at $p^2=0$. We expect that a more sophisticated derivation of $F_\Phi$, involving the momentum-dependent two-point function changes this. 

%%%%%%%%%%%%%%%%%%%%
\section{Proper time flows and the Nielsen identity}
\label{app:NielsenPractitioner}  

In this Appendix, we derive the proper time and simplified flow equation \cite{Wetterich:2024ivi} from the PIRG flow. This also allows us to briefly discuss the differences between these flows and the minimal relevance-preserving flows. 

%%%%%%%%%%%%%%%%%%%%
\subsection{Proper time PIRG}
\label{app:ProperTimePIRG}

For this purpose we have to use regulators that depend on the full kinetic operator $\Gamma_\Phi^{(0,2)}$ in 
\labelcref{eq:CoregravityPIG},  evaluated at $\Phi=0$, 
\begin{align} 
	\Gamma^{(2)}_\Phi[g]:=  \Gamma^{(2)}_\textrm{qu}[g]+\PIG_k[g,g]\,. 
	\label{eq:FullKinetic} 
\end{align} 
The regulator is then parametrised as 
\begin{align}
	R_k\left(\Gamma_{\Phi}^{(2)}[\bar g]\right)=Z_k k^2 \,y\,r(y)\,, \quad y=\frac{1}{Z_k} \frac{\Gamma_{\Phi}^{(2)}[\bar g]}{ k^2}\,,
	\label{eq:RKinetic} 
\end{align}
where $Z_k$ denotes the wave function of the fluctuation field or that of the background metric. The inclusion of $Z_k$ ensures RG-invariance of $y$ with respect to the underlying RG-transformations of the full theory at $k=0$. Consequently, the regulators \labelcref{eq:RKinetic}  maintain the RG-invariance of the underlying theory at all $k$, see \cite{Pawlowski:2005xe}. Regulators \labelcref{eq:RKinetic} ensure, that the flow only depends on the kinetic operator and its $t$-derivative. The respective diffeomorphism-invariant PIRG is given by 
\begin{align}\nonumber 
	\partial_t \Gamma_\Phi[g]=&\, \frac{1}{2}\textrm{Tr}\,(2-\eta)\,g(y)\\[1ex]
	 &+\frac12\,\textrm{Tr}\, \left[\frac{g(y)+h(y)}{y}\right]\partial_t y +F_\Phi[g]\,,
	\label{eq:PTRG+Fphi}
\end{align}
with the kernels 
\begin{align}
 g(y)=\frac{r(y)}{1+r(y)}\,, \qquad 	h(y)=\frac{y\,r'(y)}{1+r(y)} \,.
	\label{eq:Kernelshg}
\end{align}
The proper time RG is simply given by the first term on the right-hand side in \labelcref{eq:PTRG+Fphi}. The second line has to vanish and the respective driving force is given by 
\begin{align} 
	F^\textrm{(PT)}_\Phi[g] = - \frac12\,\textrm{Tr}\, \frac{r(y)+y\,r'(y)}{y(1+r(y))} \partial_t y \,.
	\label{eq:FPhiPT}
\end{align} 
Using \labelcref{eq:FPhiPT} in 	\labelcref{eq:PTRG+Fphi} leads us to the proper time RG and simplified RG \cite{Wetterich:2024ivi}, 
\begin{align}
	\partial_t \Gamma^\textrm{(PT)}_\Phi[g]=\frac{1}{2}\textrm{Tr}\left[(2-\eta)\,\frac{r(y)}{1+r(y)}\right]\,, 
	\label{eq:PTRG-PIK}
\end{align}
with $y$ in \labelcref{eq:RKinetic}. We note in passing that in \cite{Wetterich:2024ivi} specific regulators of the form $r(y)= 1/((1+y)^n-1)$ have been used. Moreover, the map between the regulator shape function $r(y)$ and standard proper time regulators has been provided in \cite{Litim:2002hj}. 

All terms in \labelcref{eq:PTRG+Fphi,eq:FPhiPT,eq:PTRG-PIK} have a proper time representation. This facilitates the computations and we provide it here also for the sake of completeness. With the Laplace transforms  $\tilde h(s), \tilde g(s)$ together with their primitives 
\begin{align} 
\tilde H'(s)=\tilde h(s)\,,\qquad  \tilde G'(s)=\tilde g(s)\,,
\label{eq:HGPrimitives}
\end{align} 
the flow equation \labelcref{eq:PTRG-PIK} can be expressed in a proper time representation, 
	\begin{align}
	\partial_t \Gamma^\textrm{(PT)}_\Phi[g] 	=\,\frac{1}{2}\int_\mathbb{R^+}\mathrm{d}s\,\tilde{g} (s)\textrm{Tr}\,(2-\eta){e}^{-sy}\,. 
	\label{eq:FlowPTRG} 
\end{align}
The proper time driving force \labelcref{eq:FPhiPT} is given by 
\begin{align}
	F_\Phi^{\textrm{(PT)}}[g]=\frac{1}{2}\int\limits_\mathbb{R^+}\frac{\mathrm{d}s}{s}\Bigl[\tilde H(s)+\tilde G(s)\Bigr]\partial_t \,e^{-sy}\,.
	\label{eq:FPhiPTRep}
\end{align}
The proper time PIRG is provided by the combination of the proper time flow \labelcref{eq:PTRG-PIK} or \labelcref{eq:FlowPTRG} with the proper time driving force \labelcref{eq:FPhiPT} or \labelcref{eq:FPhiPTRep} and the respective $\dot\Phi^\textrm{(PT)}$ solving the PIK equation \labelcref{eq:PIKequation}.

%%%%%%%%%%%%%%%%%%%%
\subsection{Proper time PIRG and relevance preservation}
\label{app:PTPIRG-RP}  

Now we compare the proper time driving force \labelcref{eq:FPhiPT} with the minimal relevance-preserving ones derived from the flow, \labelcref{eq:FPhiConstraint}, and from the Nielsen identity, \labelcref{eq:FinalDrivingNIelsen}. The total $t$-derivative form of \labelcref{eq:FlowPTRG} allows us to map it directly to the driving force from the Nielsen identity, \labelcref{eq:FinalDrivingNIelsen}. To that end we note that the driving force simply cancels the $\partial_t y$ terms in the flow equation that are proportional to $\partial_y (y r)$. We find  
\begin{align}\nonumber 
&-\frac{1}{2}\textrm{Tr}\,\partial_t\,\left[G_\Phi \frac{\delta R_\Phi}{\delta g}\right] \\[1ex] \nonumber 
	=& -\frac{1}{2}   \textrm{Tr}\,\partial_t\int_\mathbb{R^+}\mathrm{d}s[\tilde g (s)+\tilde h(s)]\frac{\delta y}{\delta g}\frac{1}{y}{e}^{-sy}\\[1ex]\nonumber 
	=&\,-\frac{1}{2}\textrm{Tr}\,\partial_t\,\int_\mathbb{R^+}\mathrm{d}s\,\partial_s[\tilde G (s)+\tilde H(s)]\,\frac{\delta y}{\delta g}\,\frac{1}{y} \,e^{-sy}\\[1ex]\nonumber 
	=&\,\frac{1}{2}\frac{\delta }{\delta  g}\textrm{Tr}\int_\mathbb{R^+}\frac{\mathrm{d}s}{s}(\tilde H(s)+\tilde G(s))\,\partial_t\,e^{-sy}\\[1ex]
	=&	\frac{\delta F^{\textrm{(PT)}}_\Phi[g]}{\delta  g}\,.
\label{eq:DrivingForcePTNielsen1} 
\end{align} 
The difference of the relevance-preserving driving forces $F_\Phi^\textrm{(a,b)}$ in \labelcref{eq:DrivingNielsen}, derived from the Nielsen identity, and the proper time driving force is given by 
\begin{align} \nonumber 
	\frac{\delta F_\Phi^\textrm{(a,b)}[g]}{\delta g}-\frac{\delta F_\Phi^\textrm{(PT)}[g]}{\delta g}& \\[1ex]
	&\hspace{-2.5cm}= - \frac{1}{2} \, \textrm{Tr}\, \partial_t \left[ G_{\Phi}[g, g]\,  S_\textrm{gauge}^{(a,b),(1,2)}[g,g]\right] \,.
	\label{eq:DeltaFphiApp}
\end{align} 
We also note that the difference may feature additional two-loop terms proportional to $\partial_t \Gamma^{(2)}_\textrm{qu}$, $\dot\phi$ and further reparametrisation terms. These terms have been dropped in the derivation of the relevance-preserving driving force from the Nielsen identity, see \Cref{app:DSE+NI} and specifically \labelcref{eq:FinalDrivingNIelsen}. Interestingly, for $\Gamma_\Phi^{(b)}$ the difference in 	\labelcref{eq:DeltaFphiApp} vanishes as $S_\textrm{gauge}^{(1,m), (b)}[g,g]\equiv 0$.

\begin{figure}[t]
	\centering
	\includegraphics[width=0.39\textwidth]{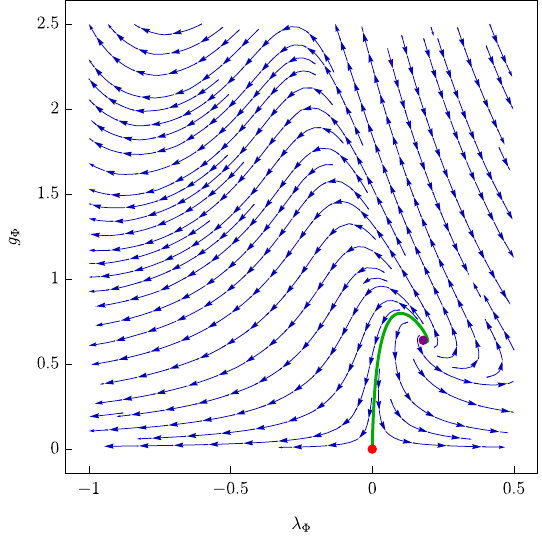}
	\caption{The phase diagram for the proper time PIRG. The red dot indicates the IR fixed point, the violet dot is the Reuter fixed point. The green line connects the Reuter fixed point with the Gaussian fixed point. The absence of the singularity is connected to the missing pole in \labelcref{eq:PTRG-PIK}  for the Litim cutoff.}	
	\label{fig:phaseDiagramSimple}
\end{figure}
The comparison of the respective fixed points reveals that the $\partial_t \Gamma^{(2)}_\textrm{qu}$ terms cannot be neglected at the fixed point. Moreover,  $S_\textrm{gauge}^{(1,2)}$ also plays a relevant role.

%%%%%%%%%%%%%%%%%%%%
\subsection{Proper time phase structure}
\label{app:PTPIRG-Comp}  

We close this Appendix with a brief discussion of the phase structure of asymptotically safe gravity derived from the proper time flow equation, e.g.~~\cite{Bonanno:2004sy, Bonanno:2019ukb}, or the simplified flow equation \cite{Wetterich:2024ivi}. The computation has been performed in the Einstein-Hilbert truncation with a Litim regulator and the phase structure shown in \Cref{fig:phaseDiagramSimple}. Note that in contradistinction to the minimally relevance-preserving PIRGs, the driving force of the proper time RG is not restricted to $\sqrt{g}$ or only the relevant operators. This leads to sizeable differences from the phase structure of the choice (b) as derived from the flow of the Nielsen identity shown in  \Cref{fig:phaseDiagramOptionB}. 

The respective Reuter fixed point is located at  
\begin{align}
	(\lambda_\Phi^\ast,g_\Phi^\ast)=(0.18, \, 0.64)\,,
\end{align}
with the critical exponents
\begin{align}
	\theta_{1,2}=1.65\pm1.99\,\mathrm{i}\,. 
\end{align}
The critical exponents form a complex-conjugate pair, a property commonly present in the background-field approximation.

%%%%%%%%%%%%%%%%%%%%
\bibliography{references}

%merlin.mbs apsrev4-1.bst 2010-07-25 4.21a (PWD, AO, DPC) hacked
%Control: key (0)
%Control: author (8) initials jnrlst
%Control: editor formatted (1) identically to author
%Control: production of article title (-1) disabled
%Control: page (0) single
%Control: year (1) truncated
%Control: production of eprint (0) enabled
\begin{thebibliography}{115}%
\makeatletter
\providecommand \@ifxundefined [1]{%
 \@ifx{#1\undefined}
}%
\providecommand \@ifnum [1]{%
 \ifnum #1\expandafter \@firstoftwo
 \else \expandafter \@secondoftwo
 \fi
}%
\providecommand \@ifx [1]{%
 \ifx #1\expandafter \@firstoftwo
 \else \expandafter \@secondoftwo
 \fi
}%
\providecommand \natexlab [1]{#1}%
\providecommand \enquote  [1]{``#1''}%
\providecommand \bibnamefont  [1]{#1}%
\providecommand \bibfnamefont [1]{#1}%
\providecommand \citenamefont [1]{#1}%
\providecommand \href@noop [0]{\@secondoftwo}%
\providecommand \href [0]{\begingroup \@sanitize@url \@href}%
\providecommand \@href[1]{\@@startlink{#1}\@@href}%
\providecommand \@@href[1]{\endgroup#1\@@endlink}%
\providecommand \@sanitize@url [0]{\catcode `\\12\catcode `\$12\catcode
  `\&12\catcode `\#12\catcode `\^12\catcode `\_12\catcode `\%12\relax}%
\providecommand \@@startlink[1]{}%
\providecommand \@@endlink[0]{}%
\providecommand \url  [0]{\begingroup\@sanitize@url \@url }%
\providecommand \@url [1]{\endgroup\@href {#1}{\urlprefix }}%
\providecommand \urlprefix  [0]{URL }%
\providecommand \Eprint [0]{\href }%
\providecommand \doibase [0]{http://dx.doi.org/}%
\providecommand \selectlanguage [0]{\@gobble}%
\providecommand \bibinfo  [0]{\@secondoftwo}%
\providecommand \bibfield  [0]{\@secondoftwo}%
\providecommand \translation [1]{[#1]}%
\providecommand \BibitemOpen [0]{}%
\providecommand \bibitemStop [0]{}%
\providecommand \bibitemNoStop [0]{.\EOS\space}%
\providecommand \EOS [0]{\spacefactor3000\relax}%
\providecommand \BibitemShut  [1]{\csname bibitem#1\endcsname}%
\let\auto@bib@innerbib\@empty
%</preamble>
\bibitem [{\citenamefont {Weinberg}(1980)}]{Weinberg:1980gg}%
  \BibitemOpen
  \bibfield  {author} {\bibinfo {author} {\bibfnamefont {S.}~\bibnamefont
  {Weinberg}},\ }\enquote {\bibinfo {title} {{Ultraviolet divergences in
  quantum theories of gravitation}},}\ in\ \href@noop {} {\emph {\bibinfo
  {booktitle} {{General Relativity}: {An Einstein Centenary Survey}}}}\
  (\bibinfo {year} {1980})\ pp.\ \bibinfo {pages} {790--831}\BibitemShut
  {NoStop}%
\bibitem [{\citenamefont {Reuter}(1998)}]{Reuter:1996cp}%
  \BibitemOpen
  \bibfield  {author} {\bibinfo {author} {\bibfnamefont {M.}~\bibnamefont
  {Reuter}},\ }\href {\doibase 10.1103/PhysRevD.57.971} {\bibfield  {journal}
  {\bibinfo  {journal} {Phys. Rev. D}\ }\textbf {\bibinfo {volume} {57}},\
  \bibinfo {pages} {971} (\bibinfo {year} {1998})},\ \Eprint
  {http://arxiv.org/abs/hep-th/9605030} {arXiv:hep-th/9605030} \BibitemShut
  {NoStop}%
\bibitem [{\citenamefont {Dupuis}\ \emph {et~al.}(2021)\citenamefont {Dupuis},
  \citenamefont {Canet}, \citenamefont {Eichhorn}, \citenamefont {Metzner},
  \citenamefont {Pawlowski}, \citenamefont {Tissier},\ and\ \citenamefont
  {Wschebor}}]{Dupuis:2020fhh}%
  \BibitemOpen
  \bibfield  {author} {\bibinfo {author} {\bibfnamefont {N.}~\bibnamefont
  {Dupuis}}, \bibinfo {author} {\bibfnamefont {L.}~\bibnamefont {Canet}},
  \bibinfo {author} {\bibfnamefont {A.}~\bibnamefont {Eichhorn}}, \bibinfo
  {author} {\bibfnamefont {W.}~\bibnamefont {Metzner}}, \bibinfo {author}
  {\bibfnamefont {J.~M.}\ \bibnamefont {Pawlowski}}, \bibinfo {author}
  {\bibfnamefont {M.}~\bibnamefont {Tissier}}, \ and\ \bibinfo {author}
  {\bibfnamefont {N.}~\bibnamefont {Wschebor}},\ }\href {\doibase
  10.1016/j.physrep.2021.01.001} {\bibfield  {journal} {\bibinfo  {journal}
  {Phys. Rept.}\ }\textbf {\bibinfo {volume} {910}},\ \bibinfo {pages} {1}
  (\bibinfo {year} {2021})},\ \Eprint {http://arxiv.org/abs/2006.04853}
  {arXiv:2006.04853 [cond-mat.stat-mech]} \BibitemShut {NoStop}%
\bibitem [{\citenamefont {Bonanno}\ \emph
  {et~al.}(2020{\natexlab{a}})\citenamefont {Bonanno}, \citenamefont
  {Eichhorn}, \citenamefont {Gies}, \citenamefont {Pawlowski}, \citenamefont
  {Percacci}, \citenamefont {Reuter}, \citenamefont {Saueressig},\ and\
  \citenamefont {Vacca}}]{Bonanno:2020bil}%
  \BibitemOpen
  \bibfield  {author} {\bibinfo {author} {\bibfnamefont {A.}~\bibnamefont
  {Bonanno}}, \bibinfo {author} {\bibfnamefont {A.}~\bibnamefont {Eichhorn}},
  \bibinfo {author} {\bibfnamefont {H.}~\bibnamefont {Gies}}, \bibinfo {author}
  {\bibfnamefont {J.~M.}\ \bibnamefont {Pawlowski}}, \bibinfo {author}
  {\bibfnamefont {R.}~\bibnamefont {Percacci}}, \bibinfo {author}
  {\bibfnamefont {M.}~\bibnamefont {Reuter}}, \bibinfo {author} {\bibfnamefont
  {F.}~\bibnamefont {Saueressig}}, \ and\ \bibinfo {author} {\bibfnamefont
  {G.~P.}\ \bibnamefont {Vacca}},\ }\href {\doibase 10.3389/fphy.2020.00269}
  {\bibfield  {journal} {\bibinfo  {journal} {Front. in Phys.}\ }\textbf
  {\bibinfo {volume} {8}},\ \bibinfo {pages} {269} (\bibinfo {year}
  {2020}{\natexlab{a}})},\ \Eprint {http://arxiv.org/abs/2004.06810}
  {arXiv:2004.06810 [gr-qc]} \BibitemShut {NoStop}%
\bibitem [{\citenamefont {Pawlowski}\ and\ \citenamefont
  {Reichert}(2021)}]{Pawlowski:2020qer}%
  \BibitemOpen
  \bibfield  {author} {\bibinfo {author} {\bibfnamefont {J.~M.}\ \bibnamefont
  {Pawlowski}}\ and\ \bibinfo {author} {\bibfnamefont {M.}~\bibnamefont
  {Reichert}},\ }\href {\doibase 10.3389/fphy.2020.551848} {\bibfield
  {journal} {\bibinfo  {journal} {Front. in Phys.}\ }\textbf {\bibinfo {volume}
  {8}},\ \bibinfo {pages} {551848} (\bibinfo {year} {2021})},\ \Eprint
  {http://arxiv.org/abs/2007.10353} {arXiv:2007.10353 [hep-th]} \BibitemShut
  {NoStop}%
\bibitem [{\citenamefont {Reichert}(2020)}]{Reichert:2020mja}%
  \BibitemOpen
  \bibfield  {author} {\bibinfo {author} {\bibfnamefont {M.}~\bibnamefont
  {Reichert}},\ }\href {\doibase 10.22323/1.384.0005} {\bibfield  {journal}
  {\bibinfo  {journal} {PoS}\ }\textbf {\bibinfo {volume} {384}},\ \bibinfo
  {pages} {005} (\bibinfo {year} {2020})}\BibitemShut {NoStop}%
\bibitem [{\citenamefont {Basile}\ \emph {et~al.}(2025)\citenamefont {Basile},
  \citenamefont {Buoninfante}, \citenamefont {Di~Filippo}, \citenamefont
  {Knorr}, \citenamefont {Platania},\ and\ \citenamefont
  {Tokareva}}]{Basile:2024oms}%
  \BibitemOpen
  \bibfield  {author} {\bibinfo {author} {\bibfnamefont {I.}~\bibnamefont
  {Basile}}, \bibinfo {author} {\bibfnamefont {L.}~\bibnamefont {Buoninfante}},
  \bibinfo {author} {\bibfnamefont {F.}~\bibnamefont {Di~Filippo}}, \bibinfo
  {author} {\bibfnamefont {B.}~\bibnamefont {Knorr}}, \bibinfo {author}
  {\bibfnamefont {A.}~\bibnamefont {Platania}}, \ and\ \bibinfo {author}
  {\bibfnamefont {A.}~\bibnamefont {Tokareva}},\ }\href {\doibase
  10.21468/SciPostPhysLectNotes.98} {\bibfield  {journal} {\bibinfo  {journal}
  {SciPost Phys. Lect. Notes}\ }\textbf {\bibinfo {volume} {98}},\ \bibinfo
  {pages} {1} (\bibinfo {year} {2025})},\ \Eprint
  {http://arxiv.org/abs/2412.08690} {arXiv:2412.08690 [hep-th]} \BibitemShut
  {NoStop}%
\bibitem [{\citenamefont {Eichhorn}(2026)}]{Eichhorn:2026uqj}%
  \BibitemOpen
  \bibfield  {author} {\bibinfo {author} {\bibfnamefont {A.}~\bibnamefont
  {Eichhorn}},\ }\href@noop {} {\  (\bibinfo {year} {2026})},\ \Eprint
  {http://arxiv.org/abs/2606.21522} {arXiv:2606.21522 [hep-th]} \BibitemShut
  {NoStop}%
\bibitem [{\citenamefont {Bambi}\ \emph {et~al.}(2024)\citenamefont {Bambi},
  \citenamefont {Modesto},\ and\ \citenamefont {Shapiro}}]{Bambi:2023jiz}%
  \BibitemOpen
  \bibinfo {editor} {\bibfnamefont {C.}~\bibnamefont {Bambi}}, \bibinfo
  {editor} {\bibfnamefont {L.}~\bibnamefont {Modesto}}, \ and\ \bibinfo
  {editor} {\bibfnamefont {I.}~\bibnamefont {Shapiro}},\ eds.,\ \href {\doibase
  10.1007/978-981-99-7681-2} {\emph {\bibinfo {title} {{Handbook of Quantum
  Gravity}}}}\ (\bibinfo  {publisher} {Springer},\ \bibinfo {year}
  {2024})\BibitemShut {NoStop}%
\bibitem [{\citenamefont {Knorr}\ \emph {et~al.}(2023)\citenamefont {Knorr},
  \citenamefont {Ripken},\ and\ \citenamefont {Saueressig}}]{Knorr:2022dsx}%
  \BibitemOpen
  \bibfield  {author} {\bibinfo {author} {\bibfnamefont {B.}~\bibnamefont
  {Knorr}}, \bibinfo {author} {\bibfnamefont {C.}~\bibnamefont {Ripken}}, \
  and\ \bibinfo {author} {\bibfnamefont {F.}~\bibnamefont {Saueressig}},\ }in\
  \href {\doibase 10.1007/978-981-19-3079-9_21-1} {\emph {\bibinfo {booktitle}
  {Handbook of Quantum Gravity}}}\ (\bibinfo  {publisher} {Springer Nature
  Singapore},\ \bibinfo {address} {Singapore},\ \bibinfo {year} {2023})\
  \Eprint {http://arxiv.org/abs/2210.16072} {arXiv:2210.16072 [hep-th]}
  \BibitemShut {NoStop}%
\bibitem [{\citenamefont {Eichhorn}\ and\ \citenamefont
  {Schiffer}(2024)}]{Eichhorn:2022gku}%
  \BibitemOpen
  \bibfield  {author} {\bibinfo {author} {\bibfnamefont {A.}~\bibnamefont
  {Eichhorn}}\ and\ \bibinfo {author} {\bibfnamefont {M.}~\bibnamefont
  {Schiffer}},\ }in\ \href {\doibase 10.1007/978-981-99-7681-2_22} {\emph
  {\bibinfo {booktitle} {Handbook of Quantum Gravity}}}\ (\bibinfo  {publisher}
  {Springer Nature Singapore},\ \bibinfo {address} {Singapore},\ \bibinfo
  {year} {2024})\ \Eprint {http://arxiv.org/abs/2212.07456} {arXiv:2212.07456
  [hep-th]} \BibitemShut {NoStop}%
\bibitem [{\citenamefont {Morris}\ and\ \citenamefont
  {Stulga}(2023)}]{Morris:2022btf}%
  \BibitemOpen
  \bibfield  {author} {\bibinfo {author} {\bibfnamefont {T.~R.}\ \bibnamefont
  {Morris}}\ and\ \bibinfo {author} {\bibfnamefont {D.}~\bibnamefont
  {Stulga}},\ }in\ \href {\doibase 10.1007/978-981-19-3079-9_19-1} {\emph
  {\bibinfo {booktitle} {Handbook of Quantum Gravity}}}\ (\bibinfo  {publisher}
  {Springer Nature Singapore},\ \bibinfo {address} {Singapore},\ \bibinfo
  {year} {2023})\ pp.\ \bibinfo {pages} {1--33},\ \Eprint
  {http://arxiv.org/abs/2210.11356} {arXiv:2210.11356 [hep-th]} \BibitemShut
  {NoStop}%
\bibitem [{\citenamefont {Wetterich}(2023)}]{Wetterich:2022ncl}%
  \BibitemOpen
  \bibfield  {author} {\bibinfo {author} {\bibfnamefont {C.}~\bibnamefont
  {Wetterich}},\ }in\ \href {\doibase 10.1007/978-981-19-3079-9_26-1} {\emph
  {\bibinfo {booktitle} {Handbook of Quantum Gravity}}}\ (\bibinfo  {publisher}
  {Springer Nature Singapore},\ \bibinfo {address} {Singapore},\ \bibinfo
  {year} {2023})\ \Eprint {http://arxiv.org/abs/2211.03596} {arXiv:2211.03596
  [gr-qc]} \BibitemShut {NoStop}%
\bibitem [{\citenamefont {Martini}\ \emph {et~al.}(2023)\citenamefont
  {Martini}, \citenamefont {Vacca},\ and\ \citenamefont
  {Zanusso}}]{Martini:2022sll}%
  \BibitemOpen
  \bibfield  {author} {\bibinfo {author} {\bibfnamefont {R.}~\bibnamefont
  {Martini}}, \bibinfo {author} {\bibfnamefont {G.~P.}\ \bibnamefont {Vacca}},
  \ and\ \bibinfo {author} {\bibfnamefont {O.}~\bibnamefont {Zanusso}},\ }in\
  \href {\doibase 10.1007/978-981-19-3079-9_25-1} {\emph {\bibinfo {booktitle}
  {Handbook of Quantum Gravity}}}\ (\bibinfo  {publisher} {Springer Nature
  Singapore},\ \bibinfo {address} {Singapore},\ \bibinfo {year} {2023})\ pp.\
  \bibinfo {pages} {1--46},\ \Eprint {http://arxiv.org/abs/2210.13910}
  {arXiv:2210.13910 [hep-th]} \BibitemShut {NoStop}%
\bibitem [{\citenamefont {Saueressig}(2023)}]{Saueressig:2023irs}%
  \BibitemOpen
  \bibfield  {author} {\bibinfo {author} {\bibfnamefont {F.}~\bibnamefont
  {Saueressig}},\ }in\ \href {\doibase 10.1007/978-981-19-3079-9_16-1} {\emph
  {\bibinfo {booktitle} {Handbook of Quantum Gravity}}}\ (\bibinfo  {publisher}
  {Springer Nature Singapore},\ \bibinfo {address} {Singapore},\ \bibinfo
  {year} {2023})\ pp.\ \bibinfo {pages} {1--33},\ \Eprint
  {http://arxiv.org/abs/2302.14152} {arXiv:2302.14152 [hep-th]} \BibitemShut
  {NoStop}%
\bibitem [{\citenamefont {Pawlowski}\ and\ \citenamefont
  {Reichert}(2023)}]{Pawlowski:2023gym}%
  \BibitemOpen
  \bibfield  {author} {\bibinfo {author} {\bibfnamefont {J.~M.}\ \bibnamefont
  {Pawlowski}}\ and\ \bibinfo {author} {\bibfnamefont {M.}~\bibnamefont
  {Reichert}},\ }in\ \href {\doibase 10.1007/978-981-19-3079-9_17-1} {\emph
  {\bibinfo {booktitle} {Handbook of Quantum Gravity}}}\ (\bibinfo  {publisher}
  {Springer Nature Singapore},\ \bibinfo {address} {Singapore},\ \bibinfo
  {year} {2023})\ \Eprint {http://arxiv.org/abs/2309.10785} {arXiv:2309.10785
  [hep-th]} \BibitemShut {NoStop}%
\bibitem [{\citenamefont {Platania}(2023)}]{Platania:2023srt}%
  \BibitemOpen
  \bibfield  {author} {\bibinfo {author} {\bibfnamefont {A.}~\bibnamefont
  {Platania}},\ }in\ \href {\doibase 10.1007/978-981-19-3079-9_24-1} {\emph
  {\bibinfo {booktitle} {Handbook of Quantum Gravity}}}\ (\bibinfo  {publisher}
  {Springer Nature Singapore},\ \bibinfo {address} {Singapore},\ \bibinfo
  {year} {2023})\ \Eprint {http://arxiv.org/abs/2302.04272} {arXiv:2302.04272
  [gr-qc]} \BibitemShut {NoStop}%
\bibitem [{\citenamefont {Bonanno}(2023)}]{Bonanno:2024xne}%
  \BibitemOpen
  \bibfield  {author} {\bibinfo {author} {\bibfnamefont {A.}~\bibnamefont
  {Bonanno}},\ }in\ \href {\doibase 10.1007/978-981-19-3079-9_23-1} {\emph
  {\bibinfo {booktitle} {Handbook of Quantum Gravity}}}\ (\bibinfo  {publisher}
  {Springer Nature Singapore},\ \bibinfo {address} {Singapore},\ \bibinfo
  {year} {2023})\BibitemShut {NoStop}%
\bibitem [{\citenamefont {Bonanno}\ and\ \citenamefont
  {Reuter}(2005)}]{Bonanno:2004sy}%
  \BibitemOpen
  \bibfield  {author} {\bibinfo {author} {\bibfnamefont {A.}~\bibnamefont
  {Bonanno}}\ and\ \bibinfo {author} {\bibfnamefont {M.}~\bibnamefont
  {Reuter}},\ }\href {\doibase 10.1088/1126-6708/2005/02/035} {\bibfield
  {journal} {\bibinfo  {journal} {JHEP}\ }\textbf {\bibinfo {volume} {02}},\
  \bibinfo {pages} {035} (\bibinfo {year} {2005})},\ \Eprint
  {http://arxiv.org/abs/hep-th/0410191} {arXiv:hep-th/0410191} \BibitemShut
  {NoStop}%
\bibitem [{\citenamefont {Bonanno}\ \emph
  {et~al.}(2020{\natexlab{b}})\citenamefont {Bonanno}, \citenamefont
  {Lippoldt}, \citenamefont {Percacci},\ and\ \citenamefont
  {Vacca}}]{Bonanno:2019ukb}%
  \BibitemOpen
  \bibfield  {author} {\bibinfo {author} {\bibfnamefont {A.}~\bibnamefont
  {Bonanno}}, \bibinfo {author} {\bibfnamefont {S.}~\bibnamefont {Lippoldt}},
  \bibinfo {author} {\bibfnamefont {R.}~\bibnamefont {Percacci}}, \ and\
  \bibinfo {author} {\bibfnamefont {G.~P.}\ \bibnamefont {Vacca}},\ }\href
  {\doibase 10.1140/epjc/s10052-020-7798-9} {\bibfield  {journal} {\bibinfo
  {journal} {Eur. Phys. J. C}\ }\textbf {\bibinfo {volume} {80}},\ \bibinfo
  {pages} {249} (\bibinfo {year} {2020}{\natexlab{b}})},\ \Eprint
  {http://arxiv.org/abs/1912.08135} {arXiv:1912.08135 [hep-th]} \BibitemShut
  {NoStop}%
\bibitem [{\citenamefont {Wetterich}(2025)}]{Wetterich:2024ivi}%
  \BibitemOpen
  \bibfield  {author} {\bibinfo {author} {\bibfnamefont {C.}~\bibnamefont
  {Wetterich}},\ }\href {\doibase 10.1016/j.physletb.2025.139435} {\bibfield
  {journal} {\bibinfo  {journal} {Phys. Lett. B}\ }\textbf {\bibinfo {volume}
  {864}},\ \bibinfo {pages} {139435} (\bibinfo {year} {2025})},\ \Eprint
  {http://arxiv.org/abs/2403.17523} {arXiv:2403.17523 [hep-th]} \BibitemShut
  {NoStop}%
\bibitem [{\citenamefont {Meibohm}\ \emph {et~al.}(2016)\citenamefont
  {Meibohm}, \citenamefont {Pawlowski},\ and\ \citenamefont
  {Reichert}}]{Meibohm:2015twa}%
  \BibitemOpen
  \bibfield  {author} {\bibinfo {author} {\bibfnamefont {J.}~\bibnamefont
  {Meibohm}}, \bibinfo {author} {\bibfnamefont {J.~M.}\ \bibnamefont
  {Pawlowski}}, \ and\ \bibinfo {author} {\bibfnamefont {M.}~\bibnamefont
  {Reichert}},\ }\href {\doibase 10.1103/PhysRevD.93.084035} {\bibfield
  {journal} {\bibinfo  {journal} {Phys. Rev. D}\ }\textbf {\bibinfo {volume}
  {93}},\ \bibinfo {pages} {084035} (\bibinfo {year} {2016})},\ \Eprint
  {http://arxiv.org/abs/1510.07018} {arXiv:1510.07018 [hep-th]} \BibitemShut
  {NoStop}%
\bibitem [{\citenamefont {Ihssen}\ and\ \citenamefont
  {Pawlowski}(2023{\natexlab{a}})}]{Ihssen:2023nqd}%
  \BibitemOpen
  \bibfield  {author} {\bibinfo {author} {\bibfnamefont {F.}~\bibnamefont
  {Ihssen}}\ and\ \bibinfo {author} {\bibfnamefont {J.~M.}\ \bibnamefont
  {Pawlowski}},\ }\href@noop {} {\  (\bibinfo {year} {2023}{\natexlab{a}})},\
  \Eprint {http://arxiv.org/abs/2305.00816} {arXiv:2305.00816 [hep-th]}
  \BibitemShut {NoStop}%
\bibitem [{\citenamefont {Ihssen}\ and\ \citenamefont
  {Pawlowski}(2025{\natexlab{a}})}]{Ihssen:2024ihp}%
  \BibitemOpen
  \bibfield  {author} {\bibinfo {author} {\bibfnamefont {F.}~\bibnamefont
  {Ihssen}}\ and\ \bibinfo {author} {\bibfnamefont {J.~M.}\ \bibnamefont
  {Pawlowski}},\ }\href {\doibase 10.1016/j.aop.2025.170177} {\bibfield
  {journal} {\bibinfo  {journal} {Annals Phys.}\ }\textbf {\bibinfo {volume}
  {481}},\ \bibinfo {pages} {170177} (\bibinfo {year} {2025}{\natexlab{a}})},\
  \Eprint {http://arxiv.org/abs/2409.13679} {arXiv:2409.13679 [hep-th]}
  \BibitemShut {NoStop}%
\bibitem [{\citenamefont {Ihssen}\ and\ \citenamefont
  {Pawlowski}(2025{\natexlab{b}})}]{Ihssen:2025cff}%
  \BibitemOpen
  \bibfield  {author} {\bibinfo {author} {\bibfnamefont {F.}~\bibnamefont
  {Ihssen}}\ and\ \bibinfo {author} {\bibfnamefont {J.~M.}\ \bibnamefont
  {Pawlowski}},\ }\href {\doibase 10.1103/jhqd-cjzk} {\bibfield  {journal}
  {\bibinfo  {journal} {Phys. Rev. D}\ }\textbf {\bibinfo {volume} {112}},\
  \bibinfo {pages} {105005} (\bibinfo {year} {2025}{\natexlab{b}})},\ \Eprint
  {http://arxiv.org/abs/2503.22638} {arXiv:2503.22638 [hep-th]} \BibitemShut
  {NoStop}%
\bibitem [{\citenamefont {Ihssen}\ and\ \citenamefont
  {Pawlowski}(2026)}]{Ihssen:2025hyl}%
  \BibitemOpen
  \bibfield  {author} {\bibinfo {author} {\bibfnamefont {F.}~\bibnamefont
  {Ihssen}}\ and\ \bibinfo {author} {\bibfnamefont {J.~M.}\ \bibnamefont
  {Pawlowski}},\ }\href {\doibase 10.1103/zy96-jg9g} {\bibfield  {journal}
  {\bibinfo  {journal} {Phys. Rev. D}\ }\textbf {\bibinfo {volume} {113}},\
  \bibinfo {pages} {076003} (\bibinfo {year} {2026})},\ \Eprint
  {http://arxiv.org/abs/2507.13011} {arXiv:2507.13011 [hep-th]} \BibitemShut
  {NoStop}%
\bibitem [{\citenamefont {Pawlowski}(2007)}]{Pawlowski:2005xe}%
  \BibitemOpen
  \bibfield  {author} {\bibinfo {author} {\bibfnamefont {J.~M.}\ \bibnamefont
  {Pawlowski}},\ }\href {\doibase 10.1016/j.aop.2007.01.007} {\bibfield
  {journal} {\bibinfo  {journal} {Annals Phys.}\ }\textbf {\bibinfo {volume}
  {322}},\ \bibinfo {pages} {2831} (\bibinfo {year} {2007})},\ \Eprint
  {http://arxiv.org/abs/hep-th/0512261} {arXiv:hep-th/0512261} \BibitemShut
  {NoStop}%
\bibitem [{\citenamefont {Baldazzi}\ \emph {et~al.}(2022)\citenamefont
  {Baldazzi}, \citenamefont {Zinati},\ and\ \citenamefont
  {Falls}}]{Baldazzi:2021ydj}%
  \BibitemOpen
  \bibfield  {author} {\bibinfo {author} {\bibfnamefont {A.}~\bibnamefont
  {Baldazzi}}, \bibinfo {author} {\bibfnamefont {R.~B.~A.}\ \bibnamefont
  {Zinati}}, \ and\ \bibinfo {author} {\bibfnamefont {K.}~\bibnamefont
  {Falls}},\ }\href {\doibase 10.21468/SciPostPhys.13.4.085} {\bibfield
  {journal} {\bibinfo  {journal} {SciPost Phys.}\ }\textbf {\bibinfo {volume}
  {13}},\ \bibinfo {pages} {085} (\bibinfo {year} {2022})},\ \Eprint
  {http://arxiv.org/abs/2105.11482} {arXiv:2105.11482 [hep-th]} \BibitemShut
  {NoStop}%
\bibitem [{\citenamefont {Knorr}(2022)}]{Knorr:2022ilz}%
  \BibitemOpen
  \bibfield  {author} {\bibinfo {author} {\bibfnamefont {B.}~\bibnamefont
  {Knorr}},\ }\href@noop {} {\  (\bibinfo {year} {2022})},\ \Eprint
  {http://arxiv.org/abs/2204.08564} {arXiv:2204.08564 [hep-th]} \BibitemShut
  {NoStop}%
\bibitem [{\citenamefont {Knorr}(2024)}]{Knorr:2023usb}%
  \BibitemOpen
  \bibfield  {author} {\bibinfo {author} {\bibfnamefont {B.}~\bibnamefont
  {Knorr}},\ }\href {\doibase 10.1103/PhysRevD.110.026001} {\bibfield
  {journal} {\bibinfo  {journal} {Phys. Rev. D}\ }\textbf {\bibinfo {volume}
  {110}},\ \bibinfo {pages} {026001} (\bibinfo {year} {2024})},\ \Eprint
  {http://arxiv.org/abs/2311.12097} {arXiv:2311.12097 [hep-th]} \BibitemShut
  {NoStop}%
\bibitem [{\citenamefont {Baldazzi}\ \emph {et~al.}(2026)\citenamefont
  {Baldazzi}, \citenamefont {Falls}, \citenamefont {Kluth},\ and\ \citenamefont
  {Knorr}}]{Baldazzi:2023pep}%
  \BibitemOpen
  \bibfield  {author} {\bibinfo {author} {\bibfnamefont {A.}~\bibnamefont
  {Baldazzi}}, \bibinfo {author} {\bibfnamefont {K.}~\bibnamefont {Falls}},
  \bibinfo {author} {\bibfnamefont {Y.}~\bibnamefont {Kluth}}, \ and\ \bibinfo
  {author} {\bibfnamefont {B.}~\bibnamefont {Knorr}},\ }\href {\doibase
  10.1103/hlrm-d4g2} {\bibfield  {journal} {\bibinfo  {journal} {Phys. Rev. D}\
  }\textbf {\bibinfo {volume} {113}},\ \bibinfo {pages} {026005} (\bibinfo
  {year} {2026})},\ \Eprint {http://arxiv.org/abs/2312.03831} {arXiv:2312.03831
  [hep-th]} \BibitemShut {NoStop}%
\bibitem [{\citenamefont {Knorr}\ and\ \citenamefont
  {Platania}(2025)}]{Knorr:2024yiu}%
  \BibitemOpen
  \bibfield  {author} {\bibinfo {author} {\bibfnamefont {B.}~\bibnamefont
  {Knorr}}\ and\ \bibinfo {author} {\bibfnamefont {A.}~\bibnamefont
  {Platania}},\ }\href {\doibase 10.1007/JHEP03(2025)003} {\bibfield  {journal}
  {\bibinfo  {journal} {JHEP}\ }\textbf {\bibinfo {volume} {03}},\ \bibinfo
  {pages} {003} (\bibinfo {year} {2025})},\ \Eprint
  {http://arxiv.org/abs/2405.08860} {arXiv:2405.08860 [hep-th]} \BibitemShut
  {NoStop}%
\bibitem [{\citenamefont {Falls}(2025{\natexlab{a}})}]{Falls:2025sxu}%
  \BibitemOpen
  \bibfield  {author} {\bibinfo {author} {\bibfnamefont {K.}~\bibnamefont
  {Falls}},\ }\href {\doibase 10.1007/JHEP11(2025)126} {\bibfield  {journal}
  {\bibinfo  {journal} {JHEP}\ }\textbf {\bibinfo {volume} {11}},\ \bibinfo
  {pages} {126} (\bibinfo {year} {2025}{\natexlab{a}})},\ \Eprint
  {http://arxiv.org/abs/2504.17851} {arXiv:2504.17851 [hep-th]} \BibitemShut
  {NoStop}%
\bibitem [{\citenamefont {Ohta}\ and\ \citenamefont
  {Yamada}(2025)}]{Ohta:2025xxo}%
  \BibitemOpen
  \bibfield  {author} {\bibinfo {author} {\bibfnamefont {N.}~\bibnamefont
  {Ohta}}\ and\ \bibinfo {author} {\bibfnamefont {M.}~\bibnamefont {Yamada}},\
  }\href {\doibase 10.1103/16c5-73q2} {\bibfield  {journal} {\bibinfo
  {journal} {Phys. Rev. D}\ }\textbf {\bibinfo {volume} {112}},\ \bibinfo
  {pages} {066013} (\bibinfo {year} {2025})},\ \Eprint
  {http://arxiv.org/abs/2506.03601} {arXiv:2506.03601 [hep-th]} \BibitemShut
  {NoStop}%
\bibitem [{\citenamefont {Wegner}(1974)}]{Wegner_1974}%
  \BibitemOpen
  \bibfield  {author} {\bibinfo {author} {\bibfnamefont {F.~J.}\ \bibnamefont
  {Wegner}},\ }\href {\doibase 10.1088/0022-3719/7/12/004} {\bibfield
  {journal} {\bibinfo  {journal} {Journal of Physics C: Solid State Physics}\
  }\textbf {\bibinfo {volume} {7}},\ \bibinfo {pages} {2098} (\bibinfo {year}
  {1974})}\BibitemShut {NoStop}%
\bibitem [{\citenamefont {Ihssen}\ and\ \citenamefont
  {Pawlowski}(2023{\natexlab{b}})}]{Ihssen:2022xjv}%
  \BibitemOpen
  \bibfield  {author} {\bibinfo {author} {\bibfnamefont {F.}~\bibnamefont
  {Ihssen}}\ and\ \bibinfo {author} {\bibfnamefont {J.~M.}\ \bibnamefont
  {Pawlowski}},\ }\href {\doibase 10.21468/SciPostPhys.15.2.074} {\bibfield
  {journal} {\bibinfo  {journal} {SciPost Phys.}\ }\textbf {\bibinfo {volume}
  {15}},\ \bibinfo {pages} {074} (\bibinfo {year} {2023}{\natexlab{b}})},\
  \Eprint {http://arxiv.org/abs/2207.10057} {arXiv:2207.10057 [hep-th]}
  \BibitemShut {NoStop}%
\bibitem [{\citenamefont {Wetterich}(2024)}]{Wetterich:2024uub}%
  \BibitemOpen
  \bibfield  {author} {\bibinfo {author} {\bibfnamefont {C.}~\bibnamefont
  {Wetterich}},\ }\href {\doibase 10.1016/j.nuclphysb.2024.116707} {\bibfield
  {journal} {\bibinfo  {journal} {Nucl. Phys. B}\ }\textbf {\bibinfo {volume}
  {1008}},\ \bibinfo {pages} {116707} (\bibinfo {year} {2024})},\ \Eprint
  {http://arxiv.org/abs/2402.04679} {arXiv:2402.04679 [hep-th]} \BibitemShut
  {NoStop}%
\bibitem [{\citenamefont {Wetterich}(1993)}]{Wetterich:1992yh}%
  \BibitemOpen
  \bibfield  {author} {\bibinfo {author} {\bibfnamefont {C.}~\bibnamefont
  {Wetterich}},\ }\href {\doibase 10.1016/0370-2693(93)90726-X} {\bibfield
  {journal} {\bibinfo  {journal} {Phys. Lett. B}\ }\textbf {\bibinfo {volume}
  {301}},\ \bibinfo {pages} {90} (\bibinfo {year} {1993})},\ \Eprint
  {http://arxiv.org/abs/1710.05815} {arXiv:1710.05815 [hep-th]} \BibitemShut
  {NoStop}%
\bibitem [{\citenamefont {Ellwanger}(1994)}]{Ellwanger:1993mw}%
  \BibitemOpen
  \bibfield  {author} {\bibinfo {author} {\bibfnamefont {U.}~\bibnamefont
  {Ellwanger}},\ }\href {\doibase 10.1007/BF01555911} {\bibfield  {journal}
  {\bibinfo  {journal} {Z. Phys. C}\ }\textbf {\bibinfo {volume} {62}},\
  \bibinfo {pages} {503} (\bibinfo {year} {1994})},\ \Eprint
  {http://arxiv.org/abs/hep-ph/9308260} {arXiv:hep-ph/9308260} \BibitemShut
  {NoStop}%
\bibitem [{\citenamefont {Morris}(1994)}]{Morris:1993qb}%
  \BibitemOpen
  \bibfield  {author} {\bibinfo {author} {\bibfnamefont {T.~R.}\ \bibnamefont
  {Morris}},\ }\href {\doibase 10.1142/S0217751X94000972} {\bibfield  {journal}
  {\bibinfo  {journal} {Int. J. Mod. Phys. A}\ }\textbf {\bibinfo {volume}
  {9}},\ \bibinfo {pages} {2411} (\bibinfo {year} {1994})},\ \Eprint
  {http://arxiv.org/abs/hep-ph/9308265} {arXiv:hep-ph/9308265} \BibitemShut
  {NoStop}%
\bibitem [{\citenamefont {Nielsen}(1975)}]{Nielsen:1975fs}%
  \BibitemOpen
  \bibfield  {author} {\bibinfo {author} {\bibfnamefont {N.~K.}\ \bibnamefont
  {Nielsen}},\ }\href {\doibase 10.1016/0550-3213(75)90301-6} {\bibfield
  {journal} {\bibinfo  {journal} {Nucl. Phys. B}\ }\textbf {\bibinfo {volume}
  {101}},\ \bibinfo {pages} {173} (\bibinfo {year} {1975})}\BibitemShut
  {NoStop}%
\bibitem [{\citenamefont {Fukuda}\ and\ \citenamefont
  {Kugo}(1976)}]{Fukuda:1975di}%
  \BibitemOpen
  \bibfield  {author} {\bibinfo {author} {\bibfnamefont {R.}~\bibnamefont
  {Fukuda}}\ and\ \bibinfo {author} {\bibfnamefont {T.}~\bibnamefont {Kugo}},\
  }\href {\doibase 10.1103/PhysRevD.13.3469} {\bibfield  {journal} {\bibinfo
  {journal} {Phys. Rev. D}\ }\textbf {\bibinfo {volume} {13}},\ \bibinfo
  {pages} {3469} (\bibinfo {year} {1976})}\BibitemShut {NoStop}%
\bibitem [{\citenamefont {Manrique}\ and\ \citenamefont
  {Reuter}(2010)}]{Manrique:2009uh}%
  \BibitemOpen
  \bibfield  {author} {\bibinfo {author} {\bibfnamefont {E.}~\bibnamefont
  {Manrique}}\ and\ \bibinfo {author} {\bibfnamefont {M.}~\bibnamefont
  {Reuter}},\ }\href {\doibase 10.1016/j.aop.2009.11.009} {\bibfield  {journal}
  {\bibinfo  {journal} {Annals Phys.}\ }\textbf {\bibinfo {volume} {325}},\
  \bibinfo {pages} {785} (\bibinfo {year} {2010})},\ \Eprint
  {http://arxiv.org/abs/0907.2617} {arXiv:0907.2617 [gr-qc]} \BibitemShut
  {NoStop}%
\bibitem [{\citenamefont {Manrique}\ \emph
  {et~al.}(2011{\natexlab{a}})\citenamefont {Manrique}, \citenamefont
  {Reuter},\ and\ \citenamefont {Saueressig}}]{Manrique:2010mq}%
  \BibitemOpen
  \bibfield  {author} {\bibinfo {author} {\bibfnamefont {E.}~\bibnamefont
  {Manrique}}, \bibinfo {author} {\bibfnamefont {M.}~\bibnamefont {Reuter}}, \
  and\ \bibinfo {author} {\bibfnamefont {F.}~\bibnamefont {Saueressig}},\
  }\href {\doibase 10.1016/j.aop.2010.11.003} {\bibfield  {journal} {\bibinfo
  {journal} {Annals Phys.}\ }\textbf {\bibinfo {volume} {326}},\ \bibinfo
  {pages} {440} (\bibinfo {year} {2011}{\natexlab{a}})},\ \Eprint
  {http://arxiv.org/abs/1003.5129} {arXiv:1003.5129 [hep-th]} \BibitemShut
  {NoStop}%
\bibitem [{\citenamefont {Manrique}\ \emph
  {et~al.}(2011{\natexlab{b}})\citenamefont {Manrique}, \citenamefont
  {Reuter},\ and\ \citenamefont {Saueressig}}]{Manrique:2010am}%
  \BibitemOpen
  \bibfield  {author} {\bibinfo {author} {\bibfnamefont {E.}~\bibnamefont
  {Manrique}}, \bibinfo {author} {\bibfnamefont {M.}~\bibnamefont {Reuter}}, \
  and\ \bibinfo {author} {\bibfnamefont {F.}~\bibnamefont {Saueressig}},\
  }\href {\doibase 10.1016/j.aop.2010.11.006} {\bibfield  {journal} {\bibinfo
  {journal} {Annals Phys.}\ }\textbf {\bibinfo {volume} {326}},\ \bibinfo
  {pages} {463} (\bibinfo {year} {2011}{\natexlab{b}})},\ \Eprint
  {http://arxiv.org/abs/1006.0099} {arXiv:1006.0099 [hep-th]} \BibitemShut
  {NoStop}%
\bibitem [{\citenamefont {Becker}\ and\ \citenamefont
  {Reuter}(2014)}]{Becker:2014qya}%
  \BibitemOpen
  \bibfield  {author} {\bibinfo {author} {\bibfnamefont {D.}~\bibnamefont
  {Becker}}\ and\ \bibinfo {author} {\bibfnamefont {M.}~\bibnamefont
  {Reuter}},\ }\href {\doibase 10.1016/j.aop.2014.07.023} {\bibfield  {journal}
  {\bibinfo  {journal} {Annals Phys.}\ }\textbf {\bibinfo {volume} {350}},\
  \bibinfo {pages} {225} (\bibinfo {year} {2014})},\ \Eprint
  {http://arxiv.org/abs/1404.4537} {arXiv:1404.4537 [hep-th]} \BibitemShut
  {NoStop}%
\bibitem [{\citenamefont {Christiansen}\ \emph {et~al.}(2014)\citenamefont
  {Christiansen}, \citenamefont {Litim}, \citenamefont {Pawlowski},\ and\
  \citenamefont {Rodigast}}]{Christiansen:2012rx}%
  \BibitemOpen
  \bibfield  {author} {\bibinfo {author} {\bibfnamefont {N.}~\bibnamefont
  {Christiansen}}, \bibinfo {author} {\bibfnamefont {D.~F.}\ \bibnamefont
  {Litim}}, \bibinfo {author} {\bibfnamefont {J.~M.}\ \bibnamefont
  {Pawlowski}}, \ and\ \bibinfo {author} {\bibfnamefont {A.}~\bibnamefont
  {Rodigast}},\ }\href {\doibase 10.1016/j.physletb.2013.11.025} {\bibfield
  {journal} {\bibinfo  {journal} {Phys. Lett. B}\ }\textbf {\bibinfo {volume}
  {728}},\ \bibinfo {pages} {114} (\bibinfo {year} {2014})},\ \Eprint
  {http://arxiv.org/abs/1209.4038} {arXiv:1209.4038 [hep-th]} \BibitemShut
  {NoStop}%
\bibitem [{\citenamefont {Codello}\ \emph {et~al.}(2014)\citenamefont
  {Codello}, \citenamefont {D'Odorico},\ and\ \citenamefont
  {Pagani}}]{Codello:2013fpa}%
  \BibitemOpen
  \bibfield  {author} {\bibinfo {author} {\bibfnamefont {A.}~\bibnamefont
  {Codello}}, \bibinfo {author} {\bibfnamefont {G.}~\bibnamefont {D'Odorico}},
  \ and\ \bibinfo {author} {\bibfnamefont {C.}~\bibnamefont {Pagani}},\ }\href
  {\doibase 10.1103/PhysRevD.89.081701} {\bibfield  {journal} {\bibinfo
  {journal} {Phys. Rev. D}\ }\textbf {\bibinfo {volume} {89}},\ \bibinfo
  {pages} {081701} (\bibinfo {year} {2014})},\ \Eprint
  {http://arxiv.org/abs/1304.4777} {arXiv:1304.4777 [gr-qc]} \BibitemShut
  {NoStop}%
\bibitem [{\citenamefont {Christiansen}\ \emph {et~al.}(2016)\citenamefont
  {Christiansen}, \citenamefont {Knorr}, \citenamefont {Pawlowski},\ and\
  \citenamefont {Rodigast}}]{Christiansen:2014raa}%
  \BibitemOpen
  \bibfield  {author} {\bibinfo {author} {\bibfnamefont {N.}~\bibnamefont
  {Christiansen}}, \bibinfo {author} {\bibfnamefont {B.}~\bibnamefont {Knorr}},
  \bibinfo {author} {\bibfnamefont {J.~M.}\ \bibnamefont {Pawlowski}}, \ and\
  \bibinfo {author} {\bibfnamefont {A.}~\bibnamefont {Rodigast}},\ }\href
  {\doibase 10.1103/PhysRevD.93.044036} {\bibfield  {journal} {\bibinfo
  {journal} {Phys. Rev. D}\ }\textbf {\bibinfo {volume} {93}},\ \bibinfo
  {pages} {044036} (\bibinfo {year} {2016})},\ \Eprint
  {http://arxiv.org/abs/1403.1232} {arXiv:1403.1232 [hep-th]} \BibitemShut
  {NoStop}%
\bibitem [{\citenamefont {Christiansen}\ \emph {et~al.}(2015)\citenamefont
  {Christiansen}, \citenamefont {Knorr}, \citenamefont {Meibohm}, \citenamefont
  {Pawlowski},\ and\ \citenamefont {Reichert}}]{Christiansen:2015rva}%
  \BibitemOpen
  \bibfield  {author} {\bibinfo {author} {\bibfnamefont {N.}~\bibnamefont
  {Christiansen}}, \bibinfo {author} {\bibfnamefont {B.}~\bibnamefont {Knorr}},
  \bibinfo {author} {\bibfnamefont {J.}~\bibnamefont {Meibohm}}, \bibinfo
  {author} {\bibfnamefont {J.~M.}\ \bibnamefont {Pawlowski}}, \ and\ \bibinfo
  {author} {\bibfnamefont {M.}~\bibnamefont {Reichert}},\ }\href {\doibase
  10.1103/PhysRevD.92.121501} {\bibfield  {journal} {\bibinfo  {journal} {Phys.
  Rev. D}\ }\textbf {\bibinfo {volume} {92}},\ \bibinfo {pages} {121501}
  (\bibinfo {year} {2015})},\ \Eprint {http://arxiv.org/abs/1506.07016}
  {arXiv:1506.07016 [hep-th]} \BibitemShut {NoStop}%
\bibitem [{\citenamefont {Meibohm}\ and\ \citenamefont
  {Pawlowski}(2016)}]{Meibohm:2016mkp}%
  \BibitemOpen
  \bibfield  {author} {\bibinfo {author} {\bibfnamefont {J.}~\bibnamefont
  {Meibohm}}\ and\ \bibinfo {author} {\bibfnamefont {J.~M.}\ \bibnamefont
  {Pawlowski}},\ }\href {\doibase 10.1140/epjc/s10052-016-4132-7} {\bibfield
  {journal} {\bibinfo  {journal} {Eur. Phys. J. C}\ }\textbf {\bibinfo {volume}
  {76}},\ \bibinfo {pages} {285} (\bibinfo {year} {2016})},\ \Eprint
  {http://arxiv.org/abs/1601.04597} {arXiv:1601.04597 [hep-th]} \BibitemShut
  {NoStop}%
\bibitem [{\citenamefont {Denz}\ \emph {et~al.}(2018)\citenamefont {Denz},
  \citenamefont {Pawlowski},\ and\ \citenamefont {Reichert}}]{Denz:2016qks}%
  \BibitemOpen
  \bibfield  {author} {\bibinfo {author} {\bibfnamefont {T.}~\bibnamefont
  {Denz}}, \bibinfo {author} {\bibfnamefont {J.~M.}\ \bibnamefont {Pawlowski}},
  \ and\ \bibinfo {author} {\bibfnamefont {M.}~\bibnamefont {Reichert}},\
  }\href {\doibase 10.1140/epjc/s10052-018-5806-0} {\bibfield  {journal}
  {\bibinfo  {journal} {Eur. Phys. J. C}\ }\textbf {\bibinfo {volume} {78}},\
  \bibinfo {pages} {336} (\bibinfo {year} {2018})},\ \Eprint
  {http://arxiv.org/abs/1612.07315} {arXiv:1612.07315 [hep-th]} \BibitemShut
  {NoStop}%
\bibitem [{\citenamefont {Christiansen}\ \emph
  {et~al.}(2018{\natexlab{a}})\citenamefont {Christiansen}, \citenamefont
  {Litim}, \citenamefont {Pawlowski},\ and\ \citenamefont
  {Reichert}}]{Christiansen:2017cxa}%
  \BibitemOpen
  \bibfield  {author} {\bibinfo {author} {\bibfnamefont {N.}~\bibnamefont
  {Christiansen}}, \bibinfo {author} {\bibfnamefont {D.~F.}\ \bibnamefont
  {Litim}}, \bibinfo {author} {\bibfnamefont {J.~M.}\ \bibnamefont
  {Pawlowski}}, \ and\ \bibinfo {author} {\bibfnamefont {M.}~\bibnamefont
  {Reichert}},\ }\href {\doibase 10.1103/PhysRevD.97.106012} {\bibfield
  {journal} {\bibinfo  {journal} {Phys. Rev. D}\ }\textbf {\bibinfo {volume}
  {97}},\ \bibinfo {pages} {106012} (\bibinfo {year} {2018}{\natexlab{a}})},\
  \Eprint {http://arxiv.org/abs/1710.04669} {arXiv:1710.04669 [hep-th]}
  \BibitemShut {NoStop}%
\bibitem [{\citenamefont {Christiansen}\ \emph
  {et~al.}(2018{\natexlab{b}})\citenamefont {Christiansen}, \citenamefont
  {Falls}, \citenamefont {Pawlowski},\ and\ \citenamefont
  {Reichert}}]{Christiansen:2017bsy}%
  \BibitemOpen
  \bibfield  {author} {\bibinfo {author} {\bibfnamefont {N.}~\bibnamefont
  {Christiansen}}, \bibinfo {author} {\bibfnamefont {K.}~\bibnamefont {Falls}},
  \bibinfo {author} {\bibfnamefont {J.~M.}\ \bibnamefont {Pawlowski}}, \ and\
  \bibinfo {author} {\bibfnamefont {M.}~\bibnamefont {Reichert}},\ }\href
  {\doibase 10.1103/PhysRevD.97.046007} {\bibfield  {journal} {\bibinfo
  {journal} {Phys. Rev. D}\ }\textbf {\bibinfo {volume} {97}},\ \bibinfo
  {pages} {046007} (\bibinfo {year} {2018}{\natexlab{b}})},\ \Eprint
  {http://arxiv.org/abs/1711.09259} {arXiv:1711.09259 [hep-th]} \BibitemShut
  {NoStop}%
\bibitem [{\citenamefont {Knorr}\ and\ \citenamefont
  {Lippoldt}(2017)}]{Knorr:2017fus}%
  \BibitemOpen
  \bibfield  {author} {\bibinfo {author} {\bibfnamefont {B.}~\bibnamefont
  {Knorr}}\ and\ \bibinfo {author} {\bibfnamefont {S.}~\bibnamefont
  {Lippoldt}},\ }\href {\doibase 10.1103/PhysRevD.96.065020} {\bibfield
  {journal} {\bibinfo  {journal} {Phys. Rev. D}\ }\textbf {\bibinfo {volume}
  {96}},\ \bibinfo {pages} {065020} (\bibinfo {year} {2017})},\ \Eprint
  {http://arxiv.org/abs/1707.01397} {arXiv:1707.01397 [hep-th]} \BibitemShut
  {NoStop}%
\bibitem [{\citenamefont {Knorr}(2018)}]{Knorr:2017mhu}%
  \BibitemOpen
  \bibfield  {author} {\bibinfo {author} {\bibfnamefont {B.}~\bibnamefont
  {Knorr}},\ }\href {\doibase 10.1088/1361-6382/aabaa0} {\bibfield  {journal}
  {\bibinfo  {journal} {Class. Quant. Grav.}\ }\textbf {\bibinfo {volume}
  {35}},\ \bibinfo {pages} {115005} (\bibinfo {year} {2018})},\ \Eprint
  {http://arxiv.org/abs/1710.07055} {arXiv:1710.07055 [hep-th]} \BibitemShut
  {NoStop}%
\bibitem [{\citenamefont {Eichhorn}\ \emph {et~al.}(2018)\citenamefont
  {Eichhorn}, \citenamefont {Labus}, \citenamefont {Pawlowski},\ and\
  \citenamefont {Reichert}}]{Eichhorn:2018akn}%
  \BibitemOpen
  \bibfield  {author} {\bibinfo {author} {\bibfnamefont {A.}~\bibnamefont
  {Eichhorn}}, \bibinfo {author} {\bibfnamefont {P.}~\bibnamefont {Labus}},
  \bibinfo {author} {\bibfnamefont {J.~M.}\ \bibnamefont {Pawlowski}}, \ and\
  \bibinfo {author} {\bibfnamefont {M.}~\bibnamefont {Reichert}},\ }\href
  {\doibase 10.21468/SciPostPhys.5.4.031} {\bibfield  {journal} {\bibinfo
  {journal} {SciPost Phys.}\ }\textbf {\bibinfo {volume} {5}},\ \bibinfo
  {pages} {031} (\bibinfo {year} {2018})},\ \Eprint
  {http://arxiv.org/abs/1804.00012} {arXiv:1804.00012 [hep-th]} \BibitemShut
  {NoStop}%
\bibitem [{\citenamefont {Eichhorn}\ \emph
  {et~al.}(2019{\natexlab{a}})\citenamefont {Eichhorn}, \citenamefont
  {Lippoldt}, \citenamefont {Pawlowski}, \citenamefont {Reichert},\ and\
  \citenamefont {Schiffer}}]{Eichhorn:2018ydy}%
  \BibitemOpen
  \bibfield  {author} {\bibinfo {author} {\bibfnamefont {A.}~\bibnamefont
  {Eichhorn}}, \bibinfo {author} {\bibfnamefont {S.}~\bibnamefont {Lippoldt}},
  \bibinfo {author} {\bibfnamefont {J.~M.}\ \bibnamefont {Pawlowski}}, \bibinfo
  {author} {\bibfnamefont {M.}~\bibnamefont {Reichert}}, \ and\ \bibinfo
  {author} {\bibfnamefont {M.}~\bibnamefont {Schiffer}},\ }\href {\doibase
  10.1016/j.physletb.2019.01.071} {\bibfield  {journal} {\bibinfo  {journal}
  {Phys. Lett. B}\ }\textbf {\bibinfo {volume} {792}},\ \bibinfo {pages} {310}
  (\bibinfo {year} {2019}{\natexlab{a}})},\ \Eprint
  {http://arxiv.org/abs/1810.02828} {arXiv:1810.02828 [hep-th]} \BibitemShut
  {NoStop}%
\bibitem [{\citenamefont {Eichhorn}\ \emph
  {et~al.}(2019{\natexlab{b}})\citenamefont {Eichhorn}, \citenamefont
  {Lippoldt},\ and\ \citenamefont {Schiffer}}]{Eichhorn:2018nda}%
  \BibitemOpen
  \bibfield  {author} {\bibinfo {author} {\bibfnamefont {A.}~\bibnamefont
  {Eichhorn}}, \bibinfo {author} {\bibfnamefont {S.}~\bibnamefont {Lippoldt}},
  \ and\ \bibinfo {author} {\bibfnamefont {M.}~\bibnamefont {Schiffer}},\
  }\href {\doibase 10.1103/PhysRevD.99.086002} {\bibfield  {journal} {\bibinfo
  {journal} {Phys. Rev. D}\ }\textbf {\bibinfo {volume} {99}},\ \bibinfo
  {pages} {086002} (\bibinfo {year} {2019}{\natexlab{b}})},\ \Eprint
  {http://arxiv.org/abs/1812.08782} {arXiv:1812.08782 [hep-th]} \BibitemShut
  {NoStop}%
\bibitem [{\citenamefont {B{\"u}rger}\ \emph {et~al.}(2019)\citenamefont
  {B{\"u}rger}, \citenamefont {Pawlowski}, \citenamefont {Reichert},\ and\
  \citenamefont {Schaefer}}]{Burger:2019upn}%
  \BibitemOpen
  \bibfield  {author} {\bibinfo {author} {\bibfnamefont {B.}~\bibnamefont
  {B{\"u}rger}}, \bibinfo {author} {\bibfnamefont {J.~M.}\ \bibnamefont
  {Pawlowski}}, \bibinfo {author} {\bibfnamefont {M.}~\bibnamefont {Reichert}},
  \ and\ \bibinfo {author} {\bibfnamefont {B.-J.}\ \bibnamefont {Schaefer}},\
  }\href@noop {} {\  (\bibinfo {year} {2019})},\ \Eprint
  {http://arxiv.org/abs/1912.01624} {arXiv:1912.01624 [hep-th]} \BibitemShut
  {NoStop}%
\bibitem [{\citenamefont {Bonanno}\ \emph {et~al.}(2022)\citenamefont
  {Bonanno}, \citenamefont {Denz}, \citenamefont {Pawlowski},\ and\
  \citenamefont {Reichert}}]{Bonanno:2021squ}%
  \BibitemOpen
  \bibfield  {author} {\bibinfo {author} {\bibfnamefont {A.}~\bibnamefont
  {Bonanno}}, \bibinfo {author} {\bibfnamefont {T.}~\bibnamefont {Denz}},
  \bibinfo {author} {\bibfnamefont {J.~M.}\ \bibnamefont {Pawlowski}}, \ and\
  \bibinfo {author} {\bibfnamefont {M.}~\bibnamefont {Reichert}},\ }\href
  {\doibase 10.21468/SciPostPhys.12.1.001} {\bibfield  {journal} {\bibinfo
  {journal} {SciPost Phys.}\ }\textbf {\bibinfo {volume} {12}},\ \bibinfo
  {pages} {001} (\bibinfo {year} {2022})},\ \Eprint
  {http://arxiv.org/abs/2102.02217} {arXiv:2102.02217 [hep-th]} \BibitemShut
  {NoStop}%
\bibitem [{\citenamefont {Knorr}\ and\ \citenamefont
  {Schiffer}(2021)}]{Knorr:2021niv}%
  \BibitemOpen
  \bibfield  {author} {\bibinfo {author} {\bibfnamefont {B.}~\bibnamefont
  {Knorr}}\ and\ \bibinfo {author} {\bibfnamefont {M.}~\bibnamefont
  {Schiffer}},\ }\href {\doibase 10.3390/universe7070216} {\bibfield  {journal}
  {\bibinfo  {journal} {Universe}\ }\textbf {\bibinfo {volume} {7}},\ \bibinfo
  {pages} {216} (\bibinfo {year} {2021})},\ \Eprint
  {http://arxiv.org/abs/2105.04566} {arXiv:2105.04566 [hep-th]} \BibitemShut
  {NoStop}%
\bibitem [{\citenamefont {Fehre}\ \emph {et~al.}(2023)\citenamefont {Fehre},
  \citenamefont {Litim}, \citenamefont {Pawlowski},\ and\ \citenamefont
  {Reichert}}]{Fehre:2021eob}%
  \BibitemOpen
  \bibfield  {author} {\bibinfo {author} {\bibfnamefont {J.}~\bibnamefont
  {Fehre}}, \bibinfo {author} {\bibfnamefont {D.~F.}\ \bibnamefont {Litim}},
  \bibinfo {author} {\bibfnamefont {J.~M.}\ \bibnamefont {Pawlowski}}, \ and\
  \bibinfo {author} {\bibfnamefont {M.}~\bibnamefont {Reichert}},\ }\href
  {\doibase 10.1103/PhysRevLett.130.081501} {\bibfield  {journal} {\bibinfo
  {journal} {Phys. Rev. Lett.}\ }\textbf {\bibinfo {volume} {130}},\ \bibinfo
  {pages} {081501} (\bibinfo {year} {2023})},\ \Eprint
  {http://arxiv.org/abs/2111.13232} {arXiv:2111.13232 [hep-th]} \BibitemShut
  {NoStop}%
\bibitem [{\citenamefont {Pastor-Guti{\'e}rrez}\ \emph
  {et~al.}(2023)\citenamefont {Pastor-Guti{\'e}rrez}, \citenamefont
  {Pawlowski},\ and\ \citenamefont {Reichert}}]{Pastor-Gutierrez:2022nki}%
  \BibitemOpen
  \bibfield  {author} {\bibinfo {author} {\bibfnamefont {{\'A}.}~\bibnamefont
  {Pastor-Guti{\'e}rrez}}, \bibinfo {author} {\bibfnamefont {J.~M.}\
  \bibnamefont {Pawlowski}}, \ and\ \bibinfo {author} {\bibfnamefont
  {M.}~\bibnamefont {Reichert}},\ }\href {\doibase
  10.21468/SciPostPhys.15.3.105} {\bibfield  {journal} {\bibinfo  {journal}
  {SciPost Phys.}\ }\textbf {\bibinfo {volume} {15}},\ \bibinfo {pages} {105}
  (\bibinfo {year} {2023})},\ \Eprint {http://arxiv.org/abs/2207.09817}
  {arXiv:2207.09817 [hep-th]} \BibitemShut {NoStop}%
\bibitem [{\citenamefont {Saueressig}\ and\ \citenamefont
  {Wang}(2023)}]{Saueressig:2023tfy}%
  \BibitemOpen
  \bibfield  {author} {\bibinfo {author} {\bibfnamefont {F.}~\bibnamefont
  {Saueressig}}\ and\ \bibinfo {author} {\bibfnamefont {J.}~\bibnamefont
  {Wang}},\ }\href {\doibase 10.1007/JHEP09(2023)064} {\bibfield  {journal}
  {\bibinfo  {journal} {JHEP}\ }\textbf {\bibinfo {volume} {09}},\ \bibinfo
  {pages} {064} (\bibinfo {year} {2023})},\ \Eprint
  {http://arxiv.org/abs/2306.10408} {arXiv:2306.10408 [hep-th]} \BibitemShut
  {NoStop}%
\bibitem [{\citenamefont {Korver}\ \emph {et~al.}(2024)\citenamefont {Korver},
  \citenamefont {Saueressig},\ and\ \citenamefont {Wang}}]{Korver:2024sam}%
  \BibitemOpen
  \bibfield  {author} {\bibinfo {author} {\bibfnamefont {G.}~\bibnamefont
  {Korver}}, \bibinfo {author} {\bibfnamefont {F.}~\bibnamefont {Saueressig}},
  \ and\ \bibinfo {author} {\bibfnamefont {J.}~\bibnamefont {Wang}},\ }\href
  {\doibase 10.1016/j.physletb.2024.138789} {\bibfield  {journal} {\bibinfo
  {journal} {Phys. Lett. B}\ }\textbf {\bibinfo {volume} {855}},\ \bibinfo
  {pages} {138789} (\bibinfo {year} {2024})},\ \Eprint
  {http://arxiv.org/abs/2402.01260} {arXiv:2402.01260 [hep-th]} \BibitemShut
  {NoStop}%
\bibitem [{\citenamefont {Pastor-Guti{\'e}rrez}\ \emph
  {et~al.}(2025)\citenamefont {Pastor-Guti{\'e}rrez}, \citenamefont
  {Pawlowski}, \citenamefont {Reichert},\ and\ \citenamefont
  {Ruisi}}]{Pastor-Gutierrez:2024sbt}%
  \BibitemOpen
  \bibfield  {author} {\bibinfo {author} {\bibfnamefont {{\'A}.}~\bibnamefont
  {Pastor-Guti{\'e}rrez}}, \bibinfo {author} {\bibfnamefont {J.~M.}\
  \bibnamefont {Pawlowski}}, \bibinfo {author} {\bibfnamefont {M.}~\bibnamefont
  {Reichert}}, \ and\ \bibinfo {author} {\bibfnamefont {G.}~\bibnamefont
  {Ruisi}},\ }\href {\doibase 10.1103/PhysRevD.111.106005} {\bibfield
  {journal} {\bibinfo  {journal} {Phys. Rev. D}\ }\textbf {\bibinfo {volume}
  {111}},\ \bibinfo {pages} {106005} (\bibinfo {year} {2025})},\ \Eprint
  {http://arxiv.org/abs/2412.13800} {arXiv:2412.13800 [hep-ph]} \BibitemShut
  {NoStop}%
\bibitem [{\citenamefont {Saueressig}\ and\ \citenamefont
  {Wang}(2025)}]{Saueressig:2025ypi}%
  \BibitemOpen
  \bibfield  {author} {\bibinfo {author} {\bibfnamefont {F.}~\bibnamefont
  {Saueressig}}\ and\ \bibinfo {author} {\bibfnamefont {J.}~\bibnamefont
  {Wang}},\ }\href {\doibase 10.1103/PhysRevD.111.106007} {\bibfield  {journal}
  {\bibinfo  {journal} {Phys. Rev. D}\ }\textbf {\bibinfo {volume} {111}},\
  \bibinfo {pages} {106007} (\bibinfo {year} {2025})},\ \Eprint
  {http://arxiv.org/abs/2501.03752} {arXiv:2501.03752 [hep-th]} \BibitemShut
  {NoStop}%
\bibitem [{\citenamefont {Kher}\ \emph {et~al.}(2025)\citenamefont {Kher},
  \citenamefont {King}, \citenamefont {Litim},\ and\ \citenamefont
  {Reichert}}]{Kher:2025rve}%
  \BibitemOpen
  \bibfield  {author} {\bibinfo {author} {\bibfnamefont {V.}~\bibnamefont
  {Kher}}, \bibinfo {author} {\bibfnamefont {B.}~\bibnamefont {King}}, \bibinfo
  {author} {\bibfnamefont {D.~F.}\ \bibnamefont {Litim}}, \ and\ \bibinfo
  {author} {\bibfnamefont {M.}~\bibnamefont {Reichert}},\ }\href@noop {} {\
  (\bibinfo {year} {2025})},\ \Eprint {http://arxiv.org/abs/2507.17862}
  {arXiv:2507.17862 [hep-th]} \BibitemShut {NoStop}%
\bibitem [{\citenamefont {Pawlowski}\ \emph {et~al.}(2026)\citenamefont
  {Pawlowski}, \citenamefont {Reichert},\ and\ \citenamefont
  {Wessely}}]{Pawlowski:2025etp}%
  \BibitemOpen
  \bibfield  {author} {\bibinfo {author} {\bibfnamefont {J.~M.}\ \bibnamefont
  {Pawlowski}}, \bibinfo {author} {\bibfnamefont {M.}~\bibnamefont {Reichert}},
  \ and\ \bibinfo {author} {\bibfnamefont {J.}~\bibnamefont {Wessely}},\ }\href
  {\doibase 10.1016/j.physletb.2026.140844} {\bibfield  {journal} {\bibinfo
  {journal} {Phys. Lett. B}\ }\textbf {\bibinfo {volume} {880}},\ \bibinfo
  {pages} {140844} (\bibinfo {year} {2026})},\ \Eprint
  {http://arxiv.org/abs/2507.22169} {arXiv:2507.22169 [hep-th]} \BibitemShut
  {NoStop}%
\bibitem [{\citenamefont {Chiesa}\ \emph {et~al.}(2026)\citenamefont {Chiesa},
  \citenamefont {Pawlowski},\ and\ \citenamefont {Reichert}}]{Chiesa:2026tlz}%
  \BibitemOpen
  \bibfield  {author} {\bibinfo {author} {\bibfnamefont {A.~P.}\ \bibnamefont
  {Chiesa}}, \bibinfo {author} {\bibfnamefont {J.~M.}\ \bibnamefont
  {Pawlowski}}, \ and\ \bibinfo {author} {\bibfnamefont {M.}~\bibnamefont
  {Reichert}},\ }\href@noop {} {\  (\bibinfo {year} {2026})},\ \Eprint
  {http://arxiv.org/abs/2603.10168} {arXiv:2603.10168 [hep-th]} \BibitemShut
  {NoStop}%
\bibitem [{\citenamefont {Knorr}(2026)}]{Knorr:2026jcg}%
  \BibitemOpen
  \bibfield  {author} {\bibinfo {author} {\bibfnamefont {B.}~\bibnamefont
  {Knorr}},\ }\href@noop {} {\  (\bibinfo {year} {2026})},\ \Eprint
  {http://arxiv.org/abs/2606.18343} {arXiv:2606.18343 [hep-th]} \BibitemShut
  {NoStop}%
\bibitem [{\citenamefont {Assant}\ \emph {et~al.}(2026)\citenamefont {Assant},
  \citenamefont {Litim},\ and\ \citenamefont {Reichert}}]{Assant:2026dca}%
  \BibitemOpen
  \bibfield  {author} {\bibinfo {author} {\bibfnamefont {G.}~\bibnamefont
  {Assant}}, \bibinfo {author} {\bibfnamefont {D.~F.}\ \bibnamefont {Litim}}, \
  and\ \bibinfo {author} {\bibfnamefont {M.}~\bibnamefont {Reichert}},\
  }\href@noop {} {\  (\bibinfo {year} {2026})},\ \Eprint
  {http://arxiv.org/abs/2606.19321} {arXiv:2606.19321 [hep-th]} \BibitemShut
  {NoStop}%
\bibitem [{\citenamefont {Vassilevich}(2003)}]{Vassilevich:2003xt}%
  \BibitemOpen
  \bibfield  {author} {\bibinfo {author} {\bibfnamefont {D.~V.}\ \bibnamefont
  {Vassilevich}},\ }\href {\doibase 10.1016/j.physrep.2003.09.002} {\bibfield
  {journal} {\bibinfo  {journal} {Phys. Rept.}\ }\textbf {\bibinfo {volume}
  {388}},\ \bibinfo {pages} {279} (\bibinfo {year} {2003})},\ \Eprint
  {http://arxiv.org/abs/hep-th/0306138} {arXiv:hep-th/0306138} \BibitemShut
  {NoStop}%
\bibitem [{\citenamefont {Reuter}\ and\ \citenamefont
  {Saueressig}(2002)}]{Reuter:2001ag}%
  \BibitemOpen
  \bibfield  {author} {\bibinfo {author} {\bibfnamefont {M.}~\bibnamefont
  {Reuter}}\ and\ \bibinfo {author} {\bibfnamefont {F.}~\bibnamefont
  {Saueressig}},\ }\href {\doibase 10.1103/PhysRevD.65.065016} {\bibfield
  {journal} {\bibinfo  {journal} {Phys. Rev. D}\ }\textbf {\bibinfo {volume}
  {65}},\ \bibinfo {pages} {065016} (\bibinfo {year} {2002})},\ \Eprint
  {http://arxiv.org/abs/hep-th/0110054} {arXiv:hep-th/0110054} \BibitemShut
  {NoStop}%
\bibitem [{\citenamefont {Lauscher}\ and\ \citenamefont
  {Reuter}(2002)}]{Lauscher:2002sq}%
  \BibitemOpen
  \bibfield  {author} {\bibinfo {author} {\bibfnamefont {O.}~\bibnamefont
  {Lauscher}}\ and\ \bibinfo {author} {\bibfnamefont {M.}~\bibnamefont
  {Reuter}},\ }\href {\doibase 10.1103/PhysRevD.66.025026} {\bibfield
  {journal} {\bibinfo  {journal} {Phys. Rev. D}\ }\textbf {\bibinfo {volume}
  {66}},\ \bibinfo {pages} {025026} (\bibinfo {year} {2002})},\ \Eprint
  {http://arxiv.org/abs/hep-th/0205062} {arXiv:hep-th/0205062} \BibitemShut
  {NoStop}%
\bibitem [{\citenamefont {Machado}\ and\ \citenamefont
  {Saueressig}(2008)}]{Machado:2007ea}%
  \BibitemOpen
  \bibfield  {author} {\bibinfo {author} {\bibfnamefont {P.~F.}\ \bibnamefont
  {Machado}}\ and\ \bibinfo {author} {\bibfnamefont {F.}~\bibnamefont
  {Saueressig}},\ }\href {\doibase 10.1103/PhysRevD.77.124045} {\bibfield
  {journal} {\bibinfo  {journal} {Phys. Rev. D}\ }\textbf {\bibinfo {volume}
  {77}},\ \bibinfo {pages} {124045} (\bibinfo {year} {2008})},\ \Eprint
  {http://arxiv.org/abs/0712.0445} {arXiv:0712.0445 [hep-th]} \BibitemShut
  {NoStop}%
\bibitem [{\citenamefont {Benedetti}\ \emph {et~al.}(2011)\citenamefont
  {Benedetti}, \citenamefont {Groh}, \citenamefont {Machado},\ and\
  \citenamefont {Saueressig}}]{Benedetti:2010nr}%
  \BibitemOpen
  \bibfield  {author} {\bibinfo {author} {\bibfnamefont {D.}~\bibnamefont
  {Benedetti}}, \bibinfo {author} {\bibfnamefont {K.}~\bibnamefont {Groh}},
  \bibinfo {author} {\bibfnamefont {P.~F.}\ \bibnamefont {Machado}}, \ and\
  \bibinfo {author} {\bibfnamefont {F.}~\bibnamefont {Saueressig}},\ }\href
  {\doibase 10.1007/JHEP06(2011)079} {\bibfield  {journal} {\bibinfo  {journal}
  {JHEP}\ }\textbf {\bibinfo {volume} {06}},\ \bibinfo {pages} {079} (\bibinfo
  {year} {2011})},\ \Eprint {http://arxiv.org/abs/1012.3081} {arXiv:1012.3081
  [hep-th]} \BibitemShut {NoStop}%
\bibitem [{\citenamefont {Falls}\ \emph {et~al.}(2016)\citenamefont {Falls},
  \citenamefont {Litim}, \citenamefont {Nikolakopoulos},\ and\ \citenamefont
  {Rahmede}}]{Falls:2014tra}%
  \BibitemOpen
  \bibfield  {author} {\bibinfo {author} {\bibfnamefont {K.}~\bibnamefont
  {Falls}}, \bibinfo {author} {\bibfnamefont {D.~F.}\ \bibnamefont {Litim}},
  \bibinfo {author} {\bibfnamefont {K.}~\bibnamefont {Nikolakopoulos}}, \ and\
  \bibinfo {author} {\bibfnamefont {C.}~\bibnamefont {Rahmede}},\ }\href
  {\doibase 10.1103/PhysRevD.93.104022} {\bibfield  {journal} {\bibinfo
  {journal} {Phys. Rev. D}\ }\textbf {\bibinfo {volume} {93}},\ \bibinfo
  {pages} {104022} (\bibinfo {year} {2016})},\ \Eprint
  {http://arxiv.org/abs/1410.4815} {arXiv:1410.4815 [hep-th]} \BibitemShut
  {NoStop}%
\bibitem [{\citenamefont {Demmel}\ \emph
  {et~al.}(2015{\natexlab{a}})\citenamefont {Demmel}, \citenamefont
  {Saueressig},\ and\ \citenamefont {Zanusso}}]{Demmel:2015oqa}%
  \BibitemOpen
  \bibfield  {author} {\bibinfo {author} {\bibfnamefont {M.}~\bibnamefont
  {Demmel}}, \bibinfo {author} {\bibfnamefont {F.}~\bibnamefont {Saueressig}},
  \ and\ \bibinfo {author} {\bibfnamefont {O.}~\bibnamefont {Zanusso}},\ }\href
  {\doibase 10.1007/JHEP08(2015)113} {\bibfield  {journal} {\bibinfo  {journal}
  {JHEP}\ }\textbf {\bibinfo {volume} {08}},\ \bibinfo {pages} {113} (\bibinfo
  {year} {2015}{\natexlab{a}})},\ \Eprint {http://arxiv.org/abs/1504.07656}
  {arXiv:1504.07656 [hep-th]} \BibitemShut {NoStop}%
\bibitem [{\citenamefont {Gies}\ \emph {et~al.}(2016)\citenamefont {Gies},
  \citenamefont {Knorr}, \citenamefont {Lippoldt},\ and\ \citenamefont
  {Saueressig}}]{Gies:2016con}%
  \BibitemOpen
  \bibfield  {author} {\bibinfo {author} {\bibfnamefont {H.}~\bibnamefont
  {Gies}}, \bibinfo {author} {\bibfnamefont {B.}~\bibnamefont {Knorr}},
  \bibinfo {author} {\bibfnamefont {S.}~\bibnamefont {Lippoldt}}, \ and\
  \bibinfo {author} {\bibfnamefont {F.}~\bibnamefont {Saueressig}},\ }\href
  {\doibase 10.1103/PhysRevLett.116.211302} {\bibfield  {journal} {\bibinfo
  {journal} {Phys. Rev. Lett.}\ }\textbf {\bibinfo {volume} {116}},\ \bibinfo
  {pages} {211302} (\bibinfo {year} {2016})},\ \Eprint
  {http://arxiv.org/abs/1601.01800} {arXiv:1601.01800 [hep-th]} \BibitemShut
  {NoStop}%
\bibitem [{\citenamefont {Falls}\ \emph {et~al.}(2018)\citenamefont {Falls},
  \citenamefont {King}, \citenamefont {Litim}, \citenamefont {Nikolakopoulos},\
  and\ \citenamefont {Rahmede}}]{Falls:2017lst}%
  \BibitemOpen
  \bibfield  {author} {\bibinfo {author} {\bibfnamefont {K.}~\bibnamefont
  {Falls}}, \bibinfo {author} {\bibfnamefont {C.~R.}\ \bibnamefont {King}},
  \bibinfo {author} {\bibfnamefont {D.~F.}\ \bibnamefont {Litim}}, \bibinfo
  {author} {\bibfnamefont {K.}~\bibnamefont {Nikolakopoulos}}, \ and\ \bibinfo
  {author} {\bibfnamefont {C.}~\bibnamefont {Rahmede}},\ }\href {\doibase
  10.1103/PhysRevD.97.086006} {\bibfield  {journal} {\bibinfo  {journal} {Phys.
  Rev. D}\ }\textbf {\bibinfo {volume} {97}},\ \bibinfo {pages} {086006}
  (\bibinfo {year} {2018})},\ \Eprint {http://arxiv.org/abs/1801.00162}
  {arXiv:1801.00162 [hep-th]} \BibitemShut {NoStop}%
\bibitem [{\citenamefont {Knorr}(2021)}]{Knorr:2021slg}%
  \BibitemOpen
  \bibfield  {author} {\bibinfo {author} {\bibfnamefont {B.}~\bibnamefont
  {Knorr}},\ }\href {\doibase 10.21468/SciPostPhysCore.4.3.020} {\bibfield
  {journal} {\bibinfo  {journal} {SciPost Phys. Core}\ }\textbf {\bibinfo
  {volume} {4}},\ \bibinfo {pages} {020} (\bibinfo {year} {2021})},\ \Eprint
  {http://arxiv.org/abs/2104.11336} {arXiv:2104.11336 [hep-th]} \BibitemShut
  {NoStop}%
\bibitem [{\citenamefont {Kluth}\ and\ \citenamefont
  {Litim}(2022)}]{Kluth:2022vnq}%
  \BibitemOpen
  \bibfield  {author} {\bibinfo {author} {\bibfnamefont {Y.}~\bibnamefont
  {Kluth}}\ and\ \bibinfo {author} {\bibfnamefont {D.~F.}\ \bibnamefont
  {Litim}},\ }\href {\doibase 10.1103/PhysRevD.106.106022} {\bibfield
  {journal} {\bibinfo  {journal} {Phys. Rev. D}\ }\textbf {\bibinfo {volume}
  {106}},\ \bibinfo {pages} {106022} (\bibinfo {year} {2022})},\ \Eprint
  {http://arxiv.org/abs/2202.10436} {arXiv:2202.10436 [hep-th]} \BibitemShut
  {NoStop}%
\bibitem [{\citenamefont {Litim}\ and\ \citenamefont
  {Pawlowski}(2002{\natexlab{a}})}]{Litim:2002ce}%
  \BibitemOpen
  \bibfield  {author} {\bibinfo {author} {\bibfnamefont {D.~F.}\ \bibnamefont
  {Litim}}\ and\ \bibinfo {author} {\bibfnamefont {J.~M.}\ \bibnamefont
  {Pawlowski}},\ }\href {\doibase 10.1088/1126-6708/2002/09/049} {\bibfield
  {journal} {\bibinfo  {journal} {JHEP}\ }\textbf {\bibinfo {volume} {09}},\
  \bibinfo {pages} {049} (\bibinfo {year} {2002}{\natexlab{a}})},\ \Eprint
  {http://arxiv.org/abs/hep-th/0203005} {arXiv:hep-th/0203005} \BibitemShut
  {NoStop}%
\bibitem [{\citenamefont {Litim}\ and\ \citenamefont
  {Pawlowski}(2002{\natexlab{b}})}]{Litim:2002xm}%
  \BibitemOpen
  \bibfield  {author} {\bibinfo {author} {\bibfnamefont {D.~F.}\ \bibnamefont
  {Litim}}\ and\ \bibinfo {author} {\bibfnamefont {J.~M.}\ \bibnamefont
  {Pawlowski}},\ }\href {\doibase 10.1103/PhysRevD.66.025030} {\bibfield
  {journal} {\bibinfo  {journal} {Phys. Rev. D}\ }\textbf {\bibinfo {volume}
  {66}},\ \bibinfo {pages} {025030} (\bibinfo {year} {2002}{\natexlab{b}})},\
  \Eprint {http://arxiv.org/abs/hep-th/0202188} {arXiv:hep-th/0202188}
  \BibitemShut {NoStop}%
\bibitem [{\citenamefont {Litim}\ and\ \citenamefont
  {Pawlowski}(2002{\natexlab{c}})}]{Litim:2002hj}%
  \BibitemOpen
  \bibfield  {author} {\bibinfo {author} {\bibfnamefont {D.~F.}\ \bibnamefont
  {Litim}}\ and\ \bibinfo {author} {\bibfnamefont {J.~M.}\ \bibnamefont
  {Pawlowski}},\ }\href {\doibase 10.1016/S0370-2693(02)02693-X} {\bibfield
  {journal} {\bibinfo  {journal} {Phys. Lett. B}\ }\textbf {\bibinfo {volume}
  {546}},\ \bibinfo {pages} {279} (\bibinfo {year} {2002}{\natexlab{c}})},\
  \Eprint {http://arxiv.org/abs/hep-th/0208216} {arXiv:hep-th/0208216}
  \BibitemShut {NoStop}%
\bibitem [{\citenamefont {Folkerts}\ \emph {et~al.}(2012)\citenamefont
  {Folkerts}, \citenamefont {Litim},\ and\ \citenamefont
  {Pawlowski}}]{Folkerts:2011jz}%
  \BibitemOpen
  \bibfield  {author} {\bibinfo {author} {\bibfnamefont {S.}~\bibnamefont
  {Folkerts}}, \bibinfo {author} {\bibfnamefont {D.~F.}\ \bibnamefont {Litim}},
  \ and\ \bibinfo {author} {\bibfnamefont {J.~M.}\ \bibnamefont {Pawlowski}},\
  }\href {\doibase 10.1016/j.physletb.2012.02.002} {\bibfield  {journal}
  {\bibinfo  {journal} {Phys. Lett. B}\ }\textbf {\bibinfo {volume} {709}},\
  \bibinfo {pages} {234} (\bibinfo {year} {2012})},\ \Eprint
  {http://arxiv.org/abs/1101.5552} {arXiv:1101.5552 [hep-th]} \BibitemShut
  {NoStop}%
\bibitem [{\citenamefont {Bridle}\ \emph {et~al.}(2014)\citenamefont {Bridle},
  \citenamefont {Dietz},\ and\ \citenamefont {Morris}}]{Bridle:2013sra}%
  \BibitemOpen
  \bibfield  {author} {\bibinfo {author} {\bibfnamefont {I.~H.}\ \bibnamefont
  {Bridle}}, \bibinfo {author} {\bibfnamefont {J.~A.}\ \bibnamefont {Dietz}}, \
  and\ \bibinfo {author} {\bibfnamefont {T.~R.}\ \bibnamefont {Morris}},\
  }\href {\doibase 10.1007/JHEP03(2014)093} {\bibfield  {journal} {\bibinfo
  {journal} {JHEP}\ }\textbf {\bibinfo {volume} {03}},\ \bibinfo {pages} {093}
  (\bibinfo {year} {2014})},\ \Eprint {http://arxiv.org/abs/1312.2846}
  {arXiv:1312.2846 [hep-th]} \BibitemShut {NoStop}%
\bibitem [{\citenamefont {Vilkovisky}(1984)}]{Vilkovisky:1984st}%
  \BibitemOpen
  \bibfield  {author} {\bibinfo {author} {\bibfnamefont {G.~A.}\ \bibnamefont
  {Vilkovisky}},\ }\href {\doibase 10.1016/0550-3213(84)90228-1} {\bibfield
  {journal} {\bibinfo  {journal} {Nucl. Phys. B}\ }\textbf {\bibinfo {volume}
  {234}},\ \bibinfo {pages} {125} (\bibinfo {year} {1984})}\BibitemShut
  {NoStop}%
\bibitem [{\citenamefont {DeWitt}(1987)}]{DeWitt:1985sg}%
  \BibitemOpen
  \bibfield  {author} {\bibinfo {author} {\bibfnamefont {B.~S.}\ \bibnamefont
  {DeWitt}},\ }in\ \href@noop {} {\emph {\bibinfo {booktitle} {{Les Houches
  School of Theoretical Physics: Architecture of Fundamental Interactions at
  Short Distances}}}}\ (\bibinfo {year} {1987})\ pp.\ \bibinfo {pages}
  {1023--1058}\BibitemShut {NoStop}%
\bibitem [{\citenamefont {Branchina}\ \emph {et~al.}(2003)\citenamefont
  {Branchina}, \citenamefont {Meissner},\ and\ \citenamefont
  {Veneziano}}]{Branchina:2003ek}%
  \BibitemOpen
  \bibfield  {author} {\bibinfo {author} {\bibfnamefont {V.}~\bibnamefont
  {Branchina}}, \bibinfo {author} {\bibfnamefont {K.~A.}\ \bibnamefont
  {Meissner}}, \ and\ \bibinfo {author} {\bibfnamefont {G.}~\bibnamefont
  {Veneziano}},\ }\href {\doibase 10.1016/j.physletb.2003.09.020} {\bibfield
  {journal} {\bibinfo  {journal} {Phys. Lett. B}\ }\textbf {\bibinfo {volume}
  {574}},\ \bibinfo {pages} {319} (\bibinfo {year} {2003})},\ \Eprint
  {http://arxiv.org/abs/hep-th/0309234} {arXiv:hep-th/0309234} \BibitemShut
  {NoStop}%
\bibitem [{\citenamefont {Pawlowski}(2003)}]{Pawlowski:2003sk}%
  \BibitemOpen
  \bibfield  {author} {\bibinfo {author} {\bibfnamefont {J.~M.}\ \bibnamefont
  {Pawlowski}},\ }\href@noop {} {\  (\bibinfo {year} {2003})},\ \Eprint
  {http://arxiv.org/abs/hep-th/0310018} {arXiv:hep-th/0310018} \BibitemShut
  {NoStop}%
\bibitem [{\citenamefont {Donkin}\ and\ \citenamefont
  {Pawlowski}(2012)}]{Donkin:2012ud}%
  \BibitemOpen
  \bibfield  {author} {\bibinfo {author} {\bibfnamefont {I.}~\bibnamefont
  {Donkin}}\ and\ \bibinfo {author} {\bibfnamefont {J.~M.}\ \bibnamefont
  {Pawlowski}},\ }\href@noop {} {\  (\bibinfo {year} {2012})},\ \Eprint
  {http://arxiv.org/abs/1203.4207} {arXiv:1203.4207 [hep-th]} \BibitemShut
  {NoStop}%
\bibitem [{\citenamefont {Falls}(2025{\natexlab{b}})}]{Falls:2025tid}%
  \BibitemOpen
  \bibfield  {author} {\bibinfo {author} {\bibfnamefont {K.}~\bibnamefont
  {Falls}},\ }\href@noop {} {\  (\bibinfo {year} {2025}{\natexlab{b}})},\
  \Eprint {http://arxiv.org/abs/2503.05869} {arXiv:2503.05869 [hep-th]}
  \BibitemShut {NoStop}%
\bibitem [{\citenamefont {Aguilar-Gutierrez}\ \emph {et~al.}(2026)\citenamefont
  {Aguilar-Gutierrez}, \citenamefont {Ferrero}, \citenamefont {Hoehn},\ and\
  \citenamefont {Marchetti}}]{Aguilar-Gutierrez:2026svf}%
  \BibitemOpen
  \bibfield  {author} {\bibinfo {author} {\bibfnamefont {S.~E.}\ \bibnamefont
  {Aguilar-Gutierrez}}, \bibinfo {author} {\bibfnamefont {R.}~\bibnamefont
  {Ferrero}}, \bibinfo {author} {\bibfnamefont {P.~A.}\ \bibnamefont {Hoehn}},
  \ and\ \bibinfo {author} {\bibfnamefont {L.}~\bibnamefont {Marchetti}},\
  }\href@noop {} {\  (\bibinfo {year} {2026})},\ \Eprint
  {http://arxiv.org/abs/2607.21463} {arXiv:2607.21463 [hep-th]} \BibitemShut
  {NoStop}%
\bibitem [{\citenamefont {Bonanno}\ \emph {et~al.}(2026)\citenamefont
  {Bonanno}, \citenamefont {Ihssen},\ and\ \citenamefont
  {Pawlowski}}]{Bonanno:2025mon}%
  \BibitemOpen
  \bibfield  {author} {\bibinfo {author} {\bibfnamefont {A.}~\bibnamefont
  {Bonanno}}, \bibinfo {author} {\bibfnamefont {F.}~\bibnamefont {Ihssen}}, \
  and\ \bibinfo {author} {\bibfnamefont {J.~M.}\ \bibnamefont {Pawlowski}},\
  }\href {\doibase 10.21468/SciPostPhysCore.9.1.005} {\bibfield  {journal}
  {\bibinfo  {journal} {SciPost Phys. Core}\ }\textbf {\bibinfo {volume} {9}},\
  \bibinfo {pages} {005} (\bibinfo {year} {2026})},\ \Eprint
  {http://arxiv.org/abs/2504.03437} {arXiv:2504.03437 [hep-th]} \BibitemShut
  {NoStop}%
\bibitem [{\citenamefont {Ihssen}\ \emph {et~al.}(2025)\citenamefont {Ihssen},
  \citenamefont {Kapust},\ and\ \citenamefont {Pawlowski}}]{Ihssen:2025ybn}%
  \BibitemOpen
  \bibfield  {author} {\bibinfo {author} {\bibfnamefont {F.}~\bibnamefont
  {Ihssen}}, \bibinfo {author} {\bibfnamefont {R.}~\bibnamefont {Kapust}}, \
  and\ \bibinfo {author} {\bibfnamefont {J.~M.}\ \bibnamefont {Pawlowski}},\
  }\href@noop {} {\  (\bibinfo {year} {2025})},\ \Eprint
  {http://arxiv.org/abs/2510.26678} {arXiv:2510.26678 [hep-lat]} \BibitemShut
  {NoStop}%
\bibitem [{\citenamefont {Ihssen}\ \emph {et~al.}(2026)\citenamefont {Ihssen},
  \citenamefont {Kapust},\ and\ \citenamefont {Pawlowski}}]{Ihssen:2026njd}%
  \BibitemOpen
  \bibfield  {author} {\bibinfo {author} {\bibfnamefont {F.}~\bibnamefont
  {Ihssen}}, \bibinfo {author} {\bibfnamefont {R.}~\bibnamefont {Kapust}}, \
  and\ \bibinfo {author} {\bibfnamefont {J.~M.}\ \bibnamefont {Pawlowski}},\
  }\href@noop {} {\  (\bibinfo {year} {2026})},\ \Eprint
  {http://arxiv.org/abs/2603.03159} {arXiv:2603.03159 [hep-lat]} \BibitemShut
  {NoStop}%
\bibitem [{\citenamefont {Baldazzi}\ and\ \citenamefont
  {Falls}(2021)}]{Baldazzi:2021orb}%
  \BibitemOpen
  \bibfield  {author} {\bibinfo {author} {\bibfnamefont {A.}~\bibnamefont
  {Baldazzi}}\ and\ \bibinfo {author} {\bibfnamefont {K.}~\bibnamefont
  {Falls}},\ }\href {\doibase 10.3390/universe7080294} {\bibfield  {journal}
  {\bibinfo  {journal} {Universe}\ }\textbf {\bibinfo {volume} {7}},\ \bibinfo
  {pages} {294} (\bibinfo {year} {2021})},\ \Eprint
  {http://arxiv.org/abs/2107.00671} {arXiv:2107.00671 [hep-th]} \BibitemShut
  {NoStop}%
\bibitem [{\citenamefont {Falls}\ and\ \citenamefont
  {Ferrero}(2025)}]{Falls:2024noj}%
  \BibitemOpen
  \bibfield  {author} {\bibinfo {author} {\bibfnamefont {K.}~\bibnamefont
  {Falls}}\ and\ \bibinfo {author} {\bibfnamefont {R.}~\bibnamefont
  {Ferrero}},\ }\href {\doibase 10.1007/JHEP08(2025)173} {\bibfield  {journal}
  {\bibinfo  {journal} {JHEP}\ }\textbf {\bibinfo {volume} {08}},\ \bibinfo
  {pages} {173} (\bibinfo {year} {2025})},\ \Eprint
  {http://arxiv.org/abs/2411.00938} {arXiv:2411.00938 [hep-th]} \BibitemShut
  {NoStop}%
\bibitem [{\citenamefont {Falls}\ \emph {et~al.}(2026)\citenamefont {Falls},
  \citenamefont {Ferrero},\ and\ \citenamefont {Oglialoro}}]{Falls:2026nuh}%
  \BibitemOpen
  \bibfield  {author} {\bibinfo {author} {\bibfnamefont {K.}~\bibnamefont
  {Falls}}, \bibinfo {author} {\bibfnamefont {R.}~\bibnamefont {Ferrero}}, \
  and\ \bibinfo {author} {\bibfnamefont {G.}~\bibnamefont {Oglialoro}},\
  }\href@noop {} {\  (\bibinfo {year} {2026})},\ \Eprint
  {http://arxiv.org/abs/2607.06657} {arXiv:2607.06657 [hep-th]} \BibitemShut
  {NoStop}%
\bibitem [{\citenamefont {Pawlowski}\ and\ \citenamefont
  {Tr{\"a}nkle}(2024)}]{Pawlowski:2023dda}%
  \BibitemOpen
  \bibfield  {author} {\bibinfo {author} {\bibfnamefont {J.~M.}\ \bibnamefont
  {Pawlowski}}\ and\ \bibinfo {author} {\bibfnamefont {J.}~\bibnamefont
  {Tr{\"a}nkle}},\ }\href {\doibase 10.1103/PhysRevD.110.086011} {\bibfield
  {journal} {\bibinfo  {journal} {Phys. Rev. D}\ }\textbf {\bibinfo {volume}
  {110}},\ \bibinfo {pages} {086011} (\bibinfo {year} {2024})},\ \Eprint
  {http://arxiv.org/abs/2309.17043} {arXiv:2309.17043 [hep-th]} \BibitemShut
  {NoStop}%
\bibitem [{\citenamefont {Litim}\ and\ \citenamefont
  {Pawlowski}(2002{\natexlab{d}})}]{Litim:2001ky}%
  \BibitemOpen
  \bibfield  {author} {\bibinfo {author} {\bibfnamefont {D.~F.}\ \bibnamefont
  {Litim}}\ and\ \bibinfo {author} {\bibfnamefont {J.~M.}\ \bibnamefont
  {Pawlowski}},\ }\href {\doibase 10.1103/PhysRevD.65.081701} {\bibfield
  {journal} {\bibinfo  {journal} {Phys. Rev. D}\ }\textbf {\bibinfo {volume}
  {65}},\ \bibinfo {pages} {081701} (\bibinfo {year} {2002}{\natexlab{d}})},\
  \Eprint {http://arxiv.org/abs/hep-th/0111191} {arXiv:hep-th/0111191}
  \BibitemShut {NoStop}%
\bibitem [{\citenamefont {Groh}\ \emph {et~al.}(2011)\citenamefont {Groh},
  \citenamefont {Saueressig},\ and\ \citenamefont {Zanusso}}]{Groh:2011dw}%
  \BibitemOpen
  \bibfield  {author} {\bibinfo {author} {\bibfnamefont {K.}~\bibnamefont
  {Groh}}, \bibinfo {author} {\bibfnamefont {F.}~\bibnamefont {Saueressig}}, \
  and\ \bibinfo {author} {\bibfnamefont {O.}~\bibnamefont {Zanusso}},\
  }\href@noop {} {\  (\bibinfo {year} {2011})},\ \Eprint
  {http://arxiv.org/abs/1112.4856} {arXiv:1112.4856 [math-ph]} \BibitemShut
  {NoStop}%
\bibitem [{\citenamefont {Codello}\ and\ \citenamefont
  {Zanusso}(2013)}]{Codello:2012kq}%
  \BibitemOpen
  \bibfield  {author} {\bibinfo {author} {\bibfnamefont {A.}~\bibnamefont
  {Codello}}\ and\ \bibinfo {author} {\bibfnamefont {O.}~\bibnamefont
  {Zanusso}},\ }\href {\doibase 10.1063/1.4776234} {\bibfield  {journal}
  {\bibinfo  {journal} {J. Math. Phys.}\ }\textbf {\bibinfo {volume} {54}},\
  \bibinfo {pages} {013513} (\bibinfo {year} {2013})},\ \Eprint
  {http://arxiv.org/abs/1203.2034} {arXiv:1203.2034 [math-ph]} \BibitemShut
  {NoStop}%
\bibitem [{\citenamefont {Kluth}\ and\ \citenamefont
  {Litim}(2020)}]{Kluth:2019vkg}%
  \BibitemOpen
  \bibfield  {author} {\bibinfo {author} {\bibfnamefont {Y.}~\bibnamefont
  {Kluth}}\ and\ \bibinfo {author} {\bibfnamefont {D.~F.}\ \bibnamefont
  {Litim}},\ }\href {\doibase 10.1140/epjc/s10052-020-7784-2} {\bibfield
  {journal} {\bibinfo  {journal} {Eur. Phys. J. C}\ }\textbf {\bibinfo {volume}
  {80}},\ \bibinfo {pages} {269} (\bibinfo {year} {2020})},\ \Eprint
  {http://arxiv.org/abs/1910.00543} {arXiv:1910.00543 [hep-th]} \BibitemShut
  {NoStop}%
\bibitem [{\citenamefont {Demmel}\ \emph
  {et~al.}(2015{\natexlab{b}})\citenamefont {Demmel}, \citenamefont
  {Saueressig},\ and\ \citenamefont {Zanusso}}]{Demmel:2014hla}%
  \BibitemOpen
  \bibfield  {author} {\bibinfo {author} {\bibfnamefont {M.}~\bibnamefont
  {Demmel}}, \bibinfo {author} {\bibfnamefont {F.}~\bibnamefont {Saueressig}},
  \ and\ \bibinfo {author} {\bibfnamefont {O.}~\bibnamefont {Zanusso}},\ }\href
  {\doibase 10.1016/j.aop.2015.04.018} {\bibfield  {journal} {\bibinfo
  {journal} {Annals Phys.}\ }\textbf {\bibinfo {volume} {359}},\ \bibinfo
  {pages} {141} (\bibinfo {year} {2015}{\natexlab{b}})},\ \Eprint
  {http://arxiv.org/abs/1412.7207} {arXiv:1412.7207 [hep-th]} \BibitemShut
  {NoStop}%
\bibitem [{\citenamefont {Dietz}\ and\ \citenamefont
  {Morris}(2015)}]{Dietz:2015owa}%
  \BibitemOpen
  \bibfield  {author} {\bibinfo {author} {\bibfnamefont {J.~A.}\ \bibnamefont
  {Dietz}}\ and\ \bibinfo {author} {\bibfnamefont {T.~R.}\ \bibnamefont
  {Morris}},\ }\href {\doibase 10.1007/JHEP04(2015)118} {\bibfield  {journal}
  {\bibinfo  {journal} {JHEP}\ }\textbf {\bibinfo {volume} {04}},\ \bibinfo
  {pages} {118} (\bibinfo {year} {2015})},\ \Eprint
  {http://arxiv.org/abs/1502.07396} {arXiv:1502.07396 [hep-th]} \BibitemShut
  {NoStop}%
\bibitem [{\citenamefont {Safari}\ and\ \citenamefont
  {Vacca}(2016)}]{Safari:2016gtj}%
  \BibitemOpen
  \bibfield  {author} {\bibinfo {author} {\bibfnamefont {M.}~\bibnamefont
  {Safari}}\ and\ \bibinfo {author} {\bibfnamefont {G.~P.}\ \bibnamefont
  {Vacca}},\ }\href {\doibase 10.1007/JHEP11(2016)139} {\bibfield  {journal}
  {\bibinfo  {journal} {JHEP}\ }\textbf {\bibinfo {volume} {11}},\ \bibinfo
  {pages} {139} (\bibinfo {year} {2016})},\ \Eprint
  {http://arxiv.org/abs/1607.07074} {arXiv:1607.07074 [hep-th]} \BibitemShut
  {NoStop}%
\bibitem [{\citenamefont {Morris}\ and\ \citenamefont
  {Preston}(2016)}]{Morris:2016nda}%
  \BibitemOpen
  \bibfield  {author} {\bibinfo {author} {\bibfnamefont {T.~R.}\ \bibnamefont
  {Morris}}\ and\ \bibinfo {author} {\bibfnamefont {A.~W.~H.}\ \bibnamefont
  {Preston}},\ }\href {\doibase 10.1007/JHEP06(2016)012} {\bibfield  {journal}
  {\bibinfo  {journal} {JHEP}\ }\textbf {\bibinfo {volume} {06}},\ \bibinfo
  {pages} {012} (\bibinfo {year} {2016})},\ \Eprint
  {http://arxiv.org/abs/1602.08993} {arXiv:1602.08993 [hep-th]} \BibitemShut
  {NoStop}%
\bibitem [{\citenamefont {Wetterich}(2018)}]{Wetterich:2016ewc}%
  \BibitemOpen
  \bibfield  {author} {\bibinfo {author} {\bibfnamefont {C.}~\bibnamefont
  {Wetterich}},\ }\href {\doibase 10.1016/j.nuclphysb.2018.04.020} {\bibfield
  {journal} {\bibinfo  {journal} {Nucl. Phys. B}\ }\textbf {\bibinfo {volume}
  {931}},\ \bibinfo {pages} {262} (\bibinfo {year} {2018})},\ \Eprint
  {http://arxiv.org/abs/1607.02989} {arXiv:1607.02989 [hep-th]} \BibitemShut
  {NoStop}%
\bibitem [{\citenamefont {Falls}(2021)}]{Falls:2020tmj}%
  \BibitemOpen
  \bibfield  {author} {\bibinfo {author} {\bibfnamefont {K.}~\bibnamefont
  {Falls}},\ }\href {\doibase 10.1140/epjc/s10052-020-08803-0} {\bibfield
  {journal} {\bibinfo  {journal} {Eur. Phys. J. C}\ }\textbf {\bibinfo {volume}
  {81}},\ \bibinfo {pages} {121} (\bibinfo {year} {2021})},\ \Eprint
  {http://arxiv.org/abs/2004.11409} {arXiv:2004.11409 [hep-th]} \BibitemShut
  {NoStop}%
\bibitem [{\citenamefont {Codello}\ \emph {et~al.}(2009)\citenamefont
  {Codello}, \citenamefont {Percacci},\ and\ \citenamefont
  {Rahmede}}]{Codello:2008vh}%
  \BibitemOpen
  \bibfield  {author} {\bibinfo {author} {\bibfnamefont {A.}~\bibnamefont
  {Codello}}, \bibinfo {author} {\bibfnamefont {R.}~\bibnamefont {Percacci}}, \
  and\ \bibinfo {author} {\bibfnamefont {C.}~\bibnamefont {Rahmede}},\ }\href
  {\doibase 10.1016/j.aop.2008.08.008} {\bibfield  {journal} {\bibinfo
  {journal} {Annals Phys.}\ }\textbf {\bibinfo {volume} {324}},\ \bibinfo
  {pages} {414} (\bibinfo {year} {2009})},\ \Eprint
  {http://arxiv.org/abs/0805.2909} {arXiv:0805.2909 [hep-th]} \BibitemShut
  {NoStop}%
\bibitem [{\citenamefont {Fischer}\ and\ \citenamefont
  {Pawlowski}(2026)}]{Fischer:2026uni}%
  \BibitemOpen
  \bibfield  {author} {\bibinfo {author} {\bibfnamefont {C.~S.}\ \bibnamefont
  {Fischer}}\ and\ \bibinfo {author} {\bibfnamefont {J.~M.}\ \bibnamefont
  {Pawlowski}},\ }\href@noop {} {\  (\bibinfo {year} {2026})},\ \Eprint
  {http://arxiv.org/abs/2603.11135} {arXiv:2603.11135 [hep-ph]} \BibitemShut
  {NoStop}%
\end{thebibliography}%
%%%%%%%%%%%%%%%%%%%%

\end{document}